\documentclass{jfm}
\usepackage[none]{hyphenat}

\usepackage[hidelinks]{hyperref}
\usepackage[utf8]{inputenc}

\usepackage{graphicx}
\usepackage{epstopdf, epsfig}

\usepackage{mathchars}
\usepackage{amsmath,amsfonts,amssymb}

\usepackage{xcolor}

\usepackage{float}

\usepackage{caption}
\usepackage{subcaption}

\usepackage[sort&compress]{natbib}

\usepackage{multicol}
\usepackage{multirow}

\usepackage{setspace}

\newcommand{\LB}{\left(}
\newcommand{\RB}{\right)}

\usepackage{tikz}
\newcommand*\circled[1]{\tikz[baseline=(char.base)]{
            \node[shape=circle,draw,inner sep=2pt] (char) {#1};}}

\shorttitle{Steady-state Taylor bubble motion}
\shortauthor{H. A. Abubakar and O. K. Matar}

\title{Taylor bubble motion in stagnant and flowing liquids in vertical pipes.
Part I: Steady-states}

\author{H. A. Abubakar\aff{1,2}
 \and O. K. Matar\aff{1}
\corresp{\email{o.matar@imperial.ac.uk}} 
 }

\affiliation{\aff{1}Department of Chemical Engineering, Imperial College London, London SW7 2AZ, UK
\aff{2}Department of Chemical Engineering, Ahmadu Bello University, Zaria 810107, Nigeria}

\begin{document}

\maketitle

\begin{abstract}
Taylor bubbles are a feature of the slug flow regime in gas-liquid flows in vertical pipes. Their dynamics exhibits a number of transitions such as symmetry-breaking in the bubble shape and wake when rising in downward-flowing and stagnant liquids, respectively, as well as breakup in sufficiently turbulent environments. Motivated by the need to examine the stability of a Taylor bubble in liquids, a systematic numerical study of a steadily-moving Taylor bubble in stagnant and flowing liquids is carried out, characterised by a dimensionless inverse viscosity ($Nf$), and E\"{o}tv\"{o}s ($Eo$), and Froude ($Fr$) numbers based on the centreline liquid velocity, using a Galerkin finite-element method. A boundary-fitted domain is used to examine the dependence of the steady bubble shape on a wide range of $Nf$ and $Eo$.
Our analysis of the bubble nose and bottom curvatures shows that the intervals  $Eo = \left[ 20,30 \right)$ and $Nf=\left[60,80 \right)$ are the limits below which surface tension and viscosity, respectively, have a strong influence on the bubble shape. In the
interval  $Eo = \left(60,100 \right]$, all bubble features studied are weakly-dependent on surface tension. This is Part I of a two-part publication in which its companion paper \citep{Abubakar_Matar_2021} reports the results of a  linear stability analysis of the steady-states discussed herein. 
\end{abstract}

\section{Introduction} \label{sec:introduction}
Slug flow is a regime observed in gas-liquid flows in pipes, which is  of central importance to steam production in geothermal power plants, hydrocarbons production in oil wells and their transportation in pipelines, and emergency cooling of nuclear reactors \citep{Capponi_etal_2016,Taha_cui_2006,Fabre_Line_1992,
Mao_Dukler_1990}. This flow regime also features in geological systems such as volcanic eruptions \citep{Pering_McGonigle_2018}. 
In vertical pipes, slug flow exhibits pseudo-periodic rise of large bullet-shaped  {\it Taylor bubbles} separated by liquid slugs.  

The starting point for understanding slug flow in vertical pipes is elucidating the behaviour of a single Taylor bubble rising through a liquid, which is governed by the interaction of gravitational, interfacial, viscous, and inertial forces parameterised by a number of dimensionless groups; these include the inverse viscosity, $Nf$, E\"{o}tv\"{o}s, $Eo$, and Froude, $Fr$, numbers, defined as
\begin{align}
Nf = \frac{\rho \left( g D^3 \right) ^{\frac{1}{2}}}{\mu}, 
\qquad
Eo = \frac{\rho g D^2}{\gamma},  
\qquad  Fr = \frac{u}{\sqrt{g D}}, 
\end{align}
where $\rho$, $\mu$, and $u$ denote the density, dynamic viscosity, and a characteristic liquid speed, respectively; $D$ is the pipe diameter while $g$ is the acceleration due to gravity.  
The Froude number may be based either on the bubble rise speed, the mean speed or the pipe centreline velocity of the flowing liquid, which are represented by $U_b$, $\bar{U}_L$, and $U_m$, respectively.
Other dimensionless groups used commonly to characterise Taylor bubble behaviour are the Reynolds number, $Re$, Morton number, $Mo$, and Archimedes number, $Ar$, respectively given by:
\begin{align}
Re = Fr Nf, \qquad
Mo = {Eo^3}{Nf^{-4}}, \qquad
Ar = {Nf^2}, \qquad
\end{align}
and we can also distinguish the Reynolds numbers that are based on the average liquid speed, $Re_L$, and the bubble rise speed, $Re_b$: 
\begin{align}
Re_L = \bar{U}_L Nf,
\qquad
Re_b = U_b Nf.
\end{align} 

A Taylor bubble generally exhibits topological symmetry and its shape can be sectioned into three distinct regions corresponding to the bubble `nose', `body', and `bottom', with each region having specific features used for its characterisation. The nose region is nearly hemispherical and is characterised by its frontal radius of curvature, the magnitude of the axial component of velocity at its tip, which is the bubble rise speed, and the maximum distance ahead of the bubble nose beyond which the bubble impact is no longer felt. The body region is nearly tubular and surrounded by a thin liquid film that can be divided into developing and fully-developed parts. Features such as the length of the developing region, the film thickness, and the velocity profile of the fully-developed film, and the wall shear stress in the film, are all used to characterise the body region. Lastly, in the bottom region, the characterising features are the shape, which could be concave or convex, the radius of curvature, the maximum distance beyond which the impact of the bubble is no longer felt, and the length, volume, and the nature of the flow pattern in the wake region (if present).

Because of the numerous applications of slug flow, extensive experimental \citep{Griffith_Wallis_1961,Moissis_Griffith_1962,White_Beardmore_1962,Nicklin_etal_1962,Campos_Carvalho_1988,Bugg_Saad_2002,Nogueira_etal_2006,
Llewellin_etal_2012,Rana_etal_2015,Pringle_etal_2015,Fershtman_etal_2017}, theoretical  \citep{Dumitrescu_1943,Brown_1965,Collins_etal_1978,Funada_etal_2005,Fabre_2016}, and numerical  \citep{Mao_Dukler_1990,Mao_Dukler_1991,Bugg_Saad_2002,Taha_cui_2006,Lizarraga-Garcia_etal_2017,Anjos_etal_2014,Taha_cui_2002,Lu_Prosperetti_2009,Kang_2010} studies have been carried out to determine some of the features highlighted for all the aforementioned topological regions. 
The rise speed is the most investigated and significant feature in Taylor bubble dynamics. For sufficiently long bubbles, typically several pipe diameters in length, the bubble rise speed is independent of the bubble length \citep{Polonsky_etal_1999,Mao_Dukler_1989,Nicklin_etal_1962,Griffith_Wallis_1961}. Neglecting the effect of surface tension and assuming an inviscid flow around the bubble nose, \cite{Dumitrescu_1943} and \cite{Davies_Taylor_1950} have shown independently that the rise speed in a stagnant liquid is given by:
\begin{equation}
u_b = C_0\sqrt{g D},			\label{eq:lr_bubble_rise_vel_stag}	
\end{equation}
where $u_b$ denotes the dimensional bubble rise speed, and $C_0$ is a dimensionless proportionality constant. 
From  \eqref{eq:lr_bubble_rise_vel_stag}, the Froude number based on the bubble rise speed is a constant and equals $C_0\approx 0.351$ \citep{Dumitrescu_1943}.

\cite{White_Beardmore_1962} generated a flow map depicting regimes where the effects of surface tension, inertia, viscous or a combination of these forces on a bubble rising in a stagnant liquid can be neglected. It was established that beyond $Eo > 70$ and $Mo > 3 \times 10^5$, in an `inertia regime', surface tension and viscosity have no significant influence on the bubble rise speed, and the assumptions underlying the analytical solutions of \cite{Dumitrescu_1943} and \cite{Davies_Taylor_1950} are valid.
Later experimental, theoretical and numerical studies \citep{ Kang_2010,Lu_Prosperetti_2009,Nicklin_Yannitel_1987,Zukoski_1966,Brown_1965,Goldsmith_Mason_1962} have provided further insights into  the role of surface tension and viscosity on the rise speed in both inertia and non-inertia regimes through their influence on the radius of curvature of the bubble nose. 
Using a large pool of experimental data for $U_b$ in stagnant liquids, \cite{Viana_etal_2003} developed a correlation, recently modified by \cite{Lizarraga-Garcia_etal_2017},  for the effect of $Eo$ and $Nf$ on the rise speed taking into account pipe inclination. 

For a Taylor bubble rising in a flowing liquid, \cite{Nicklin_etal_1962} proposed a correlation, corroborated by theoretical investigations \citep{Bendiksen_1985,Collins_etal_1978}, for upward flowing liquid, which relates $U_b$ to $\bar{U}_L$ 
\begin{equation}
U_b = C_1 \bar{U}_L + C_0, \label{eq:lr_bubble_rise_vel_flow}	
\end{equation} 
with $C_0$ and  $\bar{U}_L$ retaining their earlier definitions and $C_1$ represents a dimensionless constant whose value depends on the velocity profile of the flowing liquid and is equal to the ratio of the maximum to mean liquid velocity  \citep{Bendiksen_1985,Collins_etal_1978,Clift_etal_1978, Nicklin_etal_1962}. For turbulent flow, $C_1 \approx 1.2$ increasing with decreasing $Re_L$ approaching $C_1 \approx 1.9$ at  $Re_L = 100$  \citep{Nicklin_etal_1962}. 
Other important features that have been studied experimentally, theoretically, and numerically are the film thickness and length of developing film \citep{Llewellin_etal_2012,Goldsmith_Mason_1962,Brown_1965,Batchelor_1967,
Nogueira_etal_2006a,Araujo_etal_2012,Kang_2010},  and wake 
\citep{Moissis_Griffith_1962,Maxworthy_1967,Campos_Carvalho_1988,Pinto_etal_1998,
Nogueira_etal_2006,Araujo_etal_2012}, and wall stress features \citep{Nogueira_etal_2006a,Araujo_etal_2012,Feng_2008}. 

Despite the volume of previous research, there is still a need for a systematic study 
of the influence of the fluid properties and flow conditions on the bubble behaviour. 
This is motivated by the experimental evidence for Taylor bubble feature transitions, such as a change in the flow pattern in the wake region and bubble shape from symmetric to asymmetric in downward liquid flow, as well as bubble breakup under certain conditions (most likely caused by fluctuations in a turbulent environment).  The critical conditions at which these transitions occur, and their underlying mechanisms, can be understood by examining the stability of the  axisymmetric steady-states for the corresponding parameter values. 

In the present work, we calculate the steady shape of axisymmetric Taylor bubbles, and their associated flow fields moving in stagnant and downward-flowing liquids in vertical pipes, characterised by $Nf$, $Eo$, and $Fr$. 
Plots showing the influence of these 
parameters on the Taylor bubble shape are presented and the results of the associated impact on the  steady-state features characterising the three distinct bubble regions, nose, body, and bottom,  discussed above. Comparisons are made between our numerical predictions and those based on theoretical analysis or empirical correlations; 
insights into the physical mechanisms governing the observed influence are provided.
In a companion paper \citep{Abubakar_Matar_2021}, Part II of this two-part study, the linear stability of these steady-state solutions is examined together with an energy analysis to pinpoint the destabilising mechanisms. 

The rest of this paper is organised as follows. Section \ref{sec:problem_formulation} is devoted to details of the problem formulation and the numerical simulation strategy based on the use of a finite-element technique. The results of the steady-state simulations in stagnant and downward-flowing liquids are discussed in Sections \ref{sec:steady_state_characterisations_stag} and \ref{sec:steady_state_characterisations_flow}, respectively. Finally, in section \ref{sec:steady_state_summary}, concluding remarks are provided.

\section{Problem formulation}\label{sec:problem_formulation}
\subsection{Governing equations}
\begin{figure}
\centering
	\includegraphics[width = 0.3\linewidth]{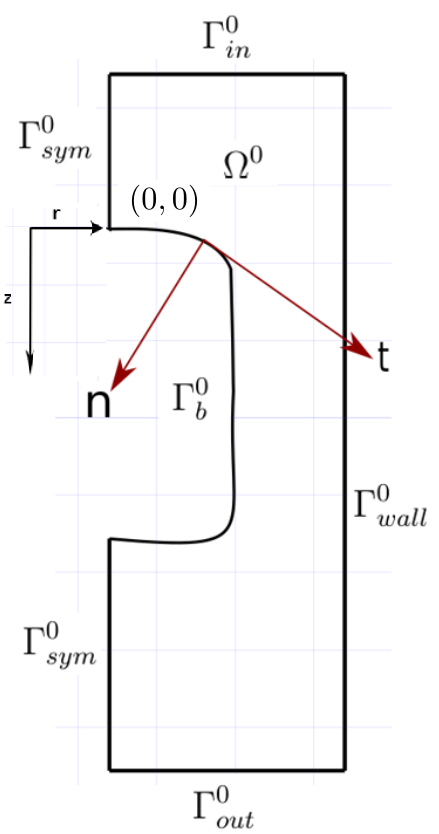}
	\caption{\label{Fig:axisymmetric_2D-domain} Schematic of the domain used to model the steady motion of an axisymmetric Taylor bubble}
\end{figure}

We consider the motion of an axisymmetric Taylor bubble of  volume, ${v_b}$, 
moving at a velocity of magnitude $u_b$ through an incompressible fluid of density $\rho$, viscosity $\mu$, and  interfacial tension $\gamma$ in a vertically-oriented, circular pipe of diameter $D$; $v_b$, $u_b$, and $\gamma$ are considered to be constants. 
In addition, we also assume that the density, $\rho_g$, and viscosity, $\mu_g$, of the gas bubble are very small as compared to those of the liquid, and that the pressure within the bubble, $p_b$, is also a constant; hence, the influence of the gas phase is restricted to the interface separating the liquid and gas phases 
\citep{Zhou_Dusek_2017,Fraggedakis_etal_2016,Tsamopoulos_etal_2008,Bae_Kim_2007,Kang_2010,Lu_Prosperetti_2009,Feng_2008}. 
A cylindrical coordinate system, $\left(r,\theta,z \right)$, is adopted so that the coordinates along and perpendicular to the axis of symmetry are $z$ and $r$, respectively, with the interface located at $(r,z)=\Gamma_b^0$, and the $z$ origin chosen to coincide with the bubble nose, as shown in Figure \ref{Fig:axisymmetric_2D-domain}. 

The Navier-Stokes and continuity equations which govern the bubble motion are rendered dimensionless by scaling the length, velocity, and pressure on $ D, \sqrt{gD} \; \mbox{and} \;  \rho g D  $, respectively. These equations, expressed in a frame of reference translating with the velocity $\mathbf{u}_b = - U_b \mathbf{i}_z$ of the bubble nose, wherein $U_b=u_b/\sqrt{gD}$, are written compactly in dimensionless forms as: 
 \begin{equation}
\frac{\partial \mathbf{u}}{\partial t} + \left(\mathbf{u} \cdot \nabla \right)\mathbf{u}
-
\nabla \cdot \mathbf{T} 
= 0, 
\label{eq:momentum}
 \end{equation}
 \begin{equation}
 \nabla \cdot \mathbf{u} = 0,  
 \label{eq:continuity} 
 \end{equation}
where $\mathbf{u}$  is the fluid velocity vector in the moving frame of reference, $t$ denotes time, $\nabla $ is the gradient operator, and $ {\mathbf{T}}$ is the stress tensor:
\begin{equation}
\mathbf{T}  = - {p} \mathbf{I} +   {Nf}^{-1}\left( \nabla \mathbf{u} + \nabla \mathbf{u}^T \right)
\label{eq:stress_tensor},
\end{equation}
in which $p$ represents the dynamic pressure, and $\mathbf{I} $ unit tensor.

In order to impose boundary conditions on the solutions of equations   \eqref{eq:momentum}-\eqref{eq:stress_tensor}, the boundary of the domain, 
$\Gamma^0$ is divided into $\Gamma^0_{\rm in}, \; \Gamma^0_{\rm out}, \; \Gamma^0_{\rm wall}$, $\Gamma^0_{\rm sym}$, and $\Gamma^0_b$, as shown in Figure \ref{Fig:axisymmetric_2D-domain}, which represent the domain inlet and outlet, the wall, and the symmetry axis, respectively. At the wall, no-slip and no-penetration boundary conditions are imposed,
%
\begin{equation}
\mathbf{u} = -\mathbf{u}_b, \quad \mathrm{on \quad   \Gamma^0_{wall} }, \label{eq:wall_bc}
\end{equation} 
while at the inlet, prescribed values, $\mathbf{u}_{in}$ are specified for the velocity: 
%
\begin{equation}
\mathbf{u} = \mathbf{u}_{in} - \mathbf{u}_b \quad \mathrm{on \quad   \Gamma^0_{in} }.  \label{eq:inlet_bc}
\end{equation} 
Along $\Gamma^0_{\rm out}$, we impose an outlet condition:
%
\begin{equation}
\mathrm{ \mathbf{n}\cdot\mathbf{T} } = 0.  \label{eq:oulet_bc}
\end{equation}
Finally, at the interface, we impose the normal stress, tangential stress, and kinematic boundary conditions, expressed respectively by 
\begin{equation}
{ \mathbf{n}\cdot\mathbf{T}\cdot{\mathbf{n}} + P_b - z  - {Eo}^{-1}} \kappa = 0,  \label{eq:normal_stress_bc}
\end{equation}
%
\begin{equation}
\mathbf{n}\cdot\mathbf{T} \times \mathbf{n}   = \mathbf{0},  \label{eq:tangential_stress_bc}
\end{equation}
%
\begin{equation}
{\frac{d\mathbf{r}_b}{dt} \cdot \mathbf{n} -\mathbf{u} \cdot \mathbf{n} }  =0,   \label{eq:kinematic_bc}
\end{equation}
where $ \kappa $ is the curvature of the interface, $ P_b = p_b/\rho g D $ denotes the dimensionless bubble pressure, $\mathbf{r}_b(t)$ represents the position vector for the location of the interface $\Gamma^0_b $, and ${\bf n}$ and ${\bf t}$ correspond to the outward-pointing unit normal and the tangent vectors to the interface, respectively. 
The $z$ term in the normal stress condition, given by equation $\left( \ref{eq:normal_stress_bc} \right) $, corresponds to the hydrostatic pressure.

In order to determine the dimensionless bubble pressure, $P_b$, a constraint of constant dimensionless bubble volume, $V_b=v_b/D^3$, is imposed: 
\begin{align}
V_b  + \frac{1}{3} \oint_{\Gamma^0_b} \left[ \mathbf{r}_b\cdot\mathbf{n} \right] d \Gamma^0_b = 0. \label{eq:volume_constraint}
\end{align}
In order to obtain a solution for the shape of the bubble of volume $V_b$, speed $U_b$, and pressure $P_b$ associated with its steady motion through a liquid of dimensionless velocity $U_m$, for given $Nf$ and $Eo$, we implemented a technique based on the kinematic update of the interface shape with an implicit treatment of the curvature \citep{Slikkeveer_etal_1996}; the numerical procedure is described next. 

\subsection{Numerical method}\label{sec:numerical_simulation}
The steady-state versions of the governing equations and boundary conditions given by \eqref{eq:momentum}-\eqref{eq:volume_constraint} are solved using a \text{consistent penalty} Galerkin finite-element method implemented within \text{FreeFem++} \citep{Hecht_2012} based on the standard Taylor-Hood element and piecewise quadratic element approximations for the flow field variables and interface deformation magnitude, respectively. The system of partial differential equations $\left( \ref{eq:momentum}\right)-\left(\ref{eq:continuity} \right)$ subject to the boundary conditions $\left( \ref{eq:wall_bc}\right)-\left(\ref{eq:volume_constraint} \right)$ are transformed into their weak forms, the dependent variables in the equations approximated using suitable basis functions. The computational domain is divided into subdomains around which the approximated variables are defined to obtain a set of nonlinear algebraic relations among the unknown parameters of the approximations. 
Due to the system nonlinearity, the set of equations was solved using Newton's method. In the determination of the interface shape, kinematic update is used based on a pseudo-time-step technique, allowing for the gradual satisfaction of the no-penetration condition on the interface.  

The numerical solution begins by providing an initial guess for the bubble steady speed, $U_b$, the flow field variables, $\LB \mathbf{u}, p \RB $, and position vector of the interface, $ \mathbf{r}_b$. For the first simulation carried out, $Nf = 40$, $Eo = 60$, and $U_{m} = 0$, corresponding to bubble rise in a stagnant liquid, $U_b $  was initially taken to be $0.35$ and the bubble interface position was assumed to be described by a quarter-circle top, a cylindrical body, and a quarter-circle bottom. The initial guess for the flow field was then obtained by solving the Stokes equation in the domain formed by the assumed bubble interface. For subsequent simulations, the previous steady-state solutions for the condition closest to the new condition was used as an initial guess.  

With a known initial guess, the solution proceeded in three stages: solution for the variables, steady bubble speed determination, and then domain deformation. In the variable solution stage,  the resulting system of linear equations in the Newton method is solved using MUltifrontal Massively Parallel sparse direct Solver (MUMPS)  to obtain updated values for the velocity, pressure, and interface deformation magnitudes. The updated velocity field is then transformed from a moving to  a fixed frame of reference from which the axial velocity at the bubble nose is extracted and set as the steady bubble speed. Using the interface deformation magnitude obtained in the variable solution stage, the magnitude of the deformation of the domain is then determined. For all other nodes in the domain, the size of their deformations is adapted to that of the interface in a way that ensures that the mesh quality does not degrade rapidly by assuming that the computational mesh is an elastic body whose interior deforms in response to the boundary deformation. This assumption forms the basis of the Elastic Mesh Update Method  of treating interior nodes which involves solving a linear elasticity equation for the mesh deformation subject to the boundary conditions that equals the desired deformation on the boundaries \citep{Johnson_Tezduyar_1994,Ganesan_Tobiska_2008}.
The iterative process is halted when the interface position vector and the values of the flow field variables, steady bubble speed and pressure no longer change, and the no-penetration condition is satisfied. The implementation details are described in \cite{Abubakar_2019}.

The numerical method was validated by simulating the experiment of \cite{Bugg_Saad_2002} where the velocity field around a Taylor bubble rising in a stagnant olive oil in a pipe of diameter $19$ mm was measured using Particle Image Velocimetry (PIV) at five different positions. The fluid properties used in the experiment and the corresponding dimensionless parameters are given in Table \ref{tab:ss_validation_fluid_ppt}.
\begin{table}
\centering
	\caption{Dimensionless parameters corresponding to the fluid properties used to validate the numerical predictions against the experimental work of \cite{Bugg_Saad_2002}. \\}
  \label{tab:ss_validation_fluid_ppt}
\begin{tabular}{cccccccc} 
\multicolumn{4}{ c }{Fluid properties} & \multicolumn{4}{ c }{ Dimensionless parameters} \\ \hline \\
$ \rho \left( {\rm kgm^{-3}}\right)$ & ${\mu} \left({\rm Nsm^{-2}}\right)$ & ${\gamma} \left( {\rm Jm^{-2}}\right)$ & ${v_b} \left( {\rm m^{3}}\right)$ & $Nf$ & $Eo$ & $U_{m}$ & $H_b$\\ 
911 & 84 $\times 10^{-3}$ & 3.28 $\times 10^{-2}$ & 10 $\times 10^{-6}$  &88.95 & 98.33 & 0 & 2.00\\  
\end{tabular}
\end{table} 
In this table, $H_b$ denotes the dimensionless height of a cylinder of the same diameter as that of the pipe used in the experiment that has the same volume as the gas phase, which is the aspect ratio for the bubble.

\begin{figure}
\centering
	\begin{subfigure}{1.0\linewidth}
	  \centering\includegraphics[width = 0.80\linewidth]{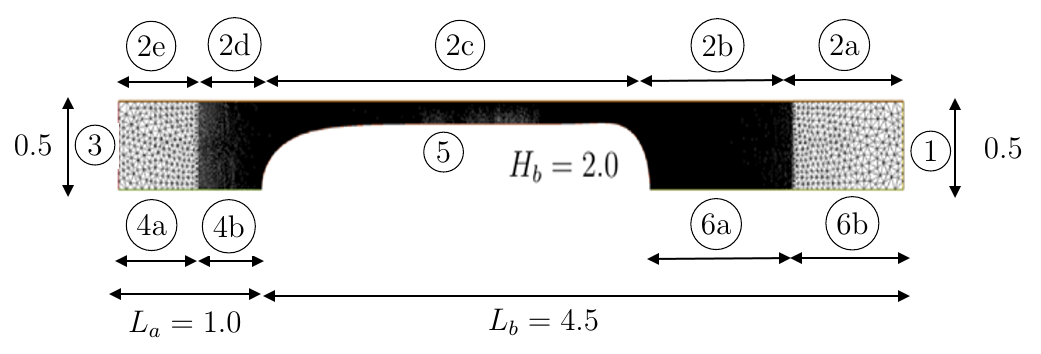}
	\caption{}
	\label{fig:ss_mesh_structure}  
	\end{subfigure}
	\qquad
	\begin{subfigure}{1.0\linewidth}
	  \centering\includegraphics[width = 0.65\linewidth,angle=180]{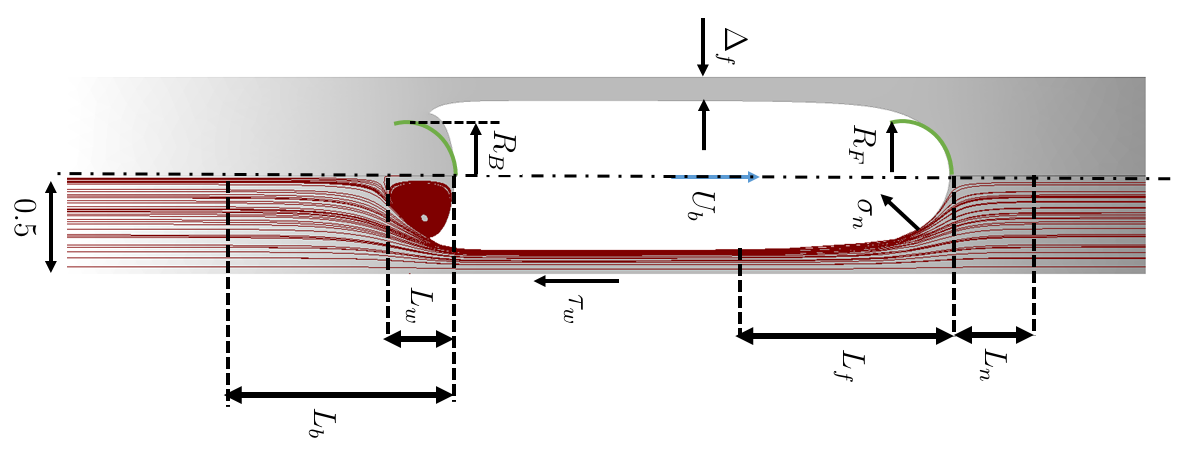}
	\caption{}
	\label{fig:ss_hydrodynamic_features}  
	\end{subfigure}
		\caption{Mesh structure around the Taylor bubble, (a), with information provided in Table \ref{tab:ss_mesh_structure_info};  schematic representation of the main hydrodynamic features of the bubble considered in the present work, (b). All features are dimensionless and are based on the characteristic scales stated in section \ref{sec:problem_formulation}. In (b),  
	$U_b$ represents the bubble rise speed, $R_F$ the average radius of curvature of the bubble nose, $L_n$ and $L_f$ the flow stabilisation lengths ahead of bubble nose and in the liquid film region, respectively; $\Delta_f$ is the equilibrium film  thickness, $\tau_w$ is the wall shear stress, $\sigma_n$ is the normal stress at the interface; $R_B$ denotes the average radius of curvature of the bubble bottom, and $L_w$ and $L_b$ the length of the wake and the flow stabilisation length below the bubble bottom, respectively.}
\end{figure}

Numerical solutions were obtained for the parameters listed in Table \ref{tab:ss_validation_fluid_ppt} using  the initial bubble shape  shown in Figure \ref{fig:ss_mesh_structure}. The mesh is boundary-fitted and structured such that regions around the bubble are finely resolved. Table \ref{tab:ss_mesh_structure_info} shows the dimensionless edge lengths of elements used for the labelled boundary regions around the  domain (see Figure \ref{fig:ss_mesh_structure}). For immediate regions around the bubble, $\left( \circled{2b}, \circled{2c}, \circled{2d}, \circled{4b} \text{and} \circled{6a}\right)$, triangles with smaller edge lengths, leading to finer mesh, are used. At the interface, we maintained a fixed range of triangles $\left(700 -800 \right)$, while the distribution  and range of triangle edge lengths are left for the automatic mesh adaptor to determine. 
\begin{table}
\centering
	\caption{Number and length of the edge of triangle elements at different sections of the domain boundaries used in mesh generation (see Figure \ref{fig:ss_mesh_structure})\\}
  \label{tab:ss_mesh_structure_info}
\begin{tabular}{cccc} 
Boundary region(s) & Triangle edge length & Boundary length &  Number of triangles \\ 
\hline 
\circled{1} and \circled{3}  & 0.5 & 0.042& 12 \\  
\circled{2a} and \circled{6b}  & varies & 0.042& varies \\  
\circled{2b} and \circled{6a}  & 1.0 & 0.004 & 250 \\  
\circled{2c}   & varies & 0.004 & varies \\ 
\circled{2d} and \circled{4b}  & 0.45 & 0.007 & 64 \\  
\circled{2e} and \circled{4a}  & 0.55 & 0.042 & 13 \\  
\circled{5}  & varies & varies & 700-800 \\
\end{tabular}
\end{table}
To guide the distribution and triangles edge length range,  the interface mesh is adapted to the curvature of the interface and flow field, and a maximum edge length of $0.06$ was stipulated. Note that in Table  \ref{tab:ss_mesh_structure_info}, the number of triangles on the boundary is calculated by  dividing the boundary length by the boundary corresponding  triangle edge length.

For the validation and the results to be discussed, a fixed dimensionless distance of $ L_a = 1.0$ and $ L_b = 4.5$ are maintained ahead and below the bubble nose, respectively. These distances and mesh structure were tested to ensure that the inlet and outlet boundaries as well as the mesh  have insignificant influence on the steady-state results.  The converged steady-state bubble shape for the validation and the flow patterns around it are shown in Figure 
\ref{fig:ss_hydrodynamic_features}. 
It should be noted that for domain length in which the inlet and outlet boundaries have no influence on the steady-state results, periodic conditions can be imposed in place of boundary conditions \eqref{eq:inlet_bc} and \eqref{eq:oulet_bc}. This approach was used by \cite{Lu_Prosperetti_2009} in their numerical study of Taylor bubble dynamics and can easily be implemented within FreeFem++. 

\begin{figure}
    \centering
    \begin{subfigure}{0.45\linewidth}
        \includegraphics[width=\linewidth]{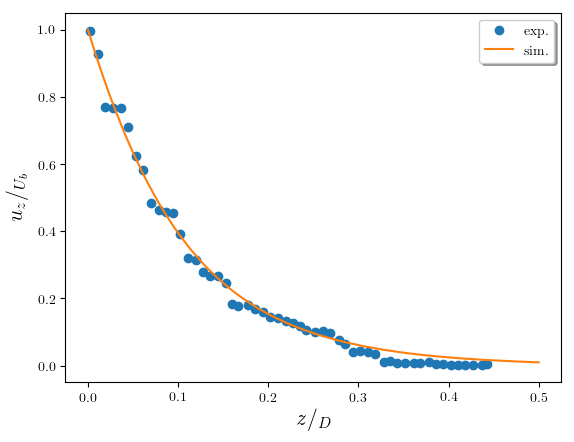}
        \caption{ }
        \label{fig:ss_pipe_axis}
    \end{subfigure}
    \quad 
    ~ 
    \begin{subfigure}{0.45\linewidth}
        \includegraphics[width=\linewidth]{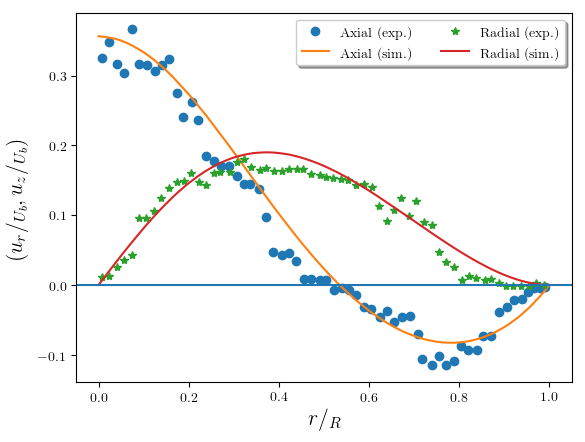}
        \caption{}
         \label{fig:ss_axial_radial_ahead}
    \end{subfigure}
    \begin{subfigure}{0.45\linewidth}
        \includegraphics[width=\linewidth]{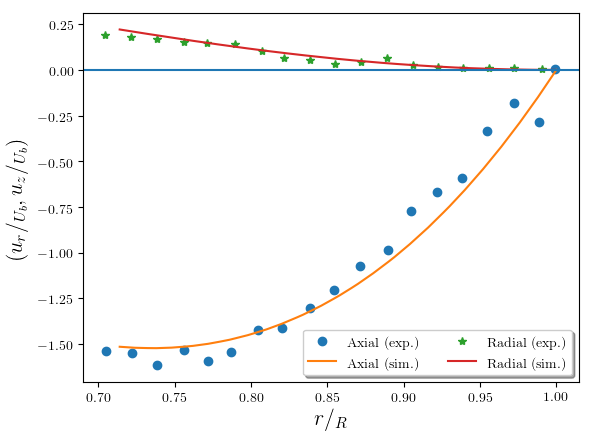}
        \caption{ }
        \label{fig:ss_axial_radial_dev_film}
    \end{subfigure}
    \quad 
    ~ 
    \begin{subfigure}{0.45\linewidth}
        \includegraphics[width=\linewidth]{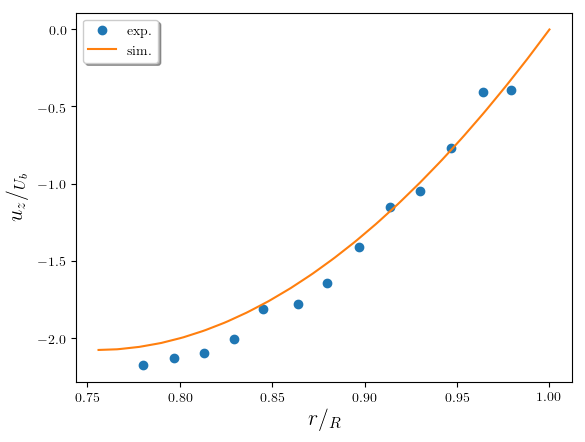}
        \caption{}
         \label{fig:ss_axial_full_film}
    \end{subfigure}
	\begin{subfigure}{0.45\linewidth}
        \includegraphics[width = \linewidth]{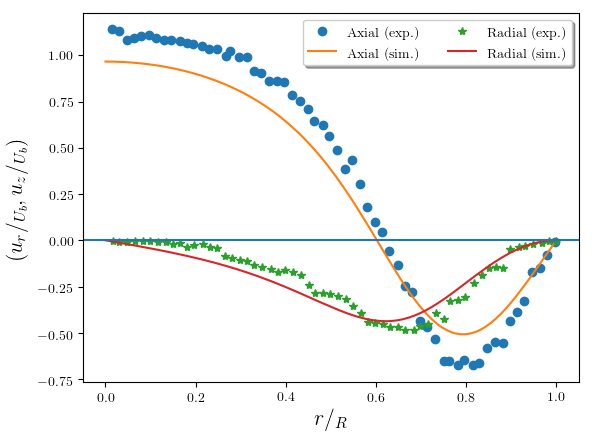}
        \caption{}
         \label{fig:ss_axial_radial_wake}
    \end{subfigure}
	\caption{Validation of the numerical predictions (lines) for the velocity profiles for the positions indicated in Fig. 2 against the PIV measurements (symbols) of \cite{Bugg_Saad_2002};
(a) axial velocity component, $u_z$, along the pipe axis \textcolor{black}{(position 1)}; (b) axial, $u_z$,  and radial, $u_r$, velocity components at $\frac{z}{D} = 0.111$ ahead of the bubble nose \textcolor{black}{(position 2)}; (c) axial and radial velocity components in the developing film at $\frac{z}{D} = 0.504$ below the bubble nose \textcolor{black}{(position 3)} and (d) axial velocity component in the fully-developed film \textcolor{black}{(position 4)}; (e)
axial and radial components of velocity at distance $\frac{z}{D} = 0.20$ below the bubble bottom \textcolor{black}{(position 5)}.}
     \label{fig:ss_velocity_ahead}
\end{figure}

The predicted dimensionless bubble rise speed is $0.2928$, corresponding to a deviation of $3.4\%$ from the experimentally measured value of $0.303$. Further comparisons with the experiment were carried out using the flow field results at five measurement positions  around the bubble.
Ahead of the bubble, velocity measurements were taken along the pipe axis and  in the radial direction at an axial distance of $0.111D$. Figures \ref{fig:ss_pipe_axis} and \ref{fig:ss_axial_radial_ahead} show the velocity profiles for these two locations and are well predicted by our simulation.
Figure \ref{fig:ss_axial_radial_dev_film} compares the velocity measurement taken at an axial distance of $0.504D$ below the bubble nose. At this point, the magnitude of the radial velocity component is still developing. When the velocity in the film is fully-developed, the magnitude of the radial velocity at all points in the radial direction is approximately zero. 
By progressively plotting the radial velocity profile at various points below the bubble nose, a point is reached at which the radial velocity becomes zero. The axial velocity profile at this location is shown in Figure \ref{fig:ss_axial_full_film} and the dimensionless film thickness was measured to be  $ 0.1235 $. Although no experimental measurement of the film thickness was reported in \cite{Bugg_Saad_2002}, the deviation of the numerical simulation results from the theoretical estimated value of \cite{Brown_1965} using \eqref{eq:ss_film_region_equilibrium_film_thickness}, which predicts the film thickness  to be $ 0.1193 $, is $ 3.52 \% $. 

As the liquid emerges from the falling film region into the wake of the bubble, the radial component of its velocity reappears in order to redirect the liquid from the film back towards the center of the pipe. Figure \ref{fig:ss_axial_radial_wake} shows the velocity profile in the wake of the Taylor
bubble at an axial distance of 
$0.2D$ below the bubble bottom.    
While the radial component of the experimental velocity profile is reasonably well predicted by the numerical simulation, it is obvious that there are larger discrepancies associated with the prediction of the axial velocity. We note that a similarly large deviation of the axial velocity was observed by \cite{Bugg_Saad_2002,Lu_Prosperetti_2009} in their numerical simulations of the same experiment. We therefore agree with \cite{Lu_Prosperetti_2009} that it is  possible that the error bars associated with the experimental data for the wake region may be relatively large. 

\section{Steady-state bubble rise in stagnant liquids ($U_m=0$)} \label{sec:steady_state_characterisations_stag}
In this section, we present a discussion of our  parametric study of a Taylor bubble of dimensionless volume   $V_B = 0.3389 \pi $, equivalent to aspect ratio $H_B = 1.3556$, in a stagnant liquid ($U_m = 0$). The effects of $Nf$ and $Eo$ on the hydrodynamic features of a steadily rising Taylor bubble depicted in Figure \ref{fig:ss_hydrodynamic_features} are examined. 
\subsection{Qualitative analysis of steady-state shapes and flow field }
\begin{figure}
\centering
    \begin{subfigure}{\linewidth}
	\centering\includegraphics[width = 0.65\linewidth, angle=0]{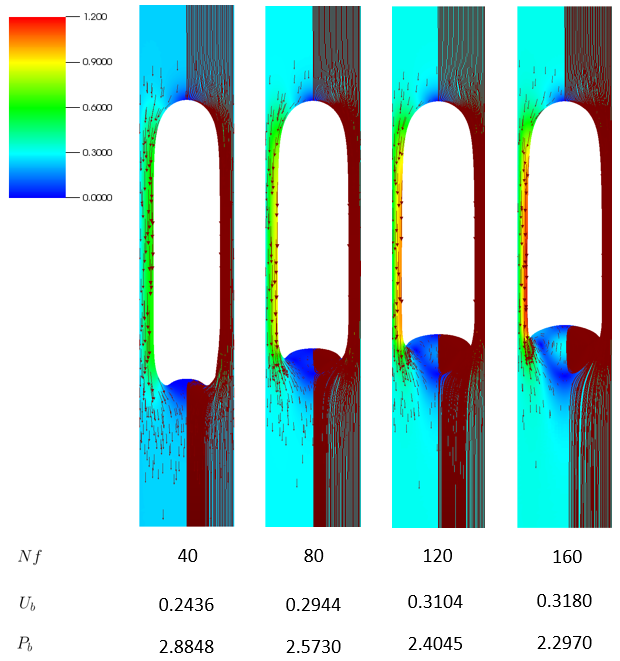}
	\caption{}
	\label{fig:ss_flowfield_eo220}
	\end{subfigure}
	\quad
	\begin{subfigure}{\linewidth}
	    \centering\includegraphics[width = 0.65\linewidth, angle=0]{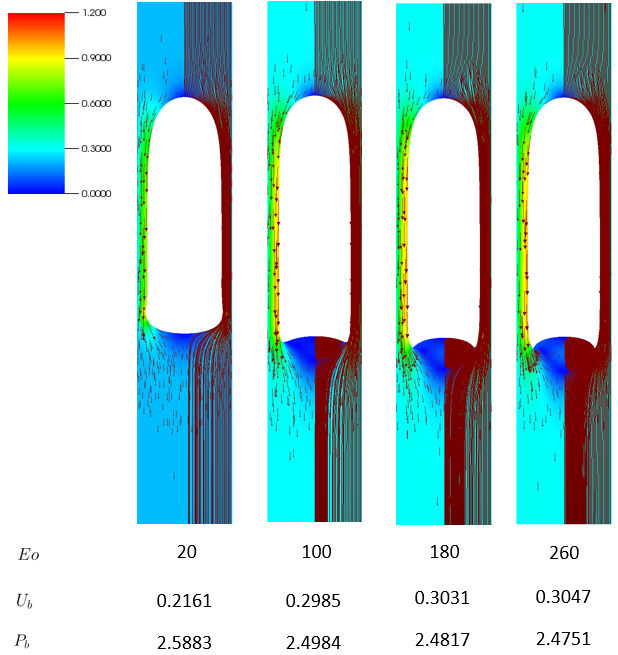}
	\caption{}
	\label{fig:ss_flowfield_nf100}
	\end{subfigure}
	\caption{Steady shapes, streamlines, and flow fields associated with bubble rise in stagnant liquids: (a) effect of $Nf$ for  $ Eo = 220$; (b) effect of $Eo$ for $Nf=100$. In each panel, we show the streamlines and vector fields superimposed on the velocity magnitude pseudocolour plot on the right and left sides of the symmetry axis, respectively. 
	For each case, we provide numerical predictions of the bubble rise speed, $U_b$, and pressure, $P_b$.
	}
\end{figure}

Inspired by \cite{Kang_2010}, for each Taylor bubble, the steady-state shape is presented as a sectional plane through the center of its three-dimensional axisymmetric shape, coloured using the velocity magnitude, with streamlines and vector fields superimposed on the left and right sides of the axis of symmetry, respectively.  
%
%
The inverse viscosity number $Nf$ is a measure of the relative importance of the magnitude of gravity to the viscous force. At constant $Eo$ and $U_m$, an increase in $Nf$ is associated with a decrease in liquid viscosity and its influence on the bubble shape and the surrounding flow field is shown in Figure \ref{fig:ss_flowfield_eo220} for $Eo = 220$ and $U_m = 0.00$. It is seen that by increasing $Nf$, the viscous drag on the bubble is reduced as reflected by an increase in the rise speed, $U_b$, whose value saturates for large $Nf$; this is in agreement with experimental observations \textcolor{black}{ \citep{Nogueira_etal_2006a,Llewellin_etal_2012,White_Beardmore_1962}}
It is also discernible from Figure \ref{fig:ss_flowfield_eo220} that the thickness of the film between the bubble and the pipe wall decreases with $Nf$ due to the decrease in viscous normal stress in this region, as expected. 
The decrease in the magnitude of the normal viscous stress component with increasing $Nf$ is also accompanied by a decrease in the bubble length as well as its pressure $P_b$.

It can also be seen from Figure \ref{fig:ss_flowfield_eo220} that the size and intensity of the counter-rotating vortices in the wake region increase with $Nf$. This is related to the adverse pressure drop that accompanies the jetting of the liquid in the film into the bottom of the bubble, leading to flow separation. The magnitude of the jetting velocity, highlighted by the colour map in this figure, increases with $Nf$, resulting in increased wake length and volume. Another effect of the increase in the intensity of the recirculation in the wake region with $Nf$ is the more pronounced dimpling of the bubble bottom.
It is anticipated that as $Nf$ is  increased further, the bubble bottom will eventually form a skirted tail and ultimately undergo breakup into small bubbles. Therefore, it is expected that at very high $Nf$ (and $Eo$), a topological transition is approached, and reaching a converged steady-state solution becomes increasingly difficult.

For a fixed value of $Nf$ and $U_m$, changes in $Eo$ are related to variations in the relative influence of  buoyancy  to surface tension forces. To assess the effect of $Eo$ on the steady-state shape and flow field around a Taylor bubble, four simulation cases with $Nf=100$ are shown in Figure \ref{fig:ss_flowfield_nf100}. Under the influence of $Eo$, changes in the concavity of the bubble bottom are most noticeable. As $Eo$  increases, the bubble bottom becomes more deformed with the tails of the Taylor bubbles becoming elongated due to the decrease in the tendency of the interface to resist deformation. 
Unlike the case of varying $Nf$, changes in $Eo$ 
result in a marginal influence on the pressure inside the bubble, and bubble length, particularly beyond $Eo = 100$, as shown in Figure \ref{fig:ss_flowfield_nf100}.  

\begin{figure}
    \centering
    \begin{subfigure}{0.45\linewidth}
        \includegraphics[width=\linewidth]{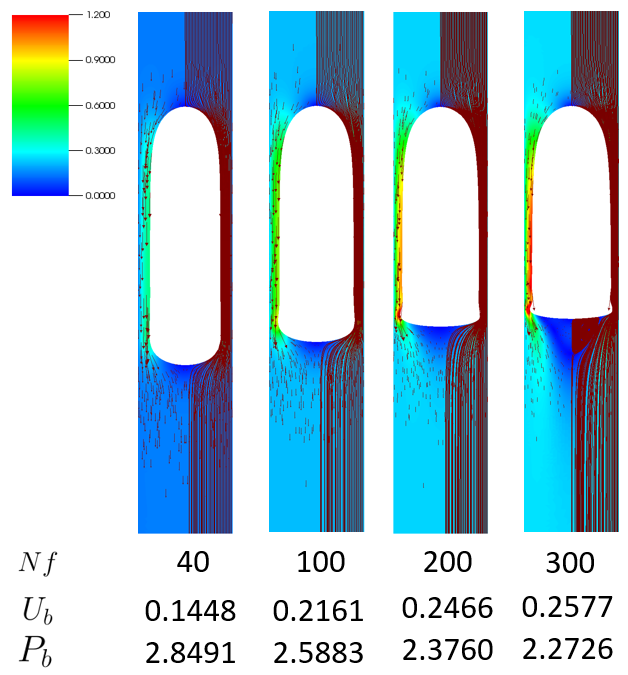}
        \caption{}
        \label{fig:ss_flowfield_eo20}
    \end{subfigure}
    \quad 
    ~ 
    \begin{subfigure}{0.365\linewidth}
        \includegraphics[width=\linewidth]{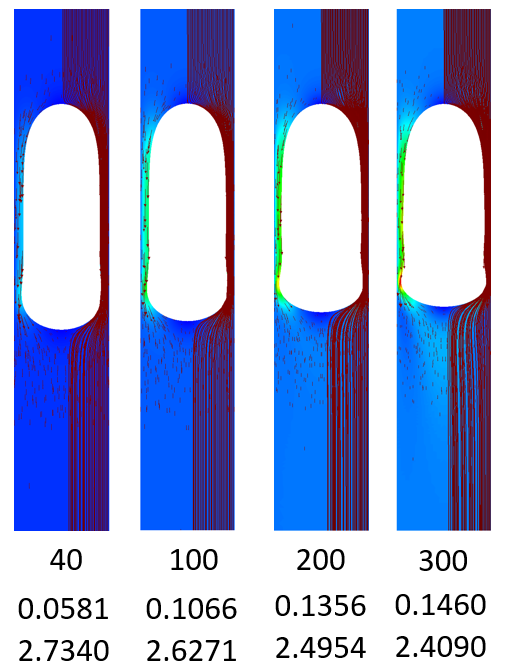}
        \caption{}
         \label{fig:ss_flowfield_eo10}
    \end{subfigure}
    \caption{Effect of variation of $Nf$ on the steady Taylor bubble shapes at low $Eo$:  (a) $Eo = 20$ and (b) $Eo = 10$. 
    } 
     \label{fig:ss_flowfield_low_eo}
\end{figure}

In Figure \ref{fig:ss_flowfield_low_eo}, we focus on the region in parameter space wherein $Eo<20$, which has been highlighted by \cite{White_Beardmore_1962} as being the one in which surface tension effects are expected to be significant; here, we show the effect of $Nf$ on the bubble steady-state shapes and flow fields at $Eo = 10$ and $20$. 
In contrast to what was observed at higher values of $Eo$ in Figure \ref{fig:ss_flowfield_eo220}, an increase in the value of $Nf$ has little influence (and this influence decreases with decreasing $Eo$) on the bubble length and deformation of the bubble bottom. What is seen instead is the emergence of a bulge in the film region close to the bubble bottom, which becomes more pronounced and appears to propagate towards the nose in the form of a capillary wave as $Nf$ and $Eo$ are increased and decreased, respectively. We now turn our attention to examining the principal regions of the bubble starting with the nose region which is discussed next. 


\subsection{The nose region }
\begin{figure}
\centering
    \begin{subfigure}{0.48\linewidth}
        \includegraphics[width = \linewidth, angle=0]{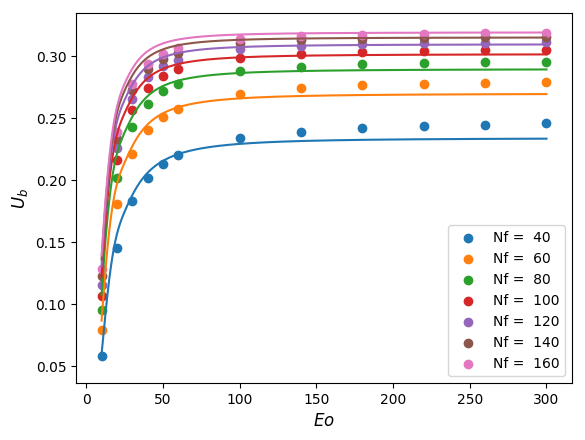}
        \caption{}
	    \label{fig:ss_nose_region_bubble_velocity}
    \end{subfigure}
    \quad
	\begin{subfigure}{0.48\linewidth}
	\includegraphics[width=\linewidth]{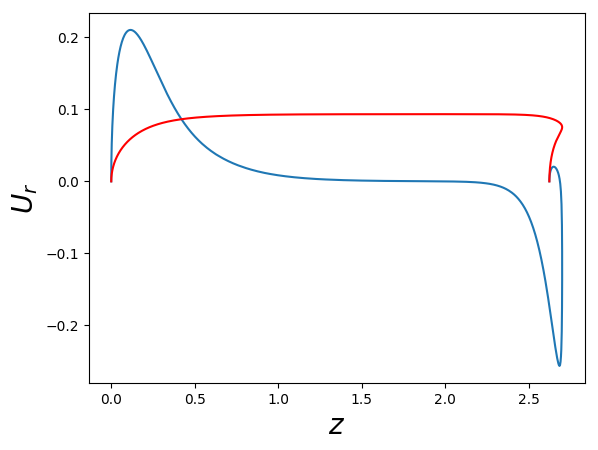}
	\caption{ }
	\label{fig:ss_nose_region_radial_velocity}
	\end{subfigure}
    \quad
	\begin{subfigure}{0.48\linewidth}
	\includegraphics[width=\linewidth]{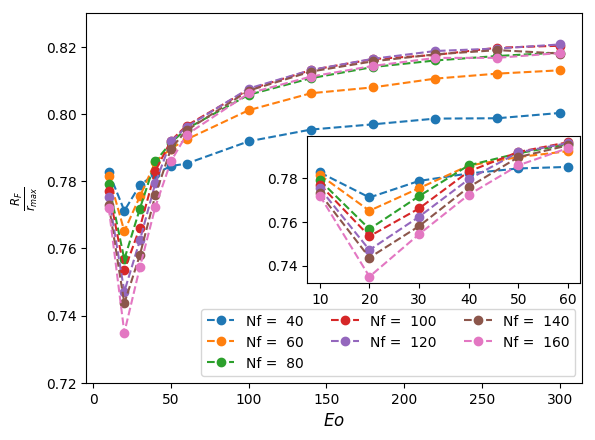}
	\caption{ }
	\label{fig:ss_nose_region_frontal_radius_rmax} 
	\end{subfigure}
	\quad
	\begin{subfigure}{0.48\linewidth}
	\includegraphics[width=\linewidth]{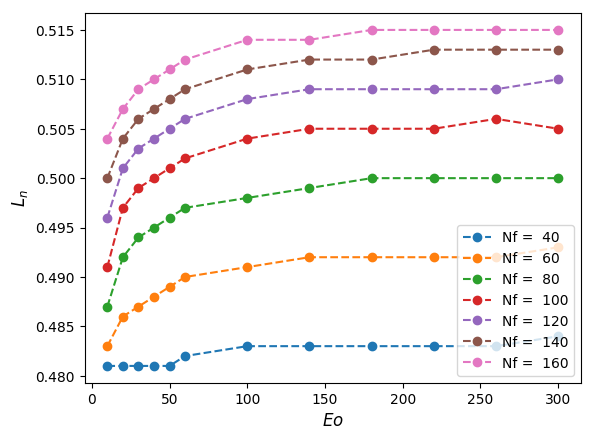}
	\caption{ }
	\label{fig:ss_nose_region_developing_length_ahead}
	\end{subfigure}
	\caption{Flow characteristics associated with the nose region for bubbles rising in stagnant liquids: (a) effect of $Nf$ and $Eo$ on steady-state bubble rise speed showing a comparison between numerical results (coloured marker symbols) and analytical prediction of equation \eqref{eq:lr_bubble_rise_vel_viana} (coloured continuous line);
	(b) typical radial velocity profile (blue) along the interface (red) \textcolor{black}{for $Nf = 80$ and $Eo = 140$}; (c) frontal radius $R_F$ normalised by the maximum Taylor bubble radius for the respective $(Nf,Eo)$ pairing $r_{max}$ showing convergence towards a constant value of $0.815$ for $Nf \geq 80$ with the inset displaying an enlarged view of the $10 \leq Eo \leq 60$ range;
	(d) effect of $Nf$ and $Eo$ on the stabilisation length ahead of the bubble nose.}
\end{figure}

The hydrodynamic features around the nose region (a precise definition of the spatial extent of this region is provided below) are the rise speed $U_b$, the distance ahead of the nose $L_n$ (in a moving frame of reference) at which the flow becomes fully-developed, and the nose curvature. 
In Figure \ref{fig:ss_nose_region_bubble_velocity}, the numerical results for $U_b$ are compared with the predictions from the empirical correlation of \cite{Viana_etal_2003} 
given by 
\begin{align}
U_b &= \frac{{0.34} \left[ 1 + \left( 14.793/Eo \right)^{3.06} \right]^{-0.58}}{\left[ 1 +  \left( {Nf} \left[ 31.08 \left( 1 + \left( 29.868/Eo \right)^{1.96} \right)^{0.49} \right]^{-1} \right)^{\Theta} \right]^{-1.0295\Theta^{-1}}}, \label{eq:lr_bubble_rise_vel_viana} 
\end{align}
where the parameter $\Theta$ is expressed by
\begin{equation}
\Theta = -1.45 \left[ 1 + \left( 24.867 \ Eo \right)^{9.93} \right]^{0.094}. \nonumber
\end{equation}

The overall agreement between the numerical predictions and those obtained from equation \eqref{eq:lr_bubble_rise_vel_viana} is satisfactory and improves with increasing $Nf$. This is because a large proportion of the data used in generating the correlation are based on experiments conducted in the inertia regime \textcolor{black}{ \citep{Viana_etal_2003}}.
It is also seen clearly from Figure \ref{fig:ss_nose_region_bubble_velocity} that for all $Nf$ values investigated, the magnitude of $U_b$ increases steeply with $Eo$  at low $Eo$ then gradually with rising $Eo$ before reaching a plateau at large $Eo$. Saturation of $U_b$ with $Nf$ is also observed at high $Nf$. 
For conditions in which $U_b$ is essentially independent of $Eo$, which can be deduced from Figure \ref{fig:ss_nose_region_bubble_velocity} to be around $Eo = 100$, the limiting value of $Nf$ and the corresponding $U_b$, as established by numerous previous studies \citep{Kang_2010,Lu_Prosperetti_2009,Viana_etal_2003, Brown_1965,Zukoski_1966,White_Beardmore_1962,Griffith_Wallis_1961,Dumitrescu_1943}, are $300$ and $0.35$, respectively, also in agreement with the numerical results shown in Figure \ref{fig:ss_nose_region_bubble_velocity}.

Figure \ref{fig:ss_nose_region_radial_velocity} shows a typical profile of the radial component of the velocity along the interface of a Taylor bubble generated with \textcolor{black}{$Nf=80$ and $Eo=140$}. Starting from the nose of the bubble, which is a stagnation point in a frame of reference that moves with the bubble rise speed, the general observation is that the radial velocity component increases until it peaks before gradually diminishing, approaching zero in the fully-developed film. The region starting from the nose and ending at the point at which the radial velocity on the interface attains its maximum value is referred to as the `nose region'. For all points in this region, the mean radius of curvature $R_m$ is related to the total curvature $\kappa$ by
\begin{equation}
\frac{2}{R_m} = 2 \kappa_m = \kappa_a + \kappa_b = \kappa, \label{eq:ss_nose_region_curvature_radius}
\end{equation} 
where $\kappa_m$ denotes the mean curvature  while $\kappa_a$ and $\kappa_b$ are the principal components of $\kappa$ in the $r-z$ and $r-\theta$ planes, respectively.
The average of the mean radius of curvature is computed and reported  as the frontal radius, $R_F$. The effects of $Nf$ and $Eo$ on $R_F$ normalised by the maximum bubble radius $r_{\rm max}$ are shown in Figure \ref{fig:ss_nose_region_frontal_radius_rmax}
from which it is seen that for $Eo < 100$, $R_F/r_{\rm max}$ is a non-monotonic function of $Eo$: it decreases with $Eo$ before increasing again beyond a certain $Eo$ value. 
This value of $Eo$, at the turning point of $R_F$, decreases with increasing $Nf$, approaching  a constant that lies between $Eo = 20$ and $Eo = 30$, probably related to the emergence of the bulge around the lower part of the film region. 
For $Eo > 100 $, the frontal radius is weakly-dependent on $Eo$, increases with $Nf$ becoming essentially independent of $Nf$ at high $Nf$. These trends are consistent with those associated with the effects of $Nf$ and $Eo$ on $U_b$ confirming the fact that the  rise speed is related to the curvature of the bubble nose. 

We also find that for $Eo > 100$ and $Nf=(40, 60, 80, 100, 120, 140,160)$, the frontal radius is $R_F= (0.2818, 0.2951, 0.3043, 0.3108, 0.3155, 0.3188, 0.3216)$, respectively, in agreement with previous studies \citep{Feng_2008,Funada_etal_2005, Bugg_etal_1998,Fabre_Line_1992}; these results suggest that the bubble nose is prolate-like rather than spherical in shape in which $R_F \approx 0.4$. Under inertial conditions, \cite{Brown_1965} demonstrated that the frontal radius of the Taylor bubbles normalised by its respective maximum bubble radius $r_{\rm max}$ is the same for all liquids and takes a value of 0.75.
The results shown in Figure \ref{fig:ss_nose_region_frontal_radius_rmax} indicate that the normalised $R_F$ approaches a value of 0.815 for $Nf > 80$ which demarcates the limit in $Nf$ at which viscosity has a strong influence on the curvature of the bubble nose.

Beyond a certain axial distance along the axis of symmetry, commonly known in the Taylor bubble literature as the  `stabilisation  length', the stagnant nature of the liquid into which the bubble is rising is attained. 
In this study, in a frame of reference moving with the bubble velocity, we define $L_n$ as the distance at which the axial velocity equals 99\% of the magnitude of the axial velocity far ahead of the bubble nose. The influence of $Nf$ and $Eo$ on $L_n$ is shown 
in Figure \ref{fig:ss_nose_region_developing_length_ahead}. Just like the bubble rise speed, $L_n$ initially increases with $Eo$ before plateauing  beyond $Eo = 100$ for all $Nf$; at constant $Eo$, $L_n$ increases with $Nf$ becoming weakly-dependent on it for sufficiently large $Nf$ values. The reason for this can be attributed to the increase in the momentum imparted on the liquid ahead of the bubble nose in a fixed frame of reference as the bubble rise speed increases with $Nf$ and some $Eo$ ranges.
\subsection{The film region }
\begin{figure}
\centering
	\begin{subfigure}{0.48\linewidth}
	\includegraphics[width=\linewidth]{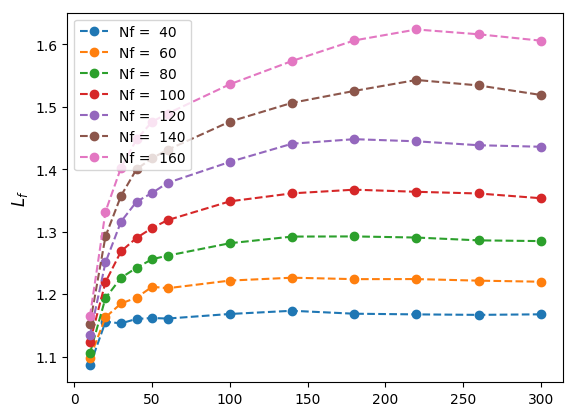}
	\caption{ }
	\label{fig:ss_film_region_developing_length_film} 	
	\end{subfigure}
	\begin{subfigure}{0.48\linewidth}
	    \includegraphics[width = \linewidth, angle=0]{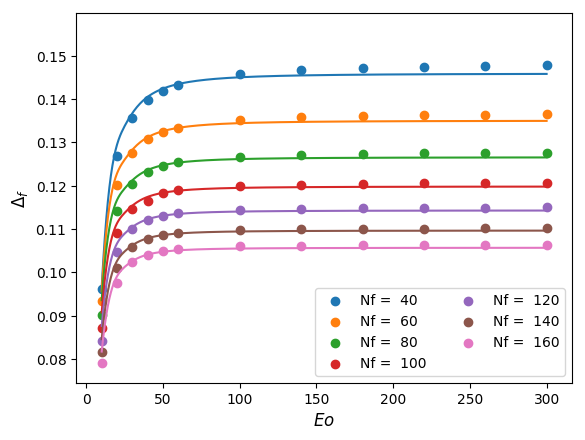}
	    \caption{}
	    \label{fig:ss_film_region_equilibrium_film_thickness}
	\end{subfigure}
	\begin{subfigure}{0.325\linewidth}
        \includegraphics[width=\linewidth]{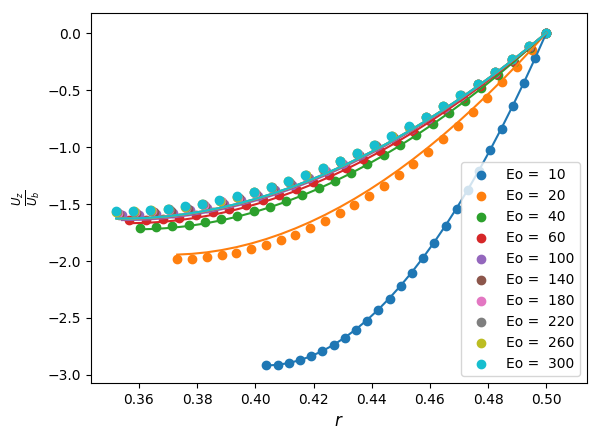}
        \caption{ }
        \label{fig:ss_film_region_flow_field_Nf40}
    \end{subfigure}
    \begin{subfigure}{0.325\linewidth}
        \includegraphics[width=\linewidth]{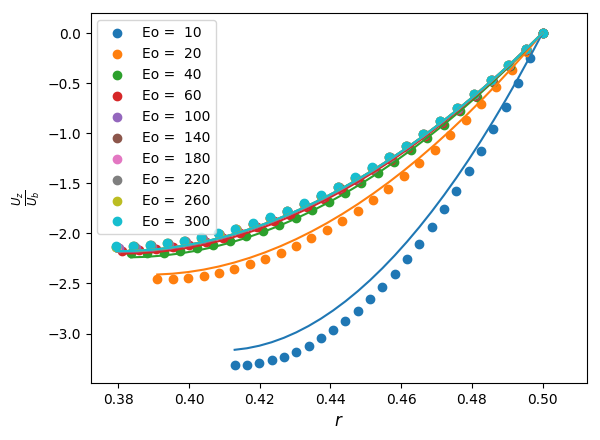}
        \caption{}
         \label{fig:ss_film_region_flow_field_Nf100}
    \end{subfigure}
    \begin{subfigure}{0.325\linewidth}
        \includegraphics[width=\linewidth]{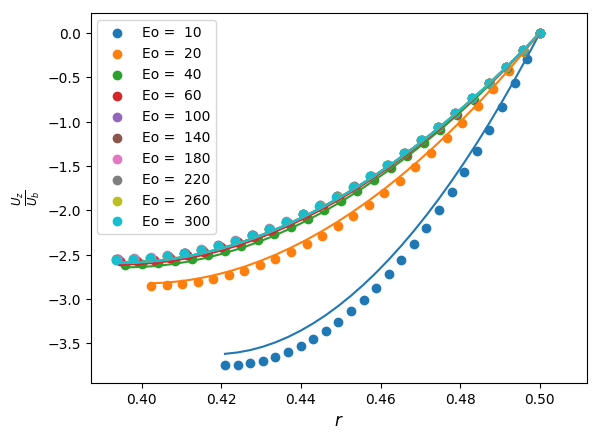}
        \caption{}
         \label{fig:ss_film_region_flow_field_Nf160}
    \end{subfigure}
    \begin{subfigure}{0.325\linewidth}
        \includegraphics[width=\linewidth]{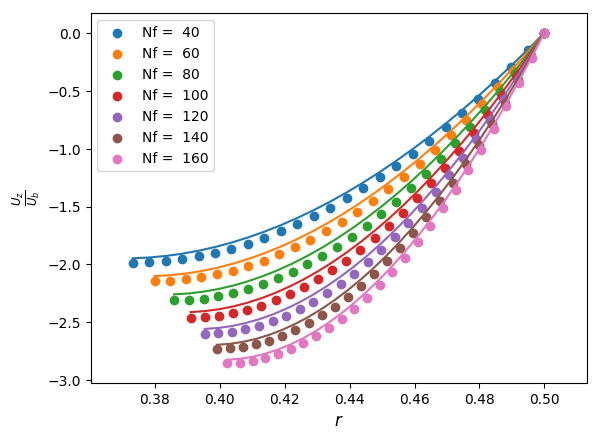}
        \caption{ }
        \label{fig:ss_film_region_flow_field_Eo20}
    \end{subfigure}
    \begin{subfigure}{0.325\linewidth}
        \includegraphics[width=\linewidth]{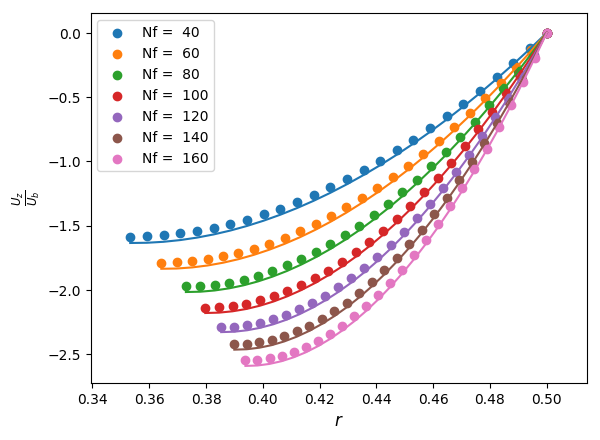}
        \caption{}
         \label{fig:ss_film_region_flow_field_Eo140}
    \end{subfigure}
    \begin{subfigure}{0.325\linewidth}
        \includegraphics[width=\linewidth]{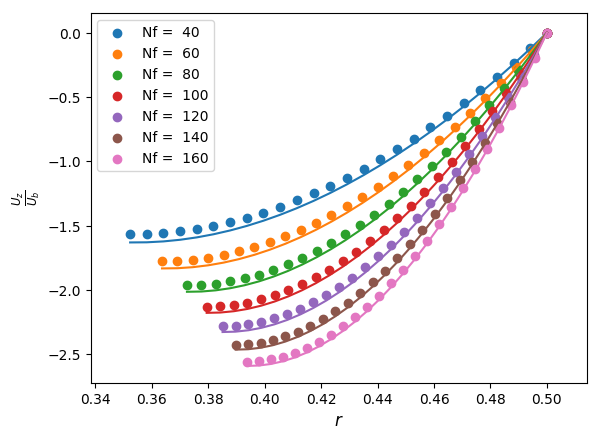}
        \caption{}
         \label{fig:ss_film_region_flow_field_Eo260}
    \end{subfigure}
	\caption{Flow characteristics associated with the film region for bubbles rising in stagnant liquids: stabilisation length $L_f$ and equilibrium film thickness $\Delta_f$, depicted in (a) and (b), respectively, showing a comparison between numerical simulations (coloured markers) and theoretical prediction using \eqref{eq:ss_film_region_equilibrium_film_thickness} and \eqref{eq:lr_bubble_rise_vel_viana} (coloured continuous solid line) for different $Nf$ and $Eo$;
	effect of  $Eo$ on the axial velocity in the fully-developed film region $u_z$ normalized by $U_b$ with $Nf = 40,100,160$ shown in (c)-(e), respectively;
	effect of $Nf$ on $u_z/U_b$ with $Eo=20,140,260$ shown in (f)-(h). In (c)-(h), the numerical simulations are represented by the coloured markers and the theoretical predictions of \eqref{eq:ss_film_region_laminar_film_velocity_profile} by coloured solid lines.
	}
\end{figure}

The features that define the hydrodynamics of the film region are the stabilisation length $L_f$, the equilibrium film thickness $\Delta_f$, and the velocity profiles in the fully-developed film.
The first two features are crucial parameters as it is expected that the flow pattern
in the wake of a Taylor bubble becomes  independent of the bubble length for bubbles of lengths greater than $L_f$ and heavily-dependent on  $\Delta_f$ \citep{Nogueira_etal_2006}. 
The stabilisation length $L_f$ is determined to be the point at which the radial velocity component, and the rate of change in the axial velocity component along the interface are less than 1\% of their maximum interfacial values. 
Figure \ref{fig:ss_film_region_developing_length_film} shows that $L_f$ increases steeply with $Eo$ before plateauing at high $Eo$ for all values of $Nf$ studied. 
For a fixed $Eo$ value, $L_f$ increases with $Nf$ indicating that the film needs to travel a longer distance below the bubble nose before it becomes fully-developed. However, unlike the dependence on $Nf$ of the bubble rise speed, or the nose stabilisation length, $L_f$ does not appear to saturate with increasing $Nf$. The results, therefore, indicate that as the viscosity is decreased, it becomes increasingly difficult to obtain a truly fully-developed film around Taylor bubbles that are not extremely long. 
%
Below the developing length in the film region, the liquid film is deemed to have attained equilibrium,  and the thickness is from there onward  constant until the Taylor bubble tail region is approached. The film thickness at the point where the equilibrium film thickness is first attained is measured from our numerical predictions and the result is compared with the theoretical prediction of \cite{Brown_1965}. From \cite{Brown_1965}, the equation that relates the equilibrium film thickness to the bubble rise speed, in dimensionless form, can be written as 
 \begin{equation}
 \frac{4 Nf}{3 U_b} \Delta_f^3 + 2 \Delta_f - 1 = 0. \label{eq:ss_film_region_equilibrium_film_thickness}
 \end{equation}
Using equation  \eqref{eq:ss_film_region_equilibrium_film_thickness} together with \eqref{eq:lr_bubble_rise_vel_viana},  $\Delta_f$ is computed for different $Nf$ and $Eo$, and the results are compared with our numerical prediction in Figure \ref{fig:ss_film_region_equilibrium_film_thickness}.
The numerical and theoretical predictions are in good agreement particularly at higher $Nf$, as expected, since  the thin liquid film assumption becomes more valid with increasing inverse viscosity number. The decline in the equilibrium film thickness with $Nf$ is due to the decrease in the magnitude of the normal stress exerted on the interface as the fluid viscosity is decreased. It is noteworthy that despite the apparent  dependence of $L_f$ on $Eo$ with increasing $Nf$, $\Delta_f$ remains almost constant beyond $Eo = 100$.

In order to obtain an approximation of the axial velocity component in the fully-developed film, $u_z$, the following reduced version of the dimensionless form of the axial momentum equation in this region is considered \citep{Brown_1965}:
\begin{equation}
\frac{1}{r} \frac{d }{d r} \left[ r \frac{d u_z}{d r}\right] = -Nf; \label{eq:ss_film_region_laminar_film_equation}
\end{equation}
the solution of equation \eqref{eq:ss_film_region_laminar_film_equation} is expressed by
\begin{equation}
u_z = - Nf \left[ \left( \frac{ 0.25 - r^2}{4}\right) - \frac{1}{2}\left( 0.5 - \Delta_f\right)^2 \ln \left( \frac{0.5}{r}\right) \right]. \label{eq:ss_film_region_laminar_film_velocity_profile}
\end{equation}
The predictions from equation \eqref{eq:ss_film_region_laminar_film_velocity_profile}, scaled using the bubble rise speed and compared to our numerical results are shown in Figures \ref{fig:ss_film_region_flow_field_Nf40}-\ref{fig:ss_film_region_flow_field_Nf160} and \ref{fig:ss_film_region_flow_field_Eo20}-\ref{fig:ss_film_region_flow_field_Eo260}, which highlight the effect of $Nf$ and $Eo$ on $u_z/U_b$, respectively. The improvement in the agreement between the numerical results and the theoretical predictions is noticeable with increasing $Eo$ particularly at high $Nf$. 

\subsection{Hydrodynamic features at the wall and interface }
\subsubsection{Wall shear stress}

\begin{figure}
    \centering
    \begin{subfigure}{0.325\linewidth}
        \includegraphics[width=\linewidth]{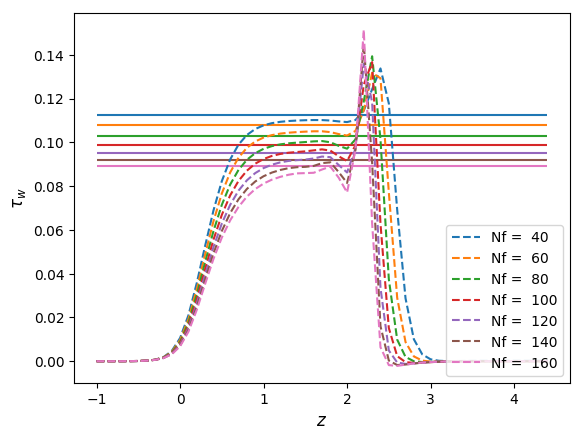}
        \caption{ }
        \label{fig:ss_wall_region_wall_shear_stressEo20}
    \end{subfigure}
    \begin{subfigure}{0.325\linewidth}
        \includegraphics[width=\linewidth]{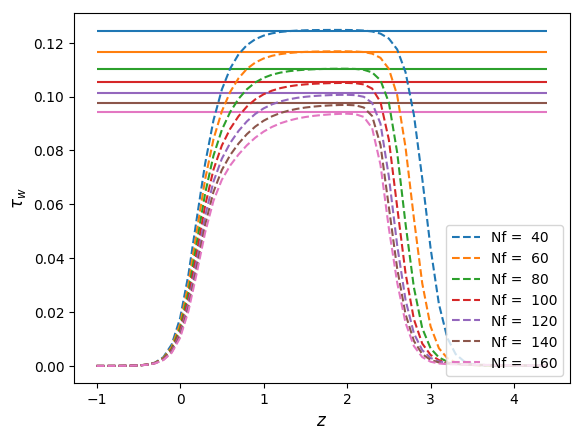}
        \caption{}
         \label{fig:ss_wall_region_wall_shear_stressEo140}
    \end{subfigure}
    \begin{subfigure}{0.325\linewidth}
        \includegraphics[width=\linewidth]{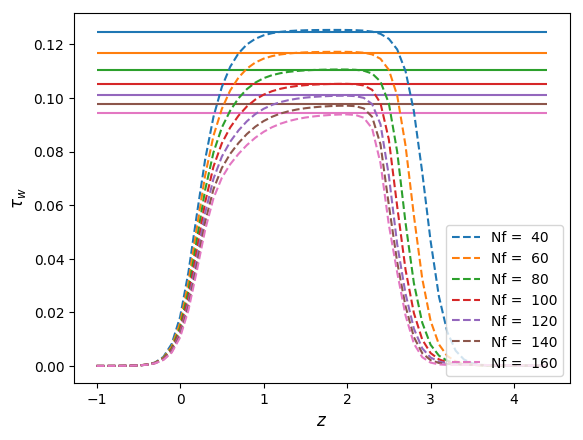}
        \caption{}
         \label{fig:ss_wall_region_wall_shear_stressEo260}
    \end{subfigure}
    \begin{subfigure}{0.325\linewidth}
        \includegraphics[width=\linewidth]{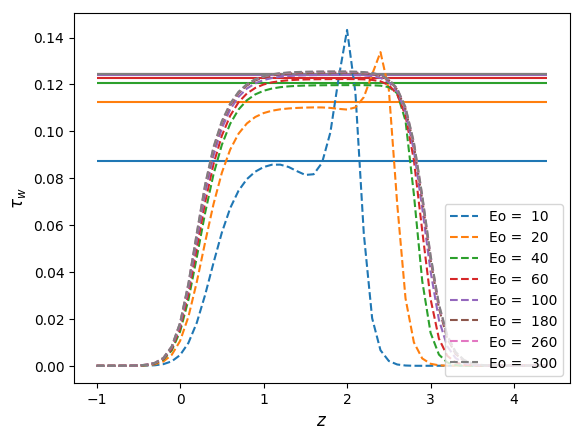}
        \caption{ }
        \label{fig:ss_wall_region_wall_shear_stressNf40}
    \end{subfigure}
    \begin{subfigure}{0.325\linewidth}
        \includegraphics[width=\linewidth]{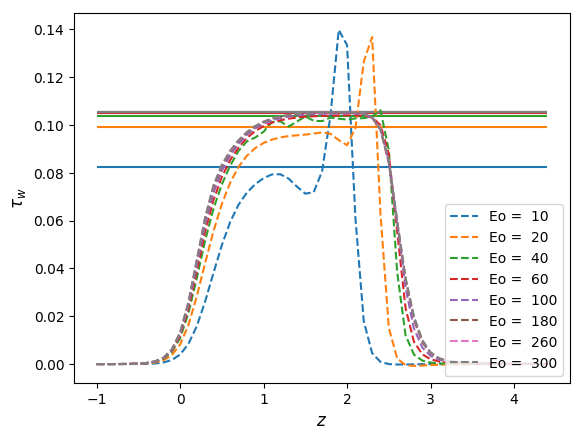}
        \caption{}
         \label{fig:ss_wall_region_wall_shear_stressNf100}
    \end{subfigure}
    \begin{subfigure}{0.325\linewidth}
        \includegraphics[width=\linewidth]{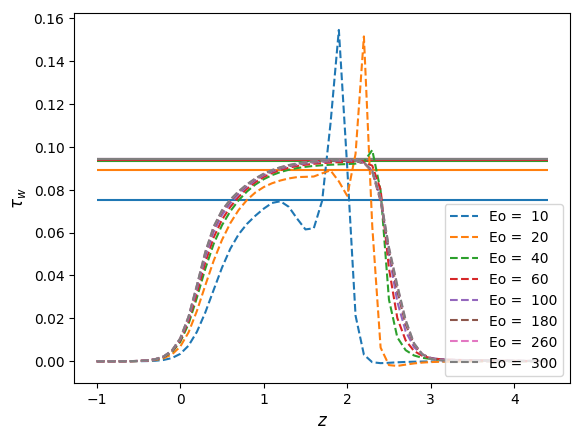}
        \caption{}
         \label{fig:ss_wall_region_wall_shear_stressNf160}
    \end{subfigure}
    \begin{subfigure}{0.8\linewidth}
    \centering\includegraphics[width = 0.60\linewidth, angle=0]{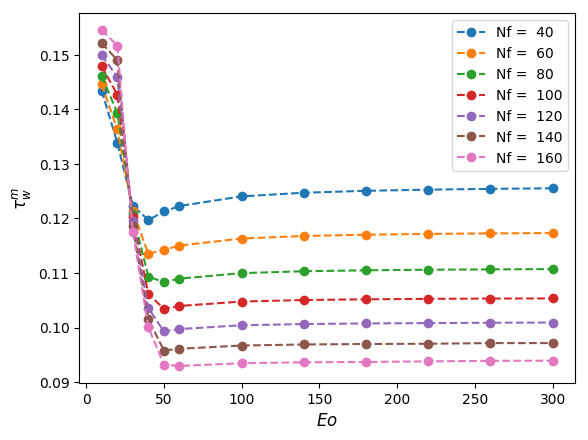}
    \caption{}
    \label{fig:ss_wall_region_max_wall_shear_stress}
    \end{subfigure}
    \caption{Shear stress at the wall boundary: effect of $Nf$ with $Eo = 20,140,260$ shown in (a)-(c), respectively; effect of $Eo$ with $Nf=40,100,160$ shown in (d)-(f), respectively;
    (g) effects of $Nf$ and $Eo$ on the maximum wall shear stress. In (a)-(f), our numerical results are shown using broken lines and the predictions of equation \eqref{eq:ss_shear_stress_brown} in the fully-developed film region using solid lines.} 
\end{figure}

From equation \eqref{eq:tangential_stress_bc}, the shear stress at any boundary is defined as
\begin{equation}
 \text{\boldmath ${\tau}$} = \mathbf{n} \cdot \mathbf{T} \times \mathbf{n}. \label{eq:ss_shear_stress}
\end{equation} 
For an axisymmetric boundary, the nonzero component of equation (\ref{eq:ss_shear_stress}) simplifies to 
\begin{equation}
\tau = Nf^{-1} \left[ \mathbf{n} \cdot \frac{d \mathbf{u}}{d s} + \mathbf{t} \cdot \frac{d \mathbf{u}}{d n} \right],
\end{equation}
which when evaluated at the wall, gives
\begin{equation}
\tau_w = -Nf^{-1} \frac{d u_z}{d r}, \label{eq:ss_shear_stress_wall}
\end{equation}
where $\tau_w$ denotes the dimensionless wall shear stress. 
In the fully-developed film region, using equation \eqref{eq:ss_film_region_laminar_film_velocity_profile}, $\tau_w$ reads 
\begin{equation}
\tau_w = 0.25 - \left( 0.5 - \Delta_f\right)^2,  \label{eq:ss_shear_stress_brown}
\end{equation}
which is a constant whose dependence on $Nf$ and $Eo$ enters equation
\eqref{eq:ss_shear_stress_brown} through the  variation of $\Delta_f$ with these parameters via equations  \eqref{eq:lr_bubble_rise_vel_viana}
 and \eqref{eq:ss_film_region_equilibrium_film_thickness}.
A comparison of the predictions of equation with the numerically computed results for $\tau_w$ using \eqref{eq:ss_shear_stress_wall} is shown in Figures (\ref{fig:ss_wall_region_wall_shear_stressEo20})-(\ref{fig:ss_wall_region_wall_shear_stressNf160}). Beyond the limit at which $Eo$ exerts a strong influence on the dynamics of the bubble, i.e., for $Eo \gtrsim 100$, equation \eqref{eq:ss_shear_stress_brown} adequately predicts the effect of $Nf$ and $Eo$ on $\tau_w$ in the developed film region. While an increase in $Nf$ leads to a reduction in $\tau_w$, $Eo$ has no significant impact on it beyond $Eo \sim 100$. Both effects can be related to that of the parameters on the equilibrium film thickness and its velocity profiles, shown in Figures \ref{fig:ss_film_region_equilibrium_film_thickness},   (\ref{fig:ss_film_region_flow_field_Nf40})-(\ref{fig:ss_film_region_flow_field_Nf160}), and 
(\ref{fig:ss_film_region_flow_field_Eo20})-(\ref{fig:ss_film_region_flow_field_Eo260}), respectively. The apparent peaks observed in figures (\ref{fig:ss_wall_region_wall_shear_stressEo20})-(\ref{fig:ss_wall_region_wall_shear_stressEo260}) and (\ref{fig:ss_wall_region_wall_shear_stressNf40})-(\ref{fig:ss_wall_region_wall_shear_stressNf160}) when surface tension effects are strong for small $Eo$ can be related to the undulation that appears towards the end of the liquid film, with the influence becoming more pronounced as $Nf$ is increased and  $Eo$ decreased.
Lastly, the maximum wall shear stress, $\tau_w^m$, for the combined effect of $Nf$ and $Eo$, is plotted in Figure \ref{fig:ss_wall_region_max_wall_shear_stress}.

\subsubsection{Interface normal stress}
\begin{figure}
    \centering
    \begin{subfigure}{0.325\linewidth}
        \includegraphics[width=\linewidth]{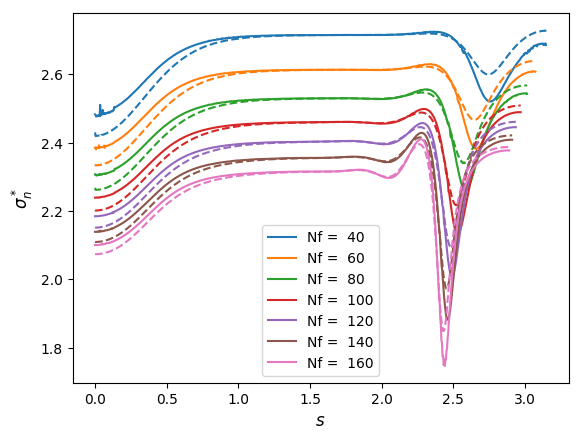}
        \caption{ }
        \label{fig:ss_interface_region_normal_stressEo20}
    \end{subfigure}
    \begin{subfigure}{0.325\linewidth}
        \includegraphics[width=\linewidth]{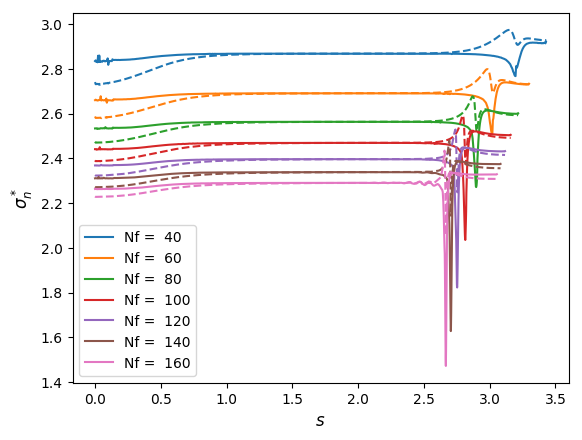}
        \caption{}
         \label{fig:ss_interface_region_normal_stressEo140}
    \end{subfigure}
    \begin{subfigure}{0.325\linewidth}
        \includegraphics[width=\linewidth]{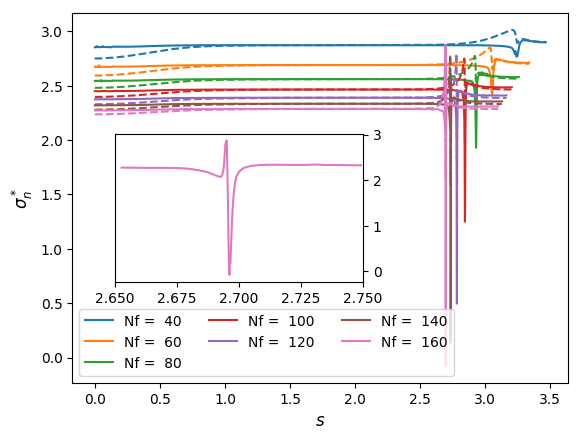}
        \caption{}
         \label{fig:ss_interface_region_normal_stressEo260}
    \end{subfigure}
        \begin{subfigure}{0.325\linewidth}
        \includegraphics[width=\linewidth]{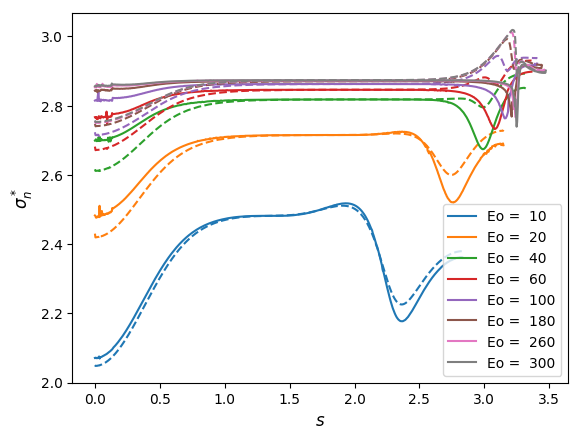}
        \caption{ }
        \label{fig:ss_interface_region_normal_stressNf40}
    \end{subfigure}
    \begin{subfigure}{0.325\linewidth}
        \includegraphics[width=\linewidth]{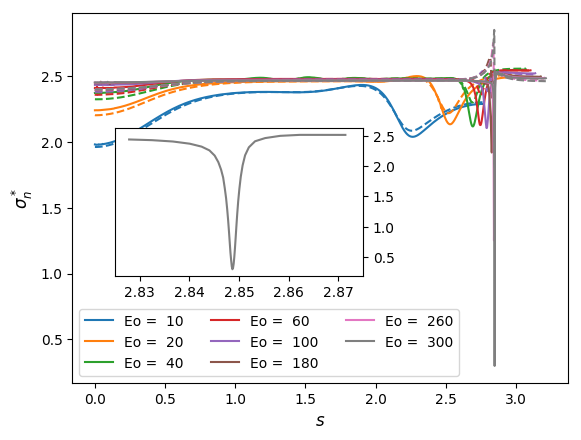}
        \caption{}
         \label{fig:ss_interface_region_normal_stressNf100}
    \end{subfigure}
    \begin{subfigure}{0.325\linewidth}
        \includegraphics[width=\linewidth]{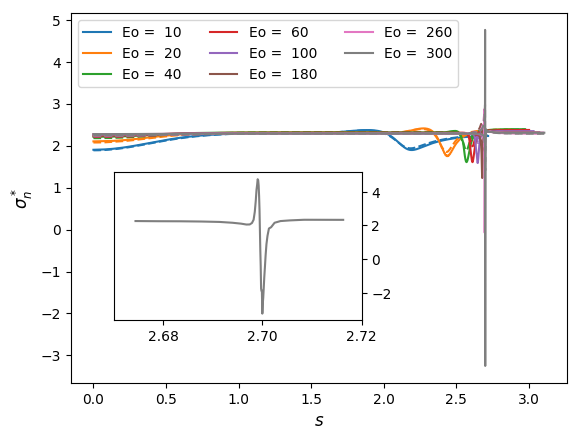}
        \caption{}
         \label{fig:ss_interface_region_normal_stressNf160}
    \end{subfigure}
    \begin{subfigure}{0.8\linewidth}
    \centering\includegraphics[width = 0.60\linewidth, angle=0]{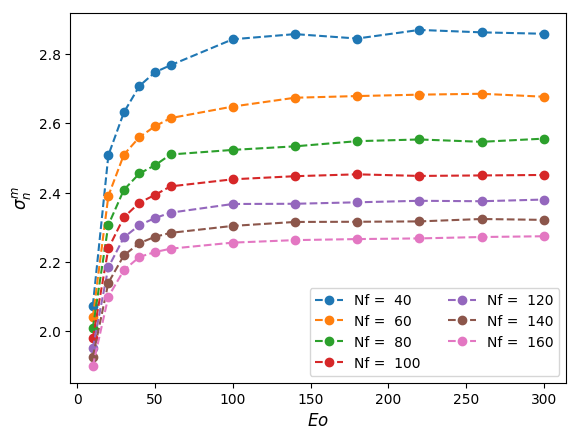}
    \caption{}
    \label{fig:ss_interface_region_max_normal_stress}
    \end{subfigure}
    \caption{Normal stress (solid lines) and total pressure (broken lines) at the interface: effect of  $Nf$ with $Eo = 20,140,260$ shown in (a)-(c), respectively; effect of  $Eo$ with $Nf = 40,100,160$ shown in (d)-(f), respectively; (g) effects of $Nf$ and $Eo$ on the maximum interface normal stress. Panels (c), (e), and (f) show an enlarged view of the curves for $Nf=160$, $Eo=300$, and $Eo=300$, respectively, for $2.5 \le s \le 3$.} 
     \label{fig:ss_interface_region_normal_stressEo}
\end{figure}

From equation \eqref{eq:normal_stress_bc}, the normal stress at the interface in the direction of unit normal to the interface is defined as
\begin{equation}
\sigma_n = - \mathbf{n} \cdot \mathbf{T} \cdot \mathbf{n} = -\left[-p + 2Nf^{-1} \mathbf{n} \cdot \frac{d \mathbf{u}}{d n} \right].
\end{equation}
Expressing the normal stress in terms of the total pressure by adding the gravity term to the hydrodynamic pressure, \eqref{eq:normal_stress_bc} becomes 
\begin{equation}
\sigma_n^* = -  \left[-p_T + 2Nf^{-1} \mathbf{n} \cdot \frac{d \mathbf{u}}{d n} \right] = P_b - {Eo}^{-1} \kappa,  \label{eq:ss_normal_stress_wrt_z}
\end{equation}
where $p_T = p + z$.
Figures (\ref{fig:ss_interface_region_normal_stressEo20})-(\ref{fig:ss_interface_region_normal_stressEo260}) and (\ref{fig:ss_interface_region_normal_stressNf40})-(\ref{fig:ss_interface_region_normal_stressNf160}) show the effects of $Nf$ and $Eo$ on the interface normal stress and total pressure.
It is apparent that the normal stress decreases with $Nf$ and it becomes weakly-dependent on $Eo$ for $Eo \gtrsim 100$. 
In the fully-developed liquid film region, both the pressure and the normal stress match in order to satisfy \eqref{eq:ss_normal_stress_wrt_z}. This is because in this region, the interface has approximately zero curvature, and $u_r =d u_r/d n = 0$, making the viscous stress and the stress due to curvature in the $r-z$ plane contributions zero. Thus, equation \eqref{eq:ss_normal_stress_wrt_z} reduces to
$
\sigma_n^* = p_T  = P_b -  {Eo}^{-1} \kappa_b \approx P_b
$.
Since the bubble pressure is a constant, the implication of this is that the viscous and curvature forces are only  important in the nose and bottom of the bubble and it is the interplay between them that determines the shape of these regions. 
For the observed sharp peaks in the interface normal stress around the bubble bottom, particularly for higher $Eo$ and $Nf$ such as the ones shown in Figures \ref{fig:ss_interface_region_normal_stressEo260}, \ref{fig:ss_interface_region_normal_stressNf100}, and \ref{fig:ss_interface_region_normal_stressNf160}, 
it is clear from Figures
\ref{fig:ss_mesh_resolution_bubble_shape}-\ref{fig:ss_mesh_resolution_bubble_mesh_tail}, \textcolor{black}{ the insets shown in these figures }
 that the bubble bottom  and the tail regions are well resolved. 
In Figure \ref{fig:ss_interface_region_max_normal_stress}, the maximum normal stress, $\sigma_n^m$ exerted on the interface was extracted to highlight its dependence on $Nf$ and $Eo$.

\subsection{Hydrodynamic features of bottom region}
The features discussed here encompass those that define the bottom of the bubble which are the shape of the bottom and the length of the developing length below the bottom, and those that define the wake, if present, which are the length of the wake and the position vector of the vortex eye. 
\subsubsection{Curvature radius of bubble bottom and shape}
\begin{figure}
	\begin{subfigure}{0.095\linewidth}
        \includegraphics[width=\linewidth]{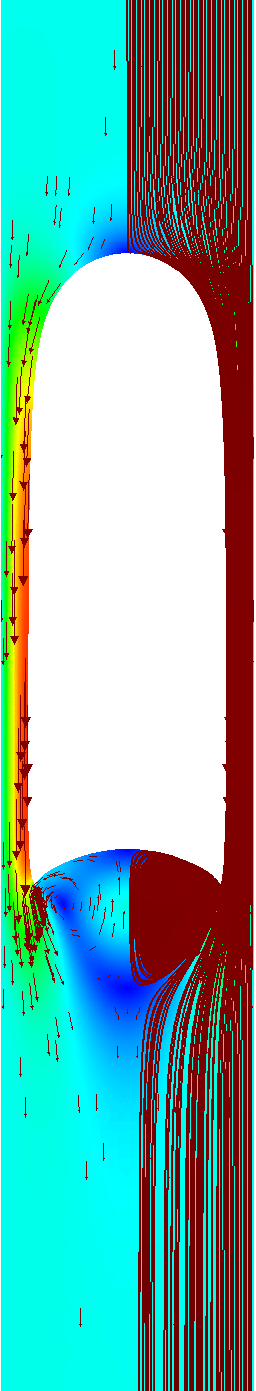}
        \caption{}
        \label{fig:ss_mesh_resolution_bubble_shape}
    \end{subfigure}
   \quad 
    \begin{subfigure}{0.05\linewidth}
        \includegraphics[width=\linewidth]{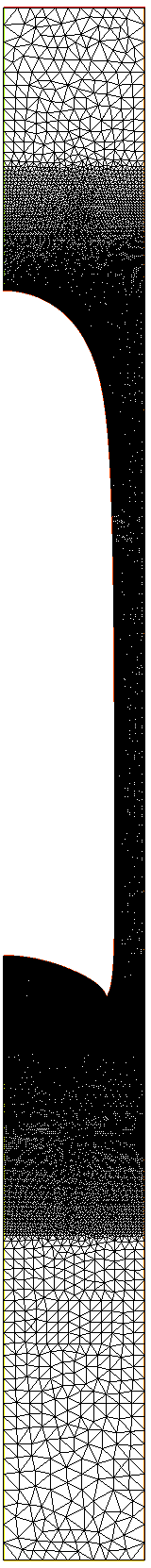}
        \caption{}
         \label{fig:ss_mesh_resolution_bubble_mesh_full}
    \end{subfigure}
    \quad 
    \begin{subfigure}{0.15\linewidth}
        \includegraphics[width=\linewidth]{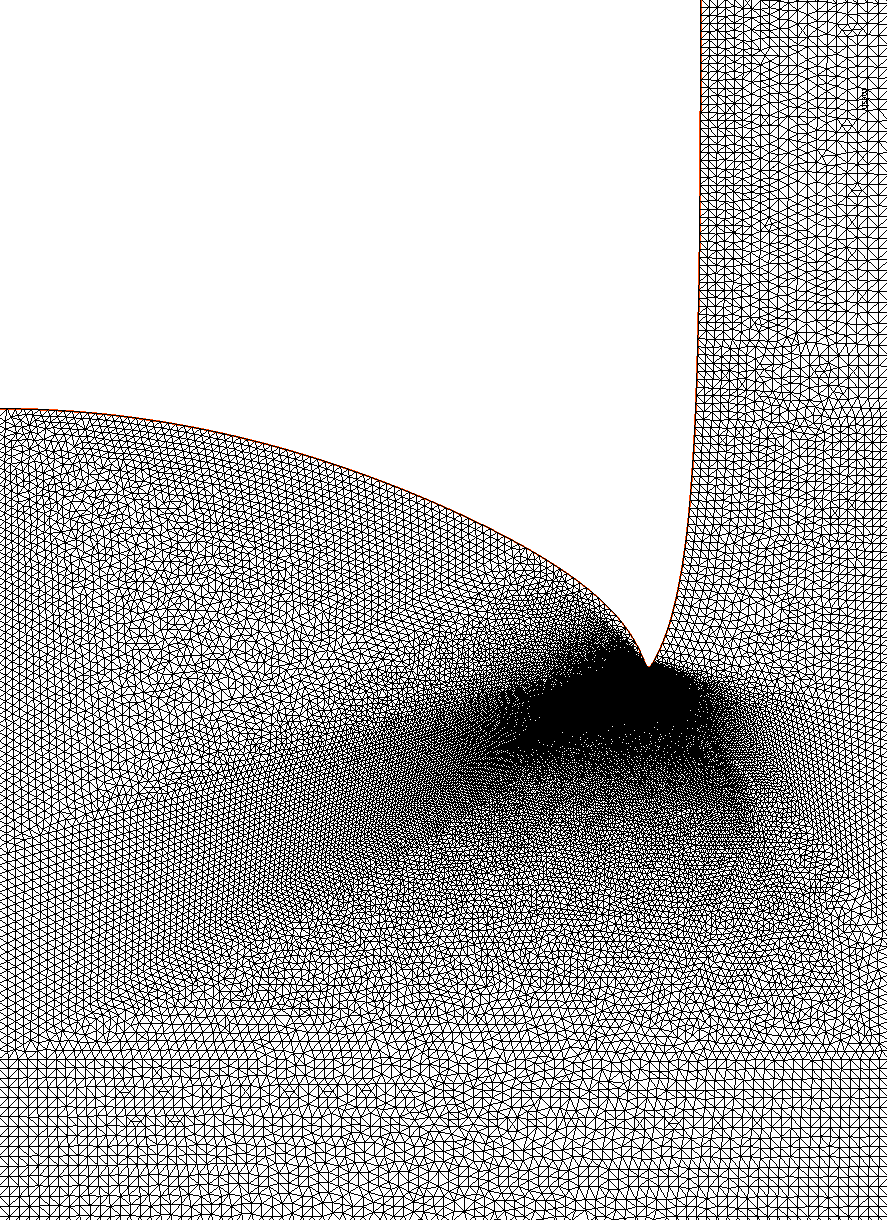}
        \caption{ }
        \label{fig:ss_mesh_resolution_bubble_mesh_bottom}
    \end{subfigure} 
    \quad  
    \begin{subfigure}{0.13\linewidth}
        \includegraphics[width=\linewidth]{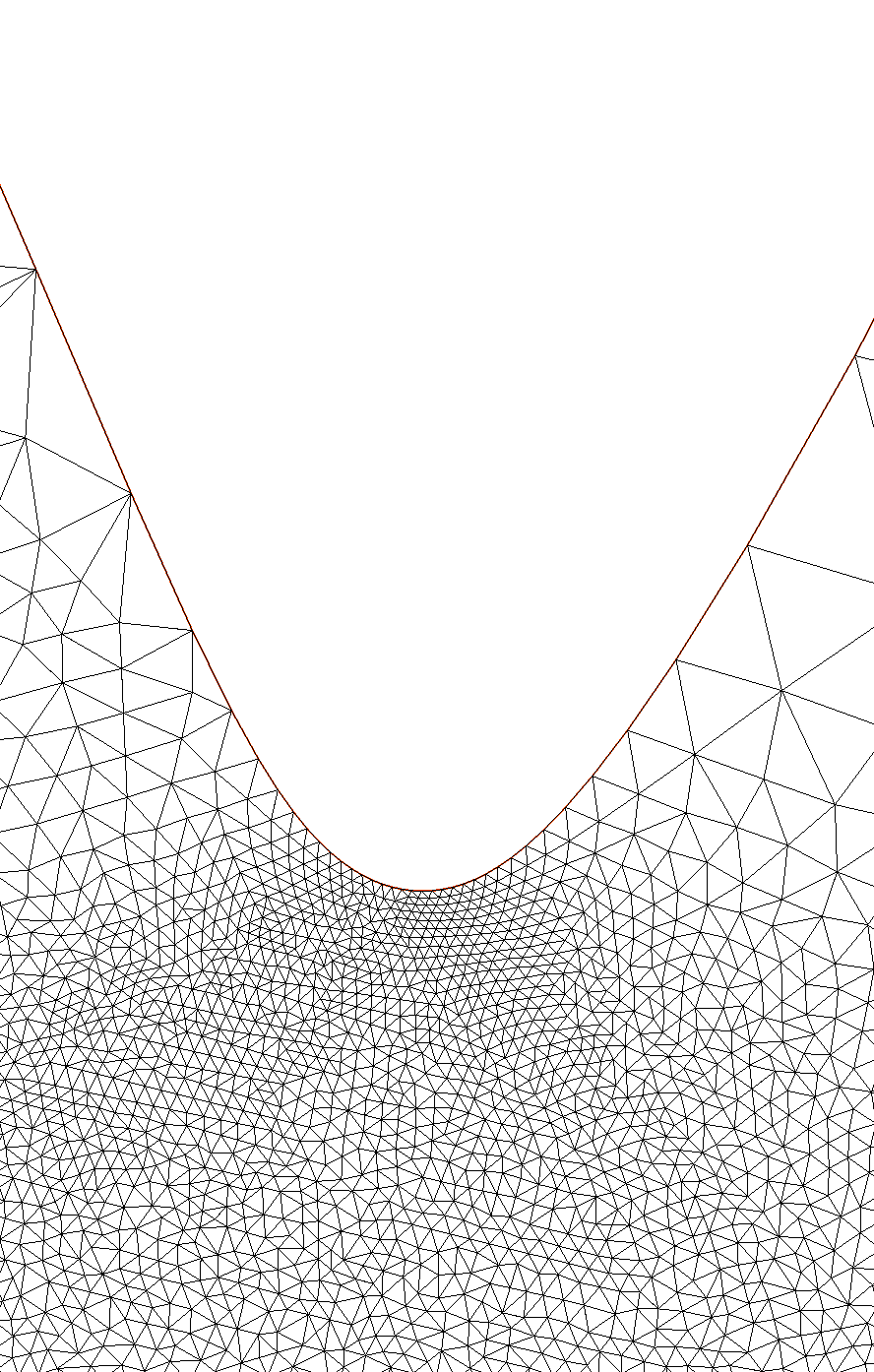}
        \caption{}
         \label{fig:ss_mesh_resolution_bubble_mesh_tail}
    \end{subfigure}
	\begin{subfigure}{0.5\linewidth}
	\centering\includegraphics[width = 0.90\linewidth, angle=0]{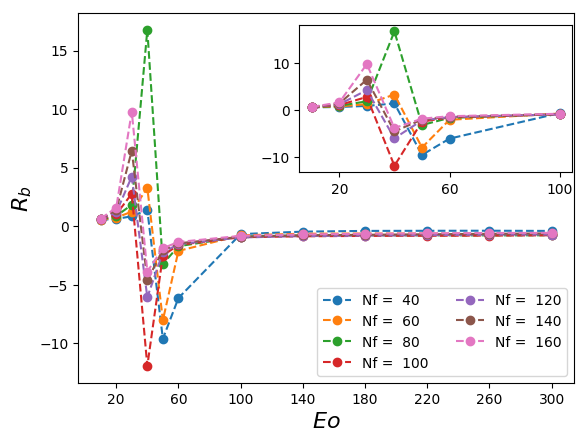}
	\caption{}
	\label{fig:ss_bottom_region_curvature_radius}
	\end{subfigure}
	\begin{subfigure}{0.8\linewidth}
	\hspace{0.5in}\includegraphics[width = \linewidth, angle=0]{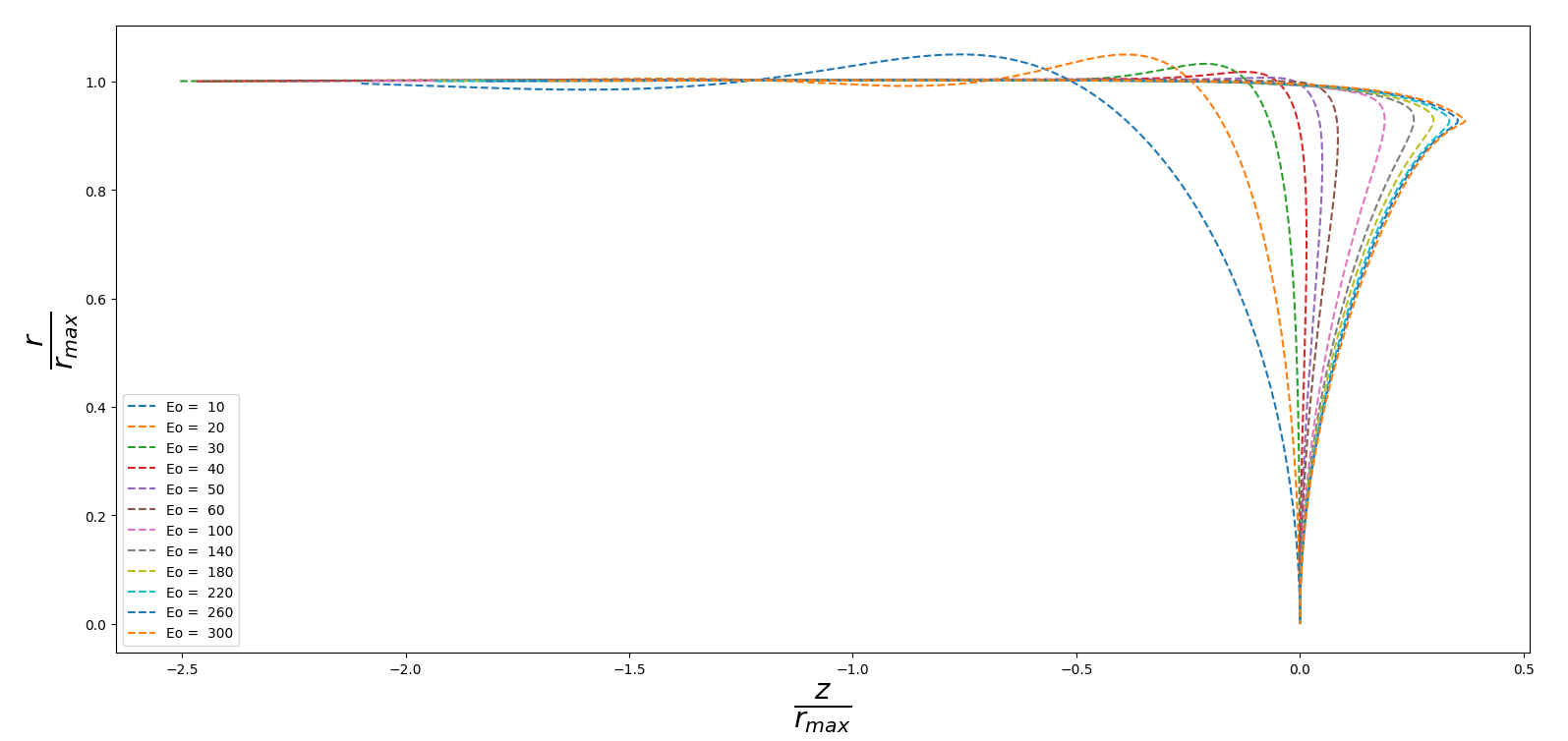}
	\caption{}
	\label{fig:ss_bottom_region_normalise_shape_bottomNf160}
	\end{subfigure}
	\begin{subfigure}{0.8\linewidth}
	\hspace{0.5in}\includegraphics[width = \linewidth, angle=0]{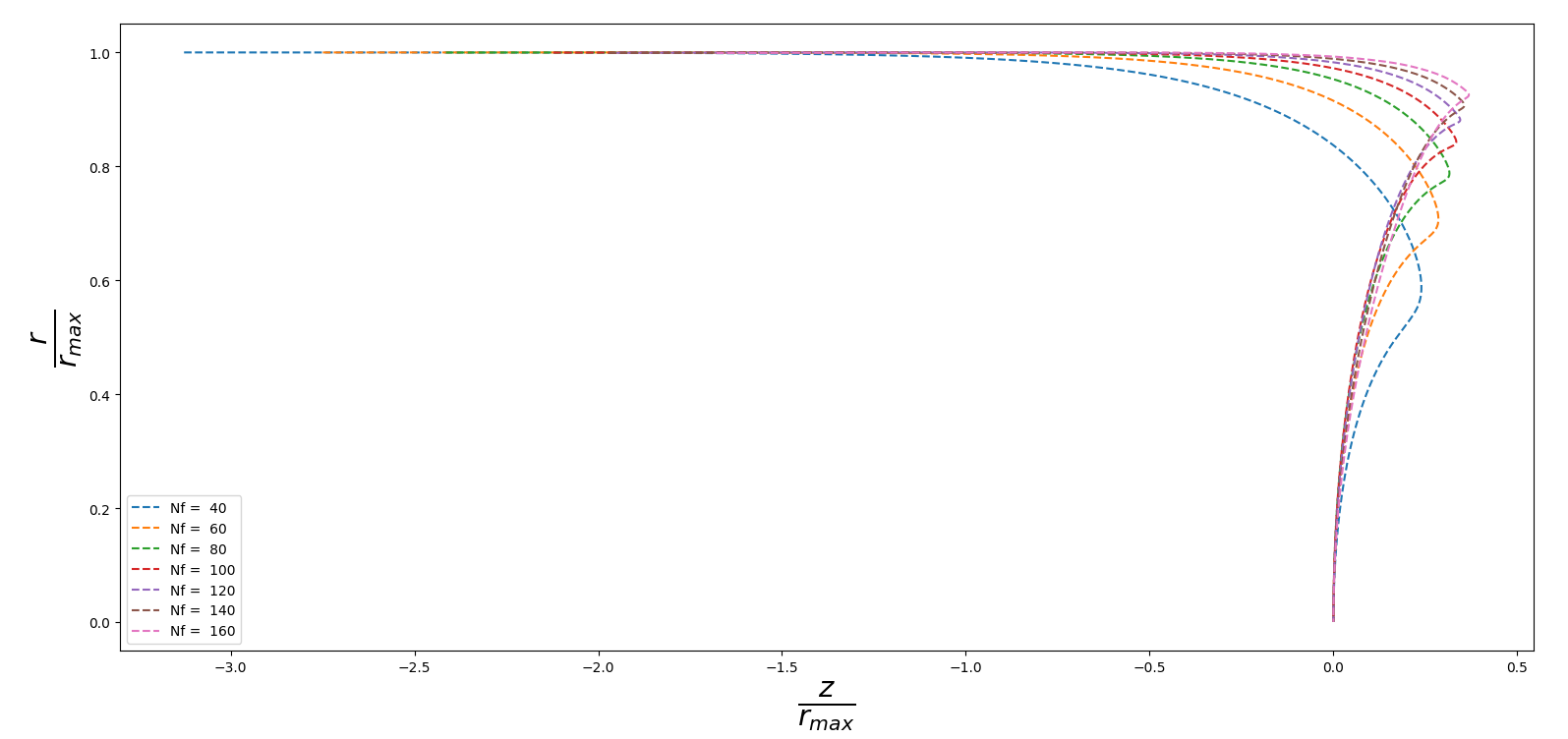}
	\caption{}
	\label{fig:ss_bottom_region_normalise_shape_bottomEo300}
	\end{subfigure}
	\caption{Flow characteristics of the bottom region for a bubble rising in a stagnant liquid: 
	shape, (a), and mesh structure, (b), for $Nf = 160$ and $Eo = 300$; enlarged views of the bottom, (c), and tail tip mesh structures, (d);
	(e) influence of $Nf$ and $Eo$ on the Taylor bubble bottom radius of curvature $R_b$; bottom deformation: influence of $Eo$  with $Nf = 160$, (f), and influence of $Nf$ for $Eo = 300$, (g).}
\end{figure}

The effects of varying flow conditions on the Taylor bubble bottom shape are quantitatively examined using  the sign of the radius of curvature. Because of the varying shapes that are associated with  bubble bottom, it is more convenient and sufficient to define the shape of the bubble bottom based on the curvature evaluated at the bottom along the axis of symmetry. Essentially, a positive (negative) radius of curvature signifies a convex (concave) bottom shape with respect to the liquid phase.  Figure \ref{fig:ss_bottom_region_curvature_radius} shows the mean radius of curvature $R_b$ for different $Nf$ and varying $Eo$.
It is clear that $R_b$ becomes independent of $Eo$ for $Eo \gtrsim 100$. For $Eo < 100$, it is seen that $R_b$ exhibits a non-monotonic dependence on $Eo$ which becomes  particularly pronounced \textcolor{black}{ for}  increasing $Nf$.
The behaviour depicted in Figure \ref{fig:ss_bottom_region_curvature_radius} is reflected in the shape of the bubble bottom and its dependence on $Eo$ and $Nf$ as illustrated in Figures \ref{fig:ss_bottom_region_normalise_shape_bottomNf160} and \ref{fig:ss_bottom_region_normalise_shape_bottomEo300}, respectively. Inspection of these figures reveals that with increasing $Nf$ and $Eo$ the bubble tail becomes more pointed. It is possible that for larger values of $Nf$ and $Eo$ a skirted bubble may form followed by the eventual breakup of the protruding tail structure into smaller bubbles. 
\subsubsection{Wake structure below bubble bottom}
\begin{figure}
\centering
	\begin{subfigure}{0.45\linewidth}
	\includegraphics[width = \linewidth, angle=0]{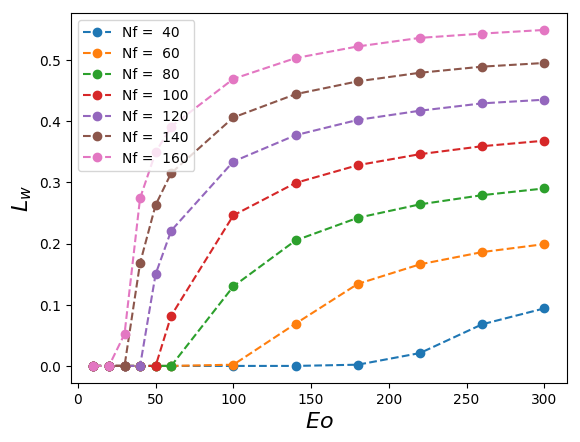}
	\caption{}
	\label{fig:ss_bottom_region_wake_length} 
	\end{subfigure}
	\begin{subfigure}{0.45\linewidth}
        \includegraphics[width=\linewidth]{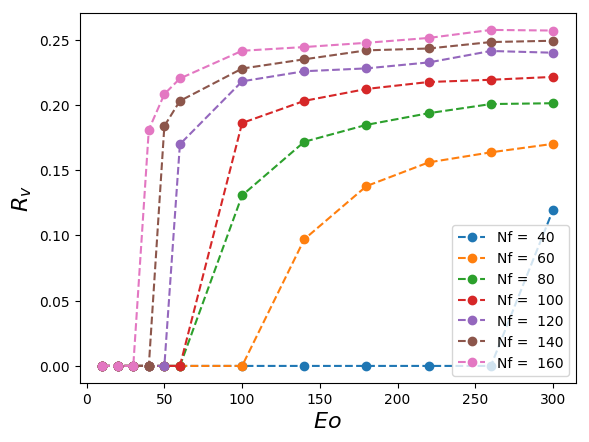}
        \caption{ }
        \label{fig:ss_bottom_region_vortex_eye_radial}
    \end{subfigure}
    \begin{subfigure}{0.45\linewidth}
        \includegraphics[width=\linewidth]{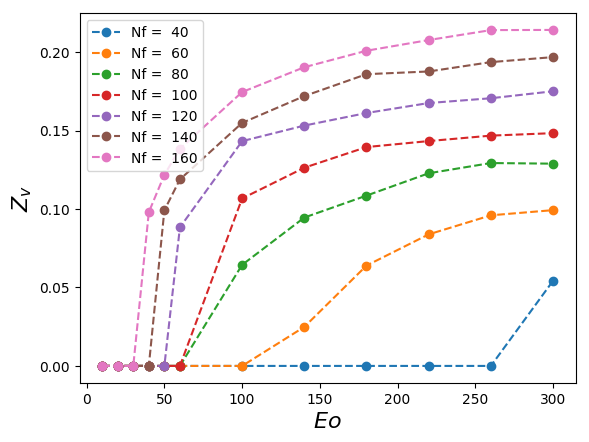}
        \caption{}
         \label{fig:ss_bottom_region_vortex_eye_axial}
    \end{subfigure}
    \begin{subfigure}{0.45\linewidth}
        \includegraphics[width = \linewidth, angle=0]{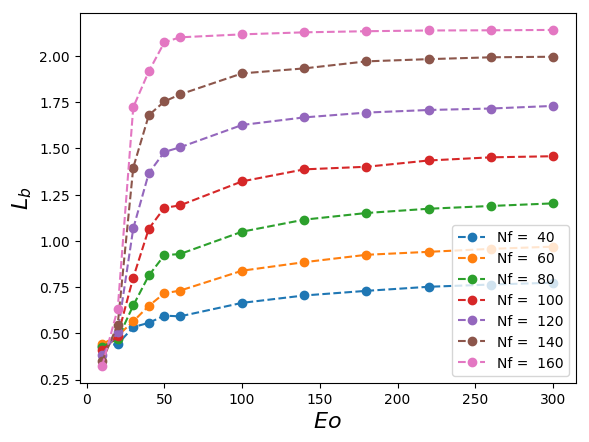}
	\caption{}
	\label{fig:ss_bottom_region_bottom_length}
    \end{subfigure}
    \caption{Characteristics of the wake region for bubble rise in stagnant liquids showing the influence of $Nf$ and $Eo$  on the wake length $L_w$, (a),  the radial and axial locations of the vortex eye with reference to bubble bottom, (c) and (d), and the stabilisation length below the bubble bottom $L_b$, (d), respectively.
	}
\end{figure}

The wake structure is characterised by its length and the position vector of the eye of the vortex, with reference to the position vector of the bubble bottom along the axis of symmetry \citep{Nogueira_etal_2006,Araujo_etal_2012}. 
The wake length $L_w$ is defined as the distance between the bottom of the bubble, along the axis of symmetry of the pipe and the stagnation point, which is the point of flow separation,  behind the bubble \citep{Nogueira_etal_2006}. Thus, $L_w$ is calculated by taking the difference between the axial position of the bubble rear and the stagnation point, and the results are shown in Figure \ref{fig:ss_bottom_region_wake_length}.
As expected, $L_w$ increases with $Nf$ for a fixed $Eo$, and at constant $Nf$ remains zero-valued over a range of $Eo$ before increasing at sufficiently large $Eo$. 
It is noticeable that the $Eo$ value at which the wake emerges depends on $Nf$, decreasing as $Nf$ is increased. 
For all $Nf$, $L_w$ becomes progressively more weakly-dependent on $Eo$ at high $Eo$. 
The dependence of $L_w$ on $Nf$ is explained by considering the fact that with increasing $Nf$ the velocity of the liquid jet emanating from liquid film into the region behind the bubble increases, making the liquid travel a longer distance before flow separation occurs.  

\begin{figure}
\centering
    \includegraphics[width = \linewidth, angle=0]{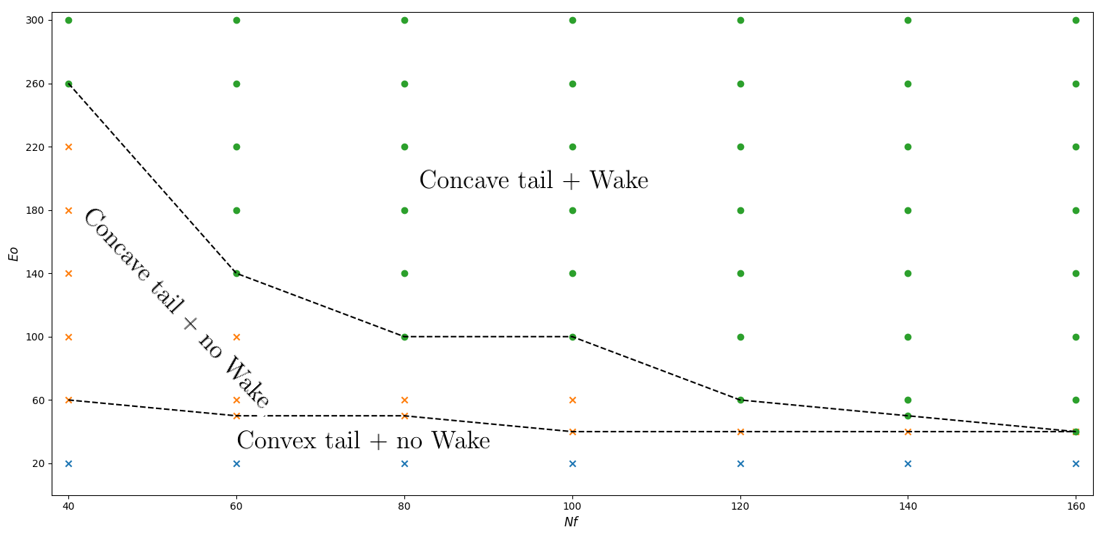}    
	\caption{ 
	Map showing the regions in $Eo$-$Nf$ space where the bubble bottom takes on a concave or convex shape and whether or not this is accompanied by wake formation.}
	\label{fig:ss_bottom_region_wake_regime_map}
\end{figure}

The location of the vortex centre was extracted from the streamline images, generated using open source visualisation tool, VisIt 2.10.3. \citep{VisIt}. For conditions where the wake structure exists, the numerical results for the dimensionless radial, $R_v$, and axial, $Z_v$, coordinates of the vortex eye are plotted as a function of $Eo$ in Figures \ref{fig:ss_bottom_region_vortex_eye_radial} and \ref{fig:ss_bottom_region_vortex_eye_axial}, respectively. The trend for all simulation sets is similar and may be closely described by a function in which the values for both  $R_v$ and  $Z_v$ eventually plateau. For a given $Nf$, these indicate that an increase in $Eo$ shifts the overall vortex center towards the tip of the tail, until no further axial or radial movement occurs. Overall, when juxtaposed with increasing the length of the wake, shown in Figure \ref{fig:ss_bottom_region_wake_length} and deformation of the bubble bottom, shown in Figures \ref{fig:ss_bottom_region_normalise_shape_bottomNf160} and \ref{fig:ss_bottom_region_normalise_shape_bottomEo300}, it appears the combined effect of increasing $Nf$ and $Eo$ is to stretch the wake structure in the axial direction about the vortex eye. 
 Utilising the information from the results of Figures \ref{fig:ss_bottom_region_curvature_radius}, \ref{fig:ss_bottom_region_vortex_eye_axial}, and \ref{fig:ss_bottom_region_vortex_eye_radial} following \cite{Araujo_etal_2012}, a map that demarcates the boundaries where the bubble bottom shape is convex or concave, and indicates whether or not the shape is associated with the presence of a wake as a function of $Nf$ and $Eo$ is shown in \textcolor{black}{Figure \ref{fig:ss_bottom_region_wake_regime_map}}. 
\subsubsection{Developing length below bubble bottom}
The dimensionless stabilisation length below the bubble bottom, $L_b$, similar to the  stabilisation length ahead of the bubble, $L_n$, refers to the distance below the bottom of the bubble in a fixed frame of reference at which the flow field far behind the bubble bottom is attained. This length, in the context of two consecutive rising bubbles, is the minimum distance below the leading bubble bottom, beyond which there is no interaction with the trailing bubble. Numerically, in a moving frame of reference, 
$L_b$ is determined as the difference between the axial locations of the bubble bottom  and the point where the magnitude of the axial velocity along the symmetry axis, starting from the far end of the bubble, is less than 99\%  of its magnitude at the far end. The computed length as a function of the model dimensionless parameters is plotted in Figure \ref{fig:ss_bottom_region_bottom_length}, displaying similar trends to those associated with the wake length $L_w$ discussed above.  
%
%
%
\section{Steady-state bubble motion in flowing liquids ($U_m \neq 0$)} \label{sec:steady_state_characterisations_flow}
In this section, we focus on situations wherein the bubble rises in flowing liquids 
in a fixed frame of reference. The flow in the liquid is characterised using a Froude number  based on the maximum liquid velocity, which corresponds to that at the pipe center. 
The focus in the literature has been on  the dynamics of Taylor bubbles rising in upwardly-flowing liquids characterised by a steady rise speed. 
In contrast, there is a relative dearth of studies concerning Taylor bubble motion in downward liquid flow, which is known to be accompanied by a transition to asymmetric bubble shapes \textcolor{black}{\citep{Martin_1976,Lu_Prosperetti_2006,Nicklin_etal_1962,Figueroa-Espinoza_Fabre_2011,Fershtman_etal_2017,Fabre_Figueroa-Espinoza_2014}}. 
\subsection{Bubble rise speed in upward liquid flow} 
\begin{figure}
\centering
    \begin{subfigure}{\linewidth}
        \centering\includegraphics[width = 0.60\linewidth, angle=0]{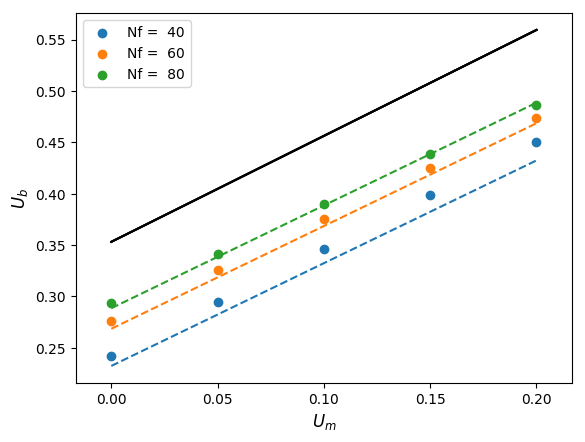}    
        \caption{}
        \label{fig:ss_upward_liq_Flow__validationEo180}
    \end{subfigure}
	\begin{subfigure}{0.48\linewidth}
        \includegraphics[width=\linewidth]{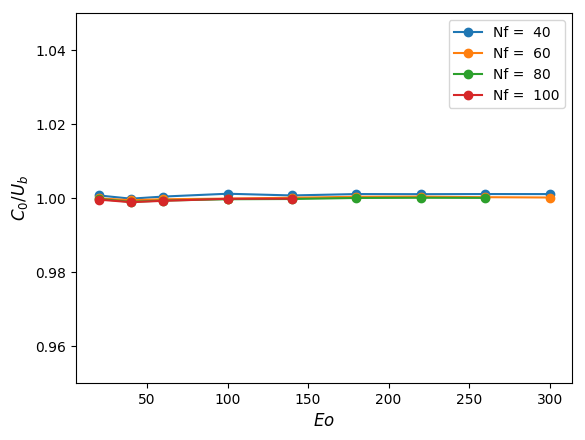}
        \caption{ }
        \label{fig:ss_upward_liq_Flow__validationC0}
    \end{subfigure}
    \begin{subfigure}{0.48\linewidth}
        \includegraphics[width=\linewidth]{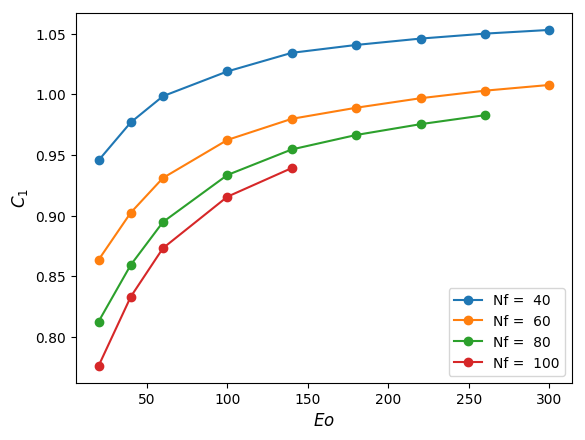}
        \caption{}
         \label{fig:ss_upward_liq_Flow__validationC1}
    \end{subfigure}
	\caption{Effect of imposed upward liquid flow speed $U_m$ on the bubble rise speed $U_b$ for varying $Nf$, (a): comparison between the numerical results (coloured markers),  predictions based on the \cite{Nicklin_etal_1962} correlation  \eqref{eq:lr_bubble_rise_vel_flow} (black solid line) with the \cite{Bendiksen_1985} relations  \eqref{eq:ss_bubble_rise_vel_bendiksen_C0} and \eqref{eq:ss_bubble_rise_vel_bendiksen_C1} used for coefficients $C_0$ and $C_1$, and predictions using the \cite{Viana_etal_2003} correlation for $C_0$ given by equation \eqref{eq:lr_bubble_rise_vel_viana} and the \cite{Bendiksen_1985} relation for $C_1$ expressed by \eqref{eq:ss_bubble_rise_vel_bendiksen_C1} (coloured dashed lines); effect of $Nf$ and $Eo$ on the numerically-generated $C_0$ (normalised by $U_b$), (b), and $C_1$, (c).}
\end{figure}

In Figure \ref{fig:ss_upward_liq_Flow__validationEo180}, the numerical simulation results for upward liquid flow are compared with predictions based on the correlation of  \cite{Nicklin_etal_1962} given by equation \eqref{eq:lr_bubble_rise_vel_flow} with expressions for $C_0$ and $C_1$ provided by \cite{Bendiksen_1985} taking into consideration the effect of $Eo$ as
\begin{equation}
C_0 = \frac{0.486}{\sqrt{2}} \sqrt{1 + 20 \left( 1 - \frac{6.8}{Eo}\right)} \left \{ \frac{1. - 0.96 e^{-0.0165Eo}}{1. - 0.52 e^{-0.0165Eo}}\right \}, \label{eq:ss_bubble_rise_vel_bendiksen_C0}
\end{equation}
\begin{equation}
C_1 = 1.145 \left[ 1 - \frac{20}{Eo} \left( 1 - e^{-0.0125 Eo}\right)\right]. \label{eq:ss_bubble_rise_vel_bendiksen_C1}
\end{equation}
It is evident that equations \eqref{eq:lr_bubble_rise_vel_flow} with \eqref{eq:ss_bubble_rise_vel_bendiksen_C0} and \eqref{eq:ss_bubble_rise_vel_bendiksen_C1} over-predict the bubble rise speed. This is because the expressions for $C_0$ and $C_1$ were derived for cases in which flow due to the bubble motion was considered to be inviscid,  an assumption that gains with increasing $Nf$. 
The agreement with the numerical results improves significantly when the correlation of \cite{Viana_etal_2003} is used to calculate $C_0$; this correlation accounts for the effects of viscosity and surface tension and the agreement improves further with increasing $Nf$. We can estimate values for  $C_0$ and $C_1$ from our numerical simulations for various $Nf$ and $Eo$, and the results are shown in Figures \ref{fig:ss_upward_liq_Flow__validationC0} and \ref{fig:ss_upward_liq_Flow__validationC1}, respectively. It is seen that $C_0/U_b$ remains approximately equal to unity over the range of $Nf$ and $Eo$ studied, while $C_1$ increases monotonically with $Eo$ for all $Nf$ considered reaching a plateau at high $Eo$. 
\subsection{Steady bubble shapes and flow fields in flowing liquids}
\begin{figure}
\centering
    \begin{subfigure}{\linewidth}
    \centering\includegraphics[width = 0.70\linewidth, angle=0]{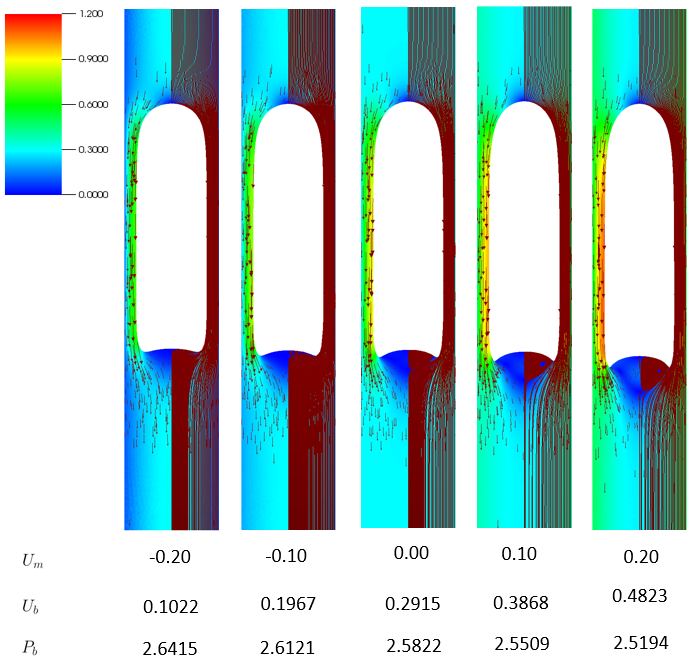}
    \caption{}
    \label{fig:ss_flowfield_nf80eo140}
    \end{subfigure}
	\begin{subfigure}{0.45\linewidth}
	\includegraphics[width = \linewidth, angle=0]{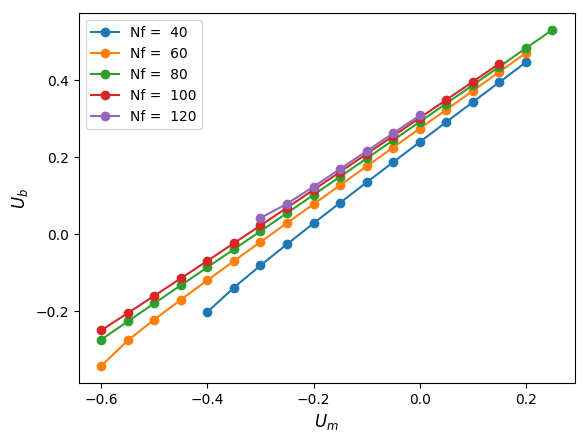}
	\caption{}
	\label{fig:ss_downward_liq_Flow_bubble_rise_velEo140}
	\end{subfigure}
	\begin{subfigure}{0.45\linewidth}
	\includegraphics[width = \linewidth, angle=0]{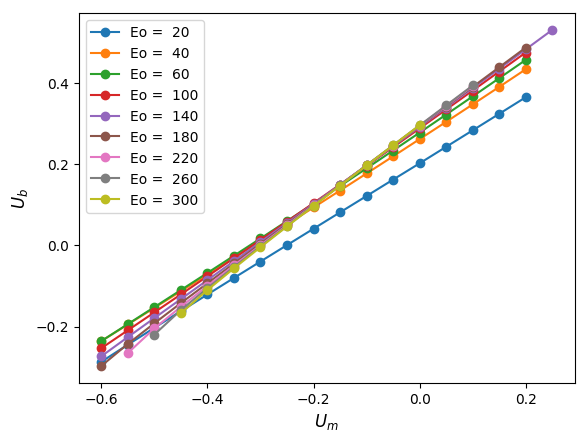}
	\caption{}
	\label{fig:ss_downward_liq_Flow_bubble_rise_velNf80}
	\end{subfigure}
	\caption{Effect of $U_m$ on the steady-state bubble shape and the surrounding flow field with $Nf = 80$ and $Eo = 140$, (a); here, the streamlines and vector fields are superimposed on velocity magnitude pseudocolour plot on the right and left sides of the symmetry axis, respectively;  variation of the steady bubble rise speed $U_b$ with $U_m$ (b); for different $Nf$ and with $Eo=140$ (c); for different $Eo$ and with $Nf=80$.}
\end{figure}

For a constant $Nf = 80$ and $Eo = 140$, the effect of imposed upward and downward liquid flow is shown in Figure \ref{fig:ss_flowfield_nf80eo140}. 
It is seen clearly that a decrease (increase) in the intensity of the wake flow, accompanied by a decrease (increase) of the concavity of the bubble bottom, is observed with an increase in the magnitude of the downward (upward) liquid flow. This, as  noted earlier when discussing the stagnant liquid case, can be linked to the decrease (increase) in the magnitude of the liquid emerging from the film into the liquid slug, which is a manifestation of the decrease (increase) in the bubble rise speed, as the downward (upward) liquid velocity is increased. Quantitatively, the effect of $U_m$ on $U_b$ is shown in Figure \ref{fig:ss_downward_liq_Flow_bubble_rise_velEo140} whence we deduce the existence of a critical $U_m$ value for downward flow that leads to bubble arrest characterised by $U_b=0$, which increases with $Nf$ and decreases (increases) with $Eo$ for $Eo \ge 100 $ ($Eo < 100$), respectively. 

\begin{figure}
\centering
	\begin{subfigure}{0.45\linewidth}
	\includegraphics[width = \linewidth, angle=0]{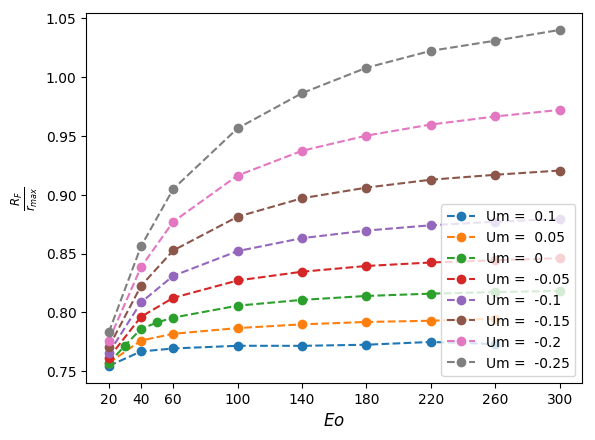}
	\caption{}
	\label{fig:ss_Liq_flow_Nose_region_frontal_radius_rmax1Nf80}
	\end{subfigure}
	\begin{subfigure}{0.45\linewidth}
	\includegraphics[width = \linewidth, angle=0]{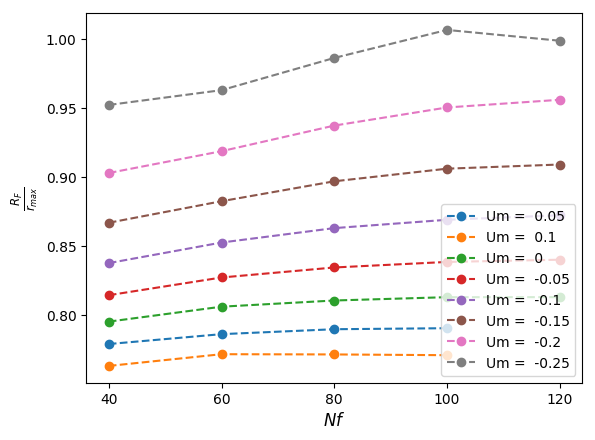}
	\caption{}
	\label{fig:ss_Liq_flow_Nose_region_frontal_radius_rmax1Eo140}
	\end{subfigure}
	\begin{subfigure}{0.45\linewidth}
	\includegraphics[width = \linewidth, angle=0]{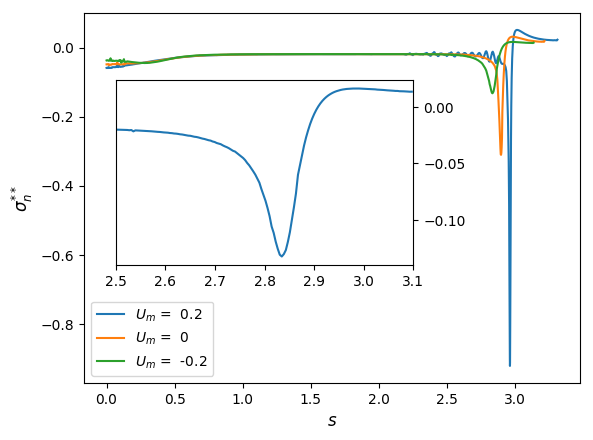}
	\caption{}
	\label{fig:ss_liq_Flow_Interface_region_normal_stress_refPb_Nf80Eo140}
	\end{subfigure}
	\caption{Effect of $U_m$ on the steady-state bubble interface features (a); variation of  frontal radius, $R_F$ with $U_m$ and $Eo$ for $Nf = 80$ (b); variation of  frontal radius, $R_F$ with $U_m$ and $Nf$ for $Eo = 140$ (c); spatial variation of the steady, modified interface normal stress $\sigma^{**}_n$ for different $U_m$ and with $Nf = 80$ and $Eo = 140$; the inset shows an enlarged view of $\sigma^{**}_n$ for $2.5 \leq s \leq 3.1$ for $U_m=0.2$ which demonstrates that this quantity is well-resolved in this boundary-like region of rapid variation.}
\end{figure}
It is also noticeable from Figure \ref{fig:ss_flowfield_nf80eo140} that there is an increase (decrease) in the radius of curvature of the bubble nose with increasing  magnitude of the downward (upward) liquid flow \textcolor{black}{(see also Figures  \ref{fig:ss_Liq_flow_Nose_region_frontal_radius_rmax1Nf80} and  \ref{fig:ss_Liq_flow_Nose_region_frontal_radius_rmax1Eo140})}. 
This flattening (sharpening) of the bubble nose can be attributed to the increase (decrease) in the normal stress exerted on the bubble nose relative to that in stagnant liquid as a result of the increased opposing (reinforcing) inertial force in the downward (upward) liquid flow.

It is clear from Figure \ref{fig:ss_liq_Flow_Interface_region_normal_stress_refPb_Nf80Eo140}
that the interface normal stress is an increasing (decreasing) function of the increased liquid velocity in the downward (upward) liquid flow. As explained in the previous section for stagnant liquids, within the equilibrium film, the normal stress, total pressure, and the bubble pressure are approximately equal, which is responsible for the observed increase (decrease) in bubble pressure with increasing downward (upward) liquid flow (see Figure \ref{fig:ss_flowfield_nf80eo140}). Also, outside the  equilibrium film region, we had stated that it is the interplay between the viscous stress and curvature that determines the shape of the regions. To buttress this claim, the normal stress is again modified by choosing the reference pressure to be the bubble pressure such that
\begin{equation}
\sigma_n^{**} = -  \left[-p_T* + 2Nf^{-1} \mathbf{n} \cdot \frac{d \mathbf{u}}{d n} \right] = - {Eo}^{-1} \kappa,
\end{equation} 
where 
$p_T* = p_T - P_b$, 
thereby making the normal stress in the equilibrium film region approximately zero as the stress due to interfacial curvature $\kappa_b$ is negligibly small. As the nose region is approached, the net effect of the viscous stress on the normal stress in downward (upward) liquid flow is to increase (decrease) the normal stress relative to that in a stagnant liquid, which in order to satisfy the normal stress balance at the interface, the curvature stress has to decrease (increase), leading to the observed increase (decrease) in the radius of curvature of the nose.


\begin{figure}
\centering
\begin{tabular}{cccccccccc} 
\includegraphics[width=0.15 \linewidth, angle=0]{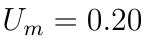} &
    \includegraphics[width= 0.03\linewidth, angle=0]{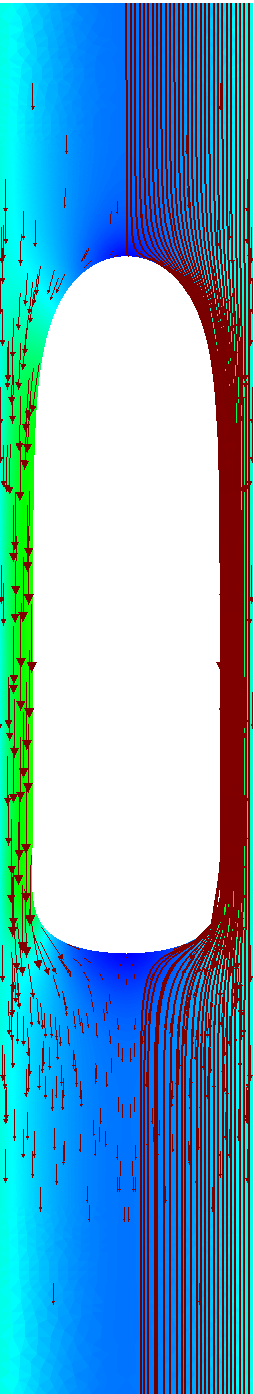} &
  	\includegraphics[width= 0.03\linewidth, angle=0]{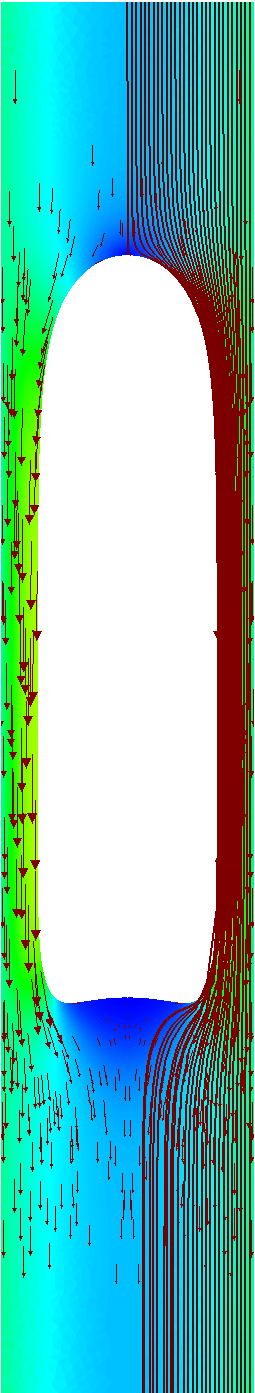} &
  	\includegraphics[width= 0.03\linewidth, angle=0]{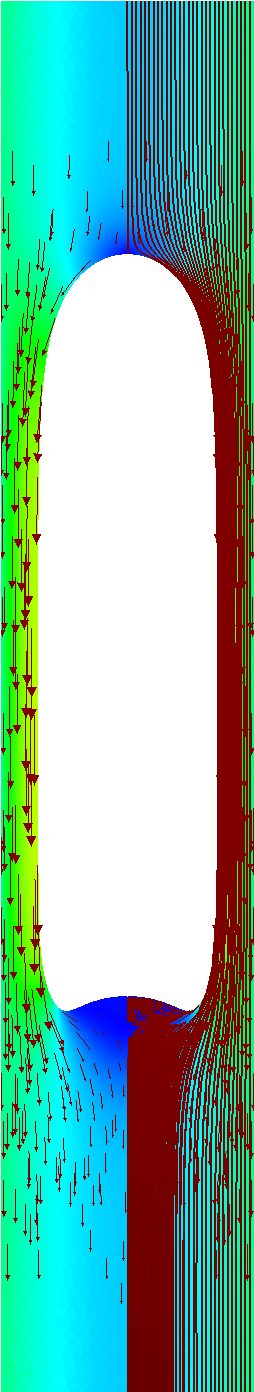} &
  	\includegraphics[width= 0.03\linewidth, angle=0]{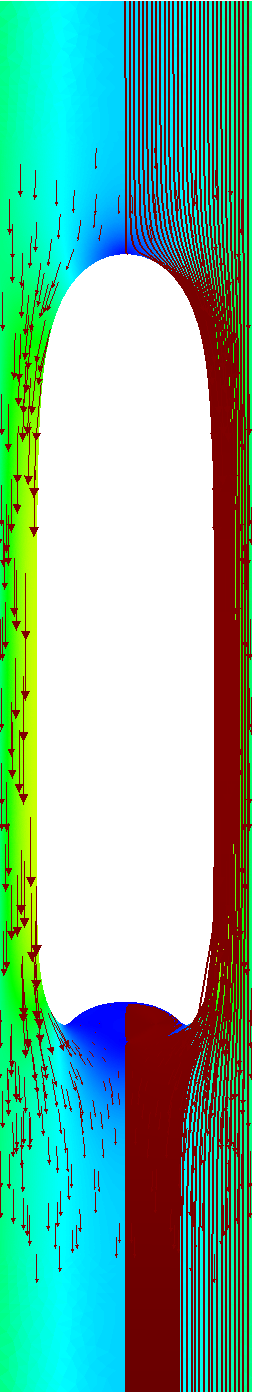} &
  	\includegraphics[width= 0.03\linewidth, angle=0]{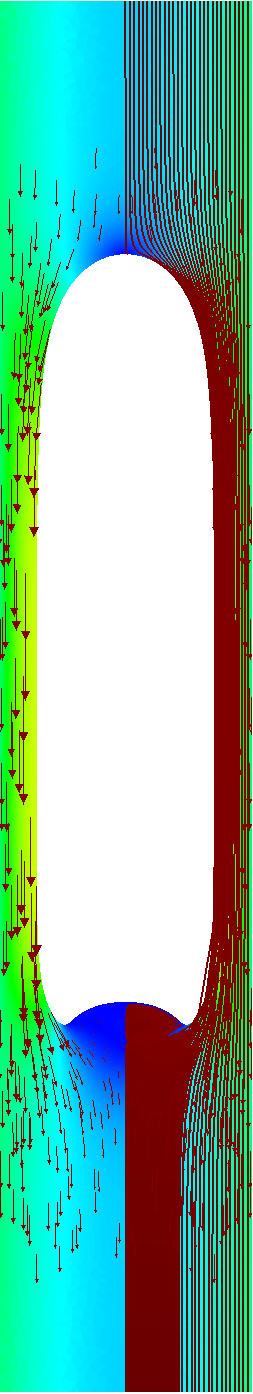} &
  	\includegraphics[width= 0.03\linewidth, angle=0]{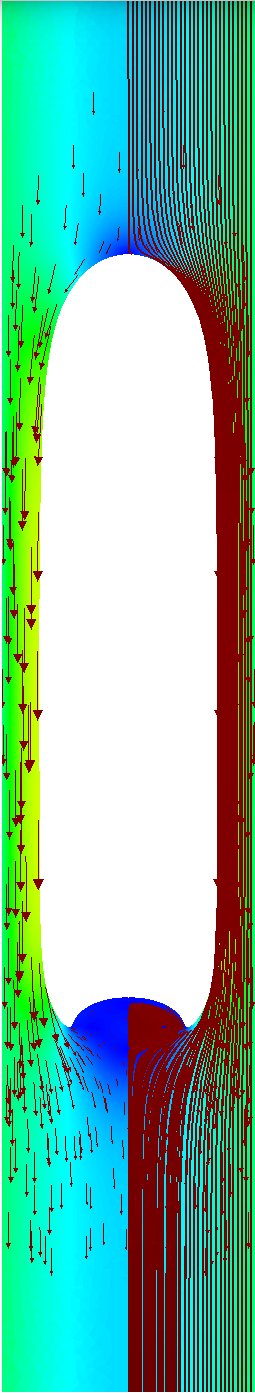} &
  	\includegraphics[width= 0.03\linewidth, angle=0]{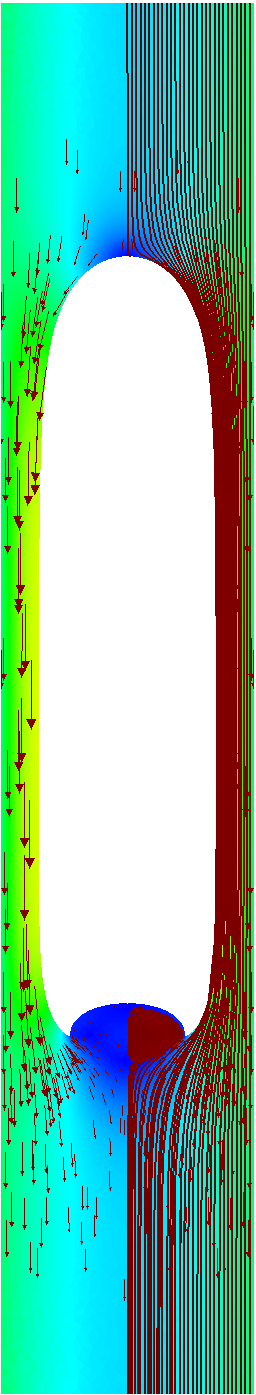} &
  	\includegraphics[width= 0.03\linewidth, angle=0]{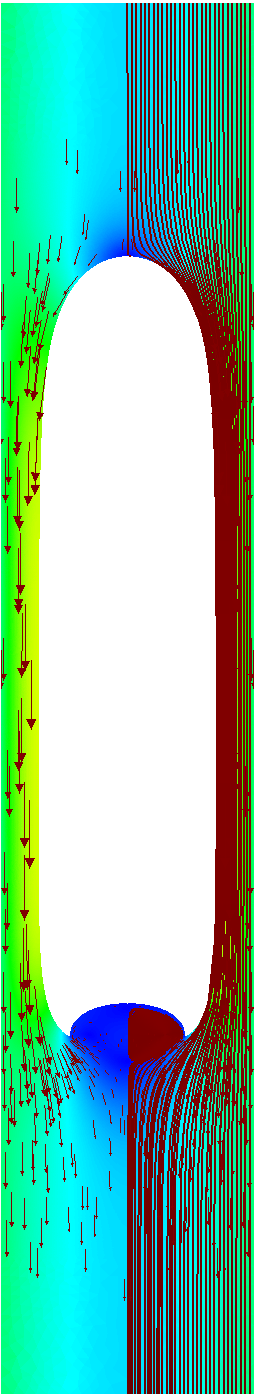} &
  	\includegraphics[width=0.08 \linewidth, angle=0]{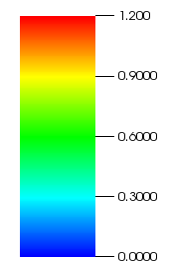}
\\ 
\includegraphics[width=0.15 \linewidth, angle=0]{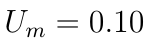} &
    \includegraphics[width= 0.03\linewidth, angle=0]{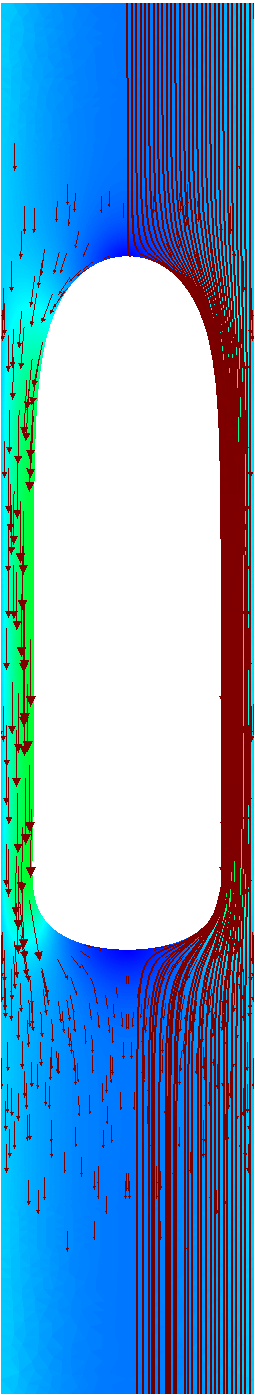} &
  	\includegraphics[width= 0.03\linewidth, angle=0]{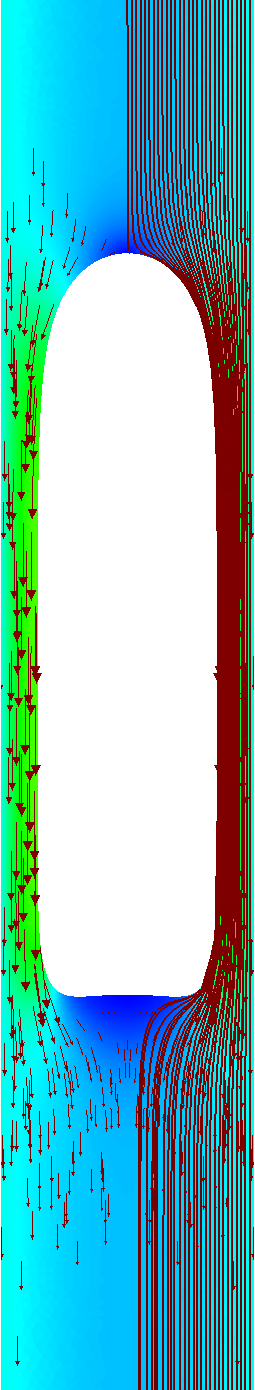} &
  	\includegraphics[width= 0.03\linewidth, angle=0]{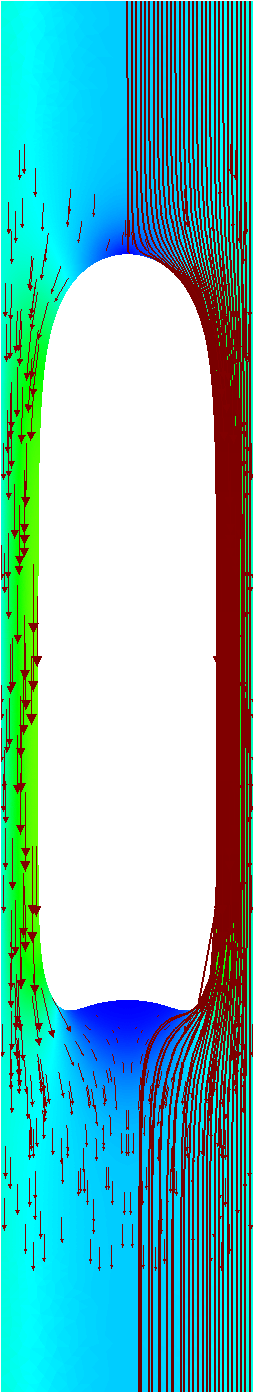} &
  	\includegraphics[width= 0.03\linewidth, angle=0]{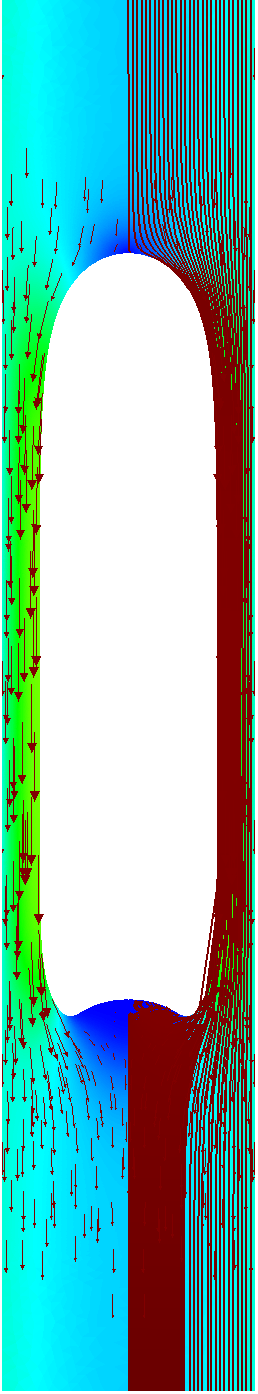} &
  	\includegraphics[width= 0.03\linewidth, angle=0]{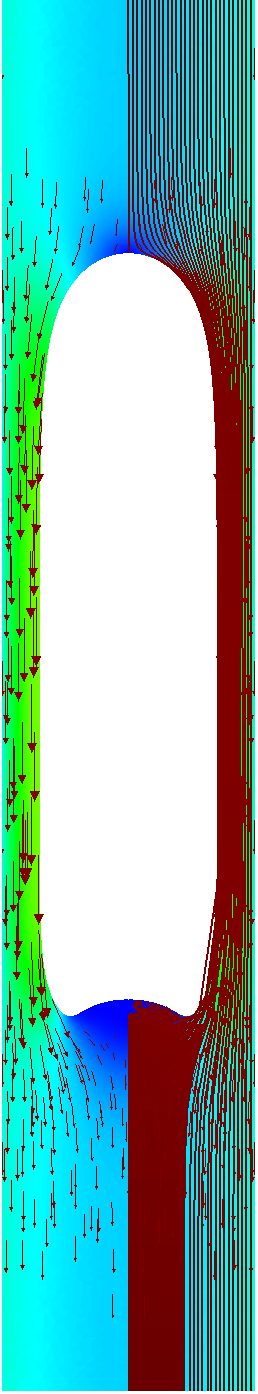} &
  	\includegraphics[width= 0.03\linewidth, angle=0]{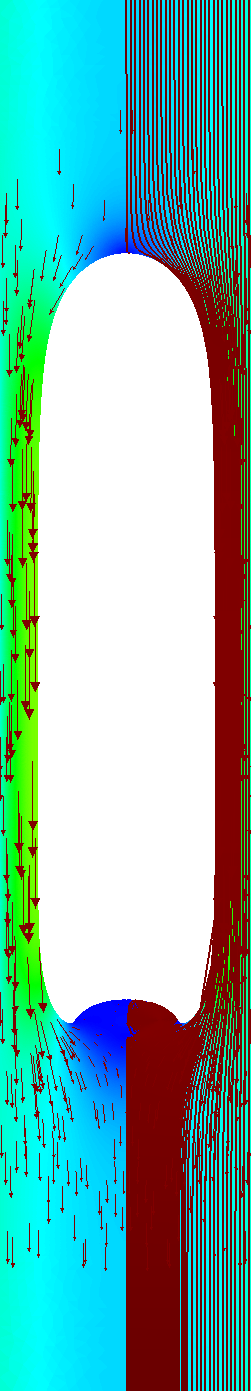} &
  	\includegraphics[width= 0.03\linewidth, angle=0]{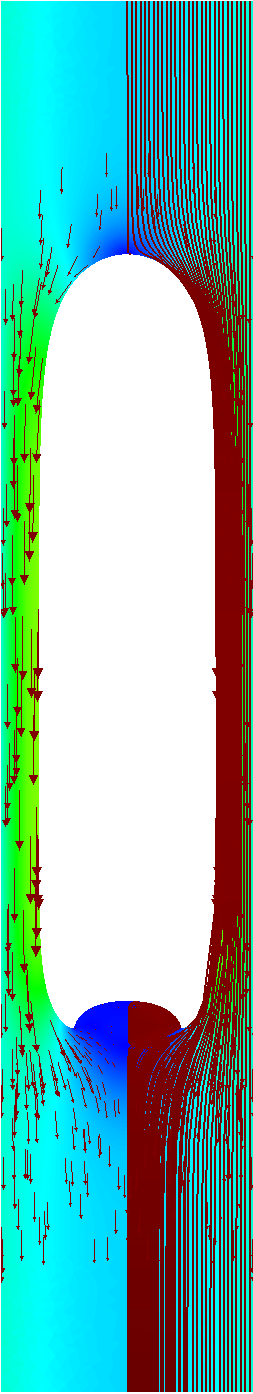} &
  	\includegraphics[width= 0.03\linewidth, angle=0]{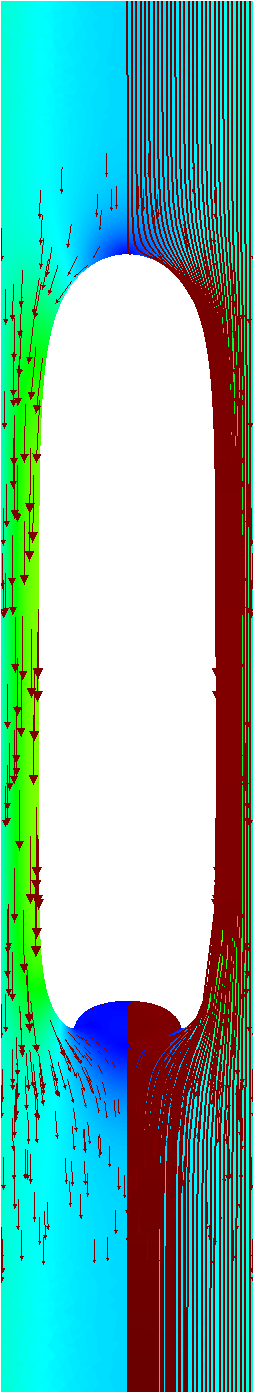} &
  	\includegraphics[width=0.08 \linewidth, angle=0]{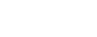}
\\ 
\includegraphics[width=0.10 \linewidth, angle=0]{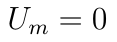} &
\includegraphics[width= 0.03\linewidth, angle=0]{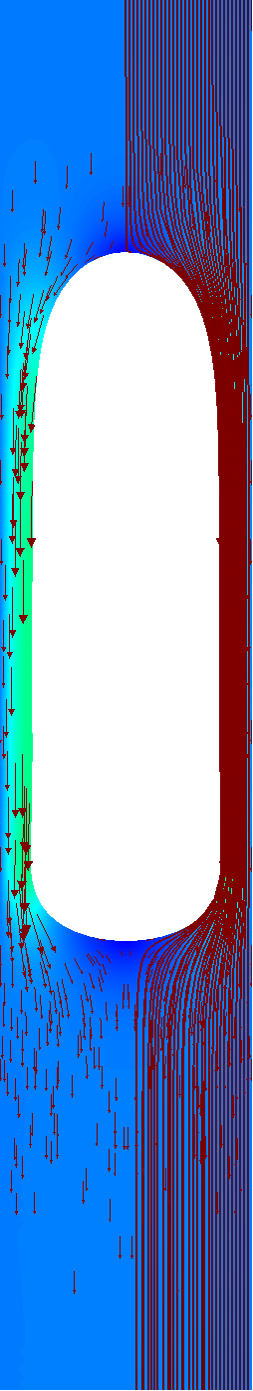} & \includegraphics[width= 0.03\linewidth, angle=0]{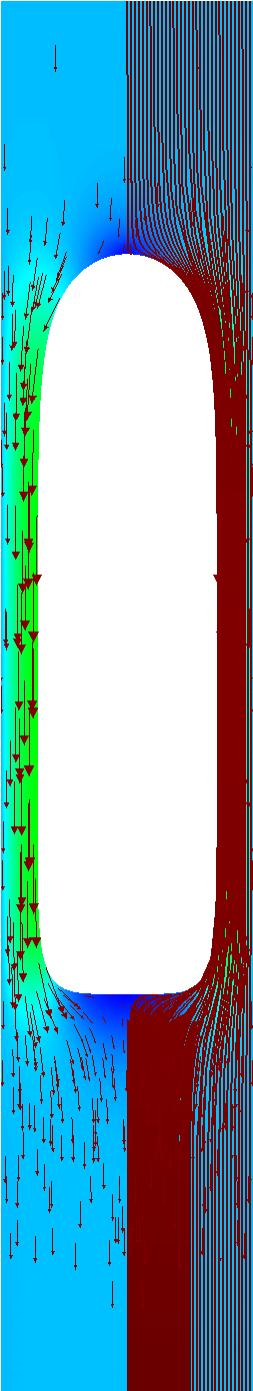} & \includegraphics[width= 0.03\linewidth, angle=0]{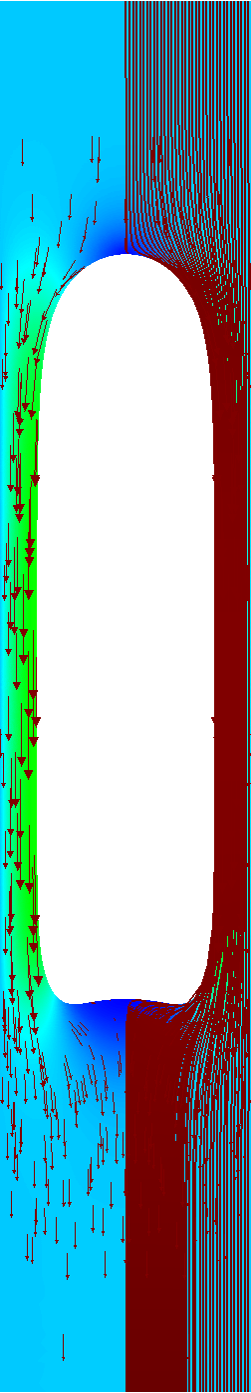} & \includegraphics[width= 0.03\linewidth, angle=0]{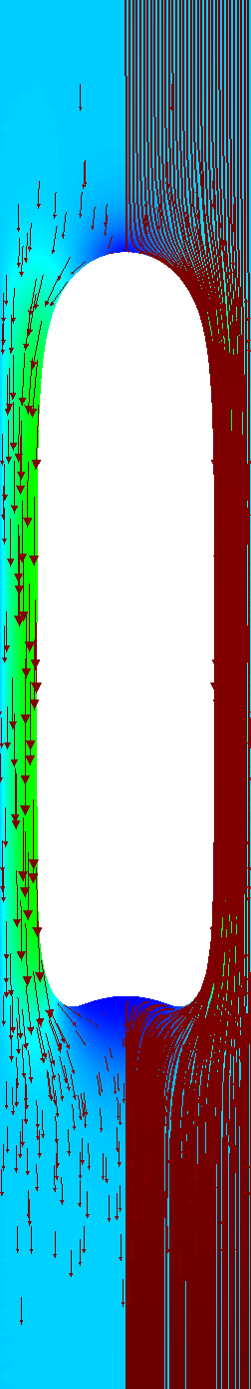} &
\includegraphics[width= 0.03\linewidth, angle=0]{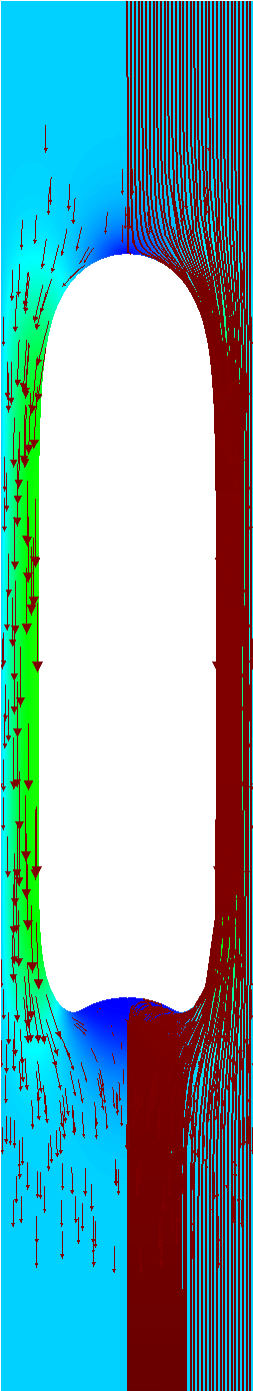} & \includegraphics[width= 0.03\linewidth, angle=0]{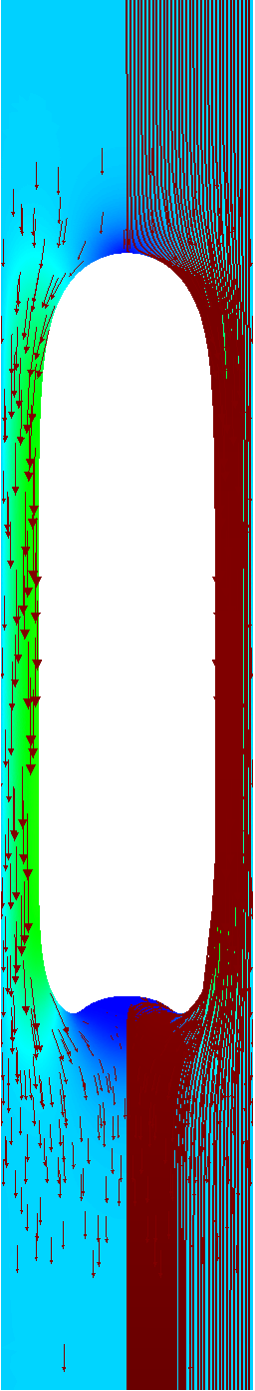} & \includegraphics[width= 0.03\linewidth, angle=0]{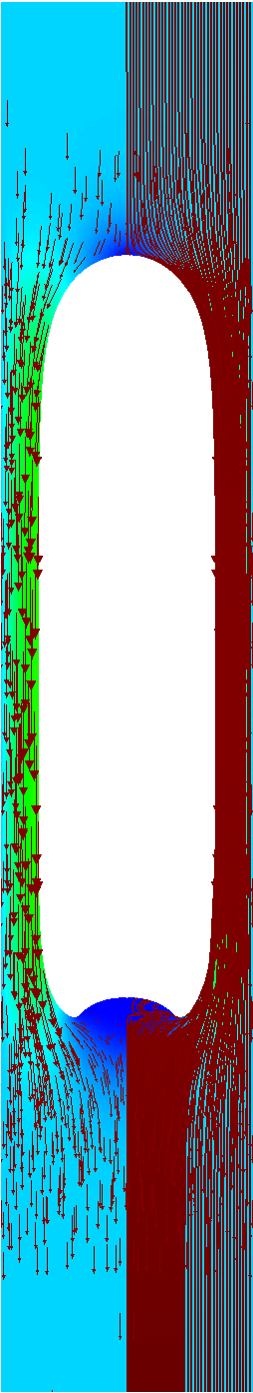} & \includegraphics[width= 0.03\linewidth, angle=0]{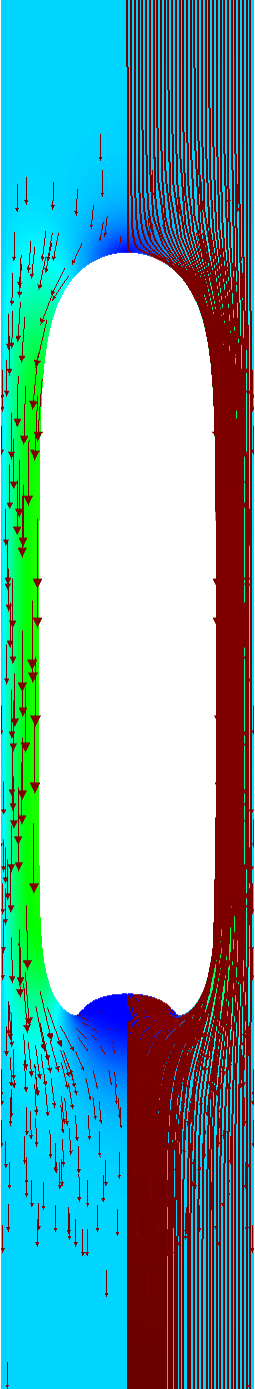} &
\includegraphics[width=0.08 \linewidth, angle=0]{Figures/Eo000}
\\ 
\includegraphics[width=0.15 \linewidth, angle=0]{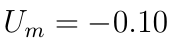} &
    \includegraphics[width= 0.03\linewidth, angle=0]{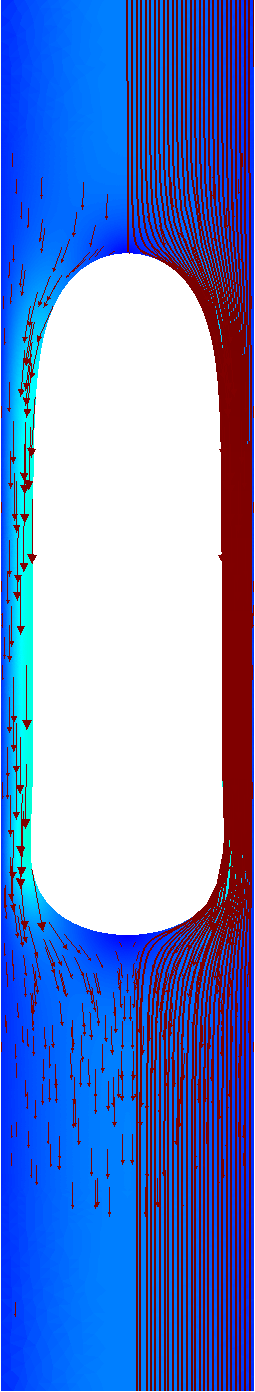} &
  	\includegraphics[width= 0.03\linewidth, angle=0]{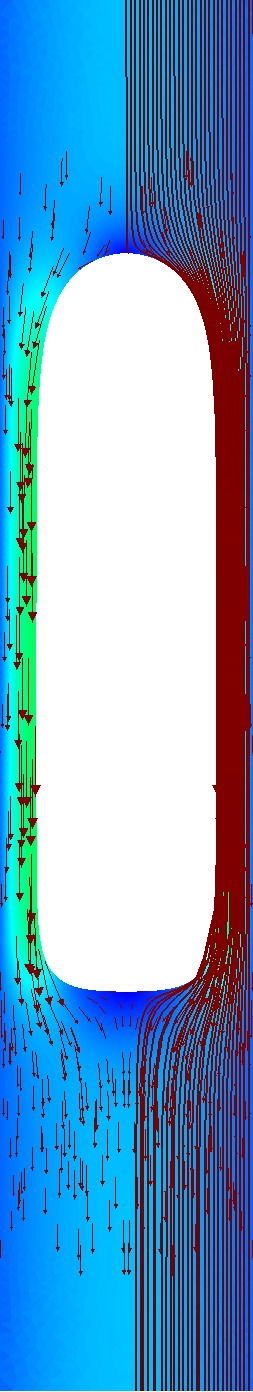} &
  	\includegraphics[width= 0.03\linewidth, angle=0]{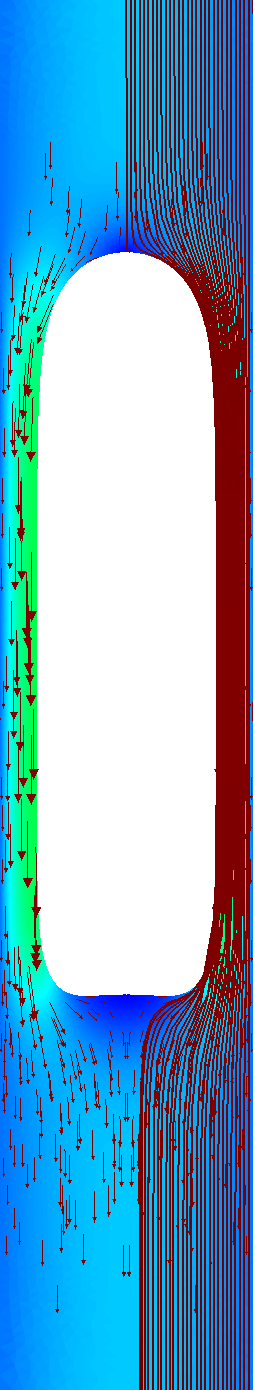} &
  	\includegraphics[width= 0.03\linewidth, angle=0]{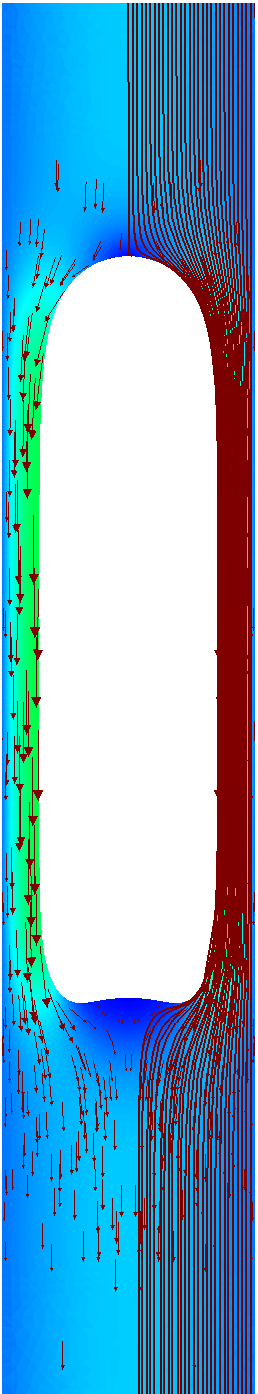} &
  	\includegraphics[width= 0.03\linewidth, angle=0]{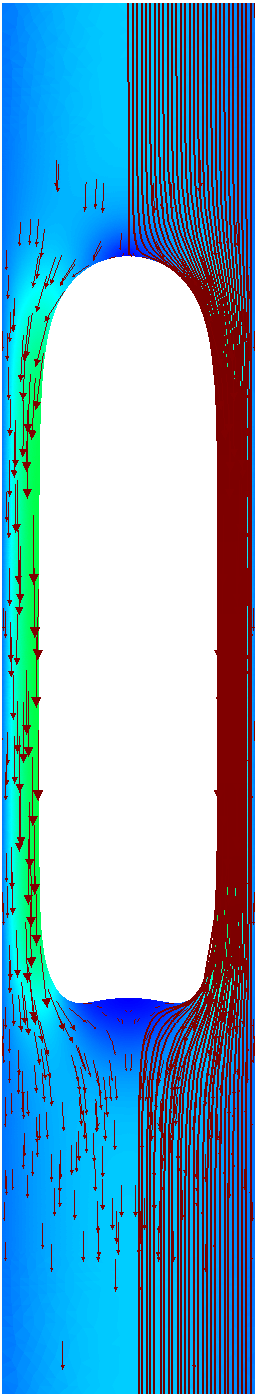} &
  	\includegraphics[width= 0.03\linewidth, angle=0]{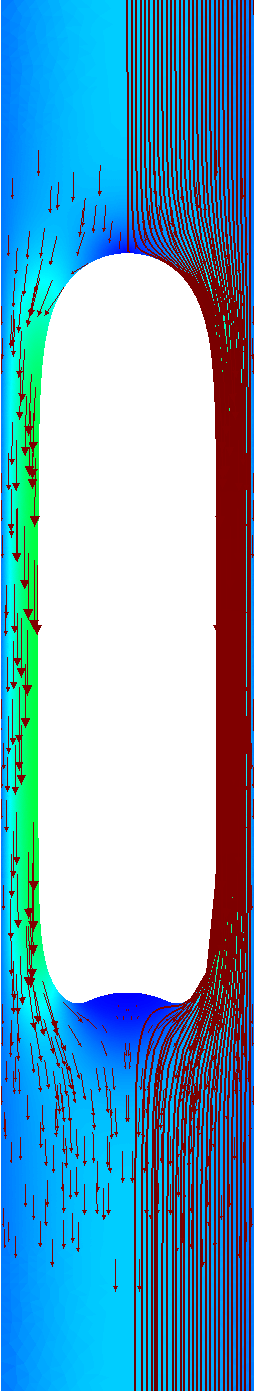} &
  	\includegraphics[width= 0.03\linewidth, angle=0]{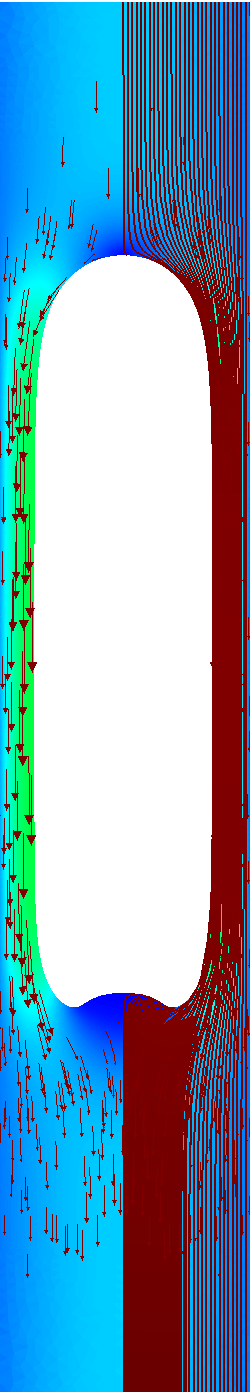} &
  	\includegraphics[width= 0.03\linewidth, angle=0]{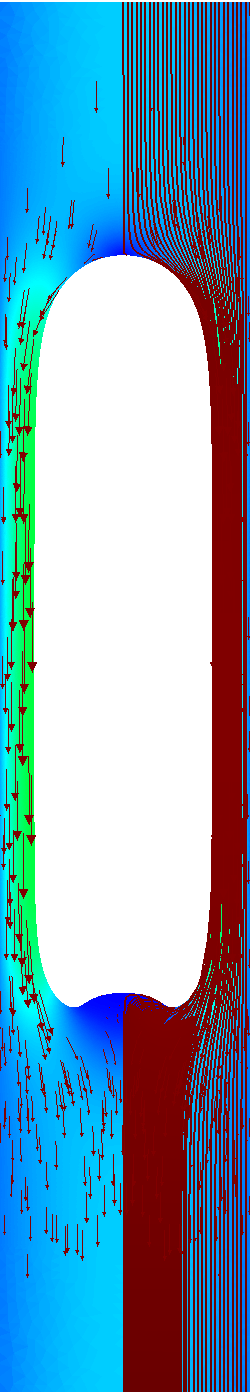} &
  	\includegraphics[width=0.08 \linewidth, angle=0]{Figures/Eo000}
\\ 
\includegraphics[width=0.15 \linewidth, angle=0]{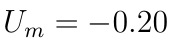} &
    \includegraphics[width= 0.03\linewidth, angle=0]{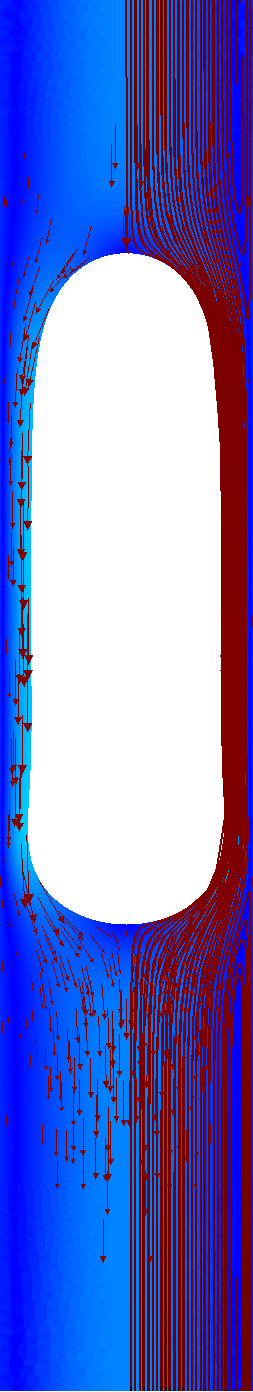} &
  	\includegraphics[width= 0.03\linewidth, angle=0]{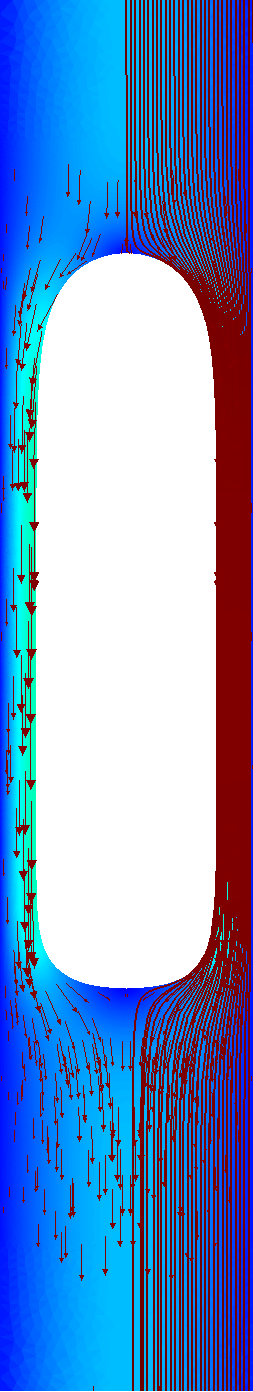} &
  	\includegraphics[width= 0.03\linewidth, angle=0]{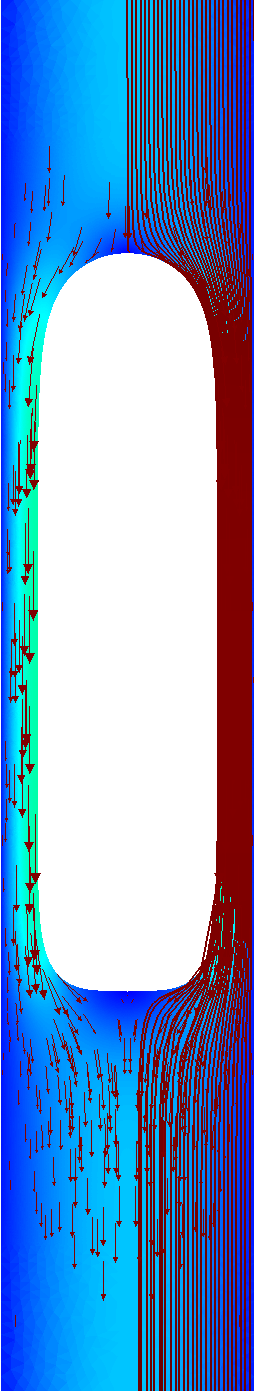} &
  	\includegraphics[width= 0.03\linewidth, angle=0]{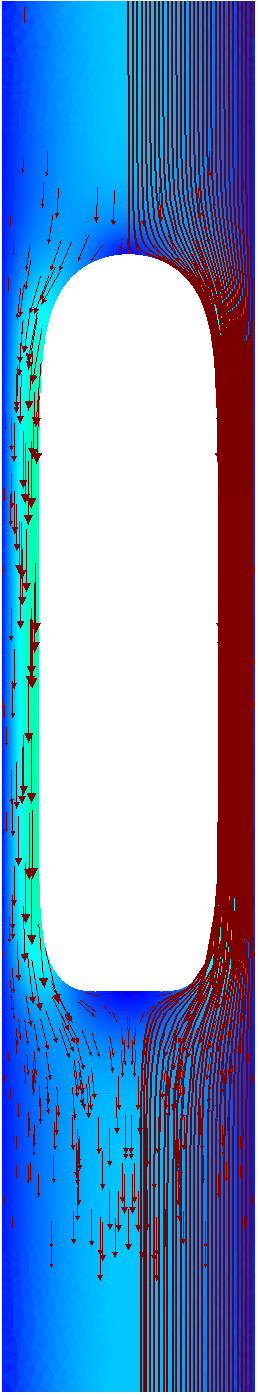} &
  	\includegraphics[width= 0.03\linewidth, angle=0]{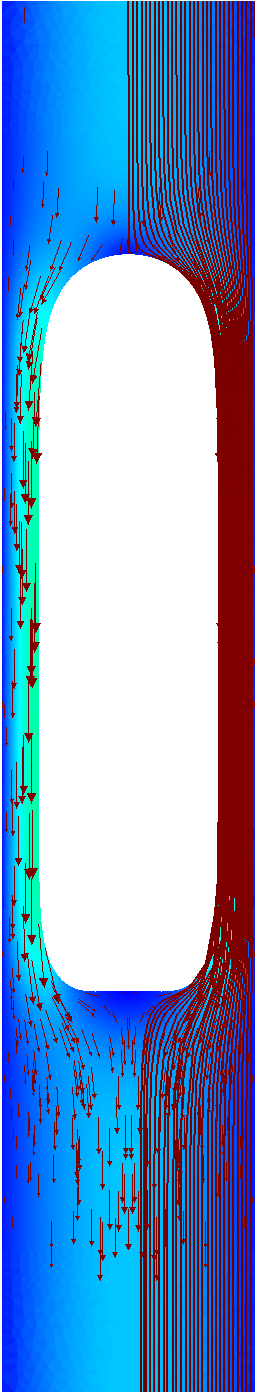} &
  	\includegraphics[width= 0.03\linewidth, angle=0]{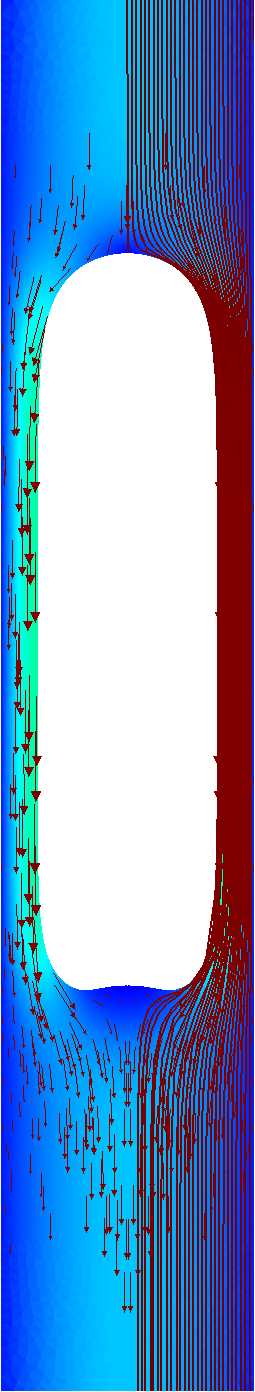} &
  	\includegraphics[width= 0.03\linewidth, angle=0]{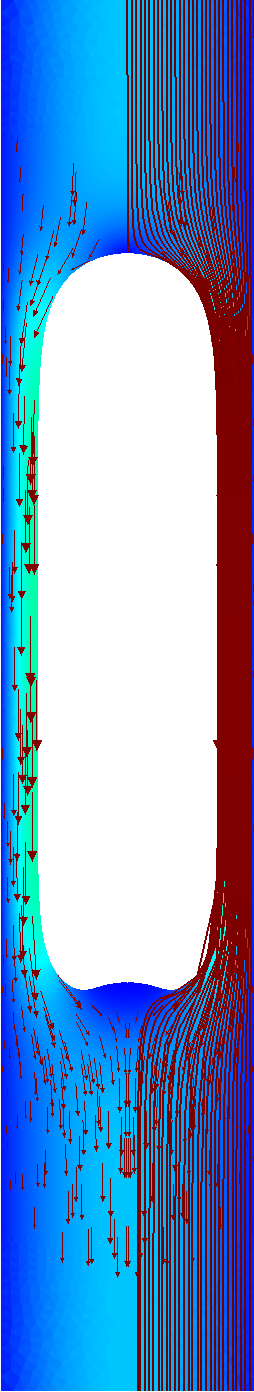} &
  	\includegraphics[width= 0.03\linewidth, angle=0]{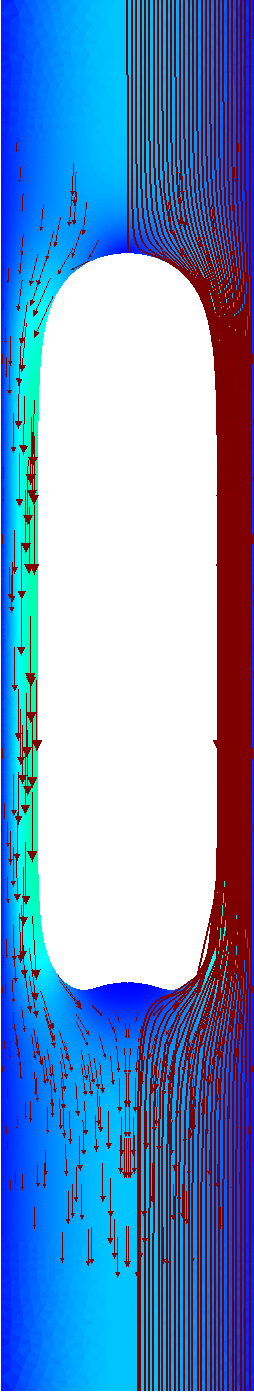} &
  	\includegraphics[width=0.08 \linewidth, angle=0]{Figures/Eo000}
\\ 
\includegraphics[width=0.15 \linewidth, angle=0]{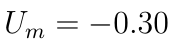} &
    \includegraphics[width= 0.03\linewidth, angle=0]{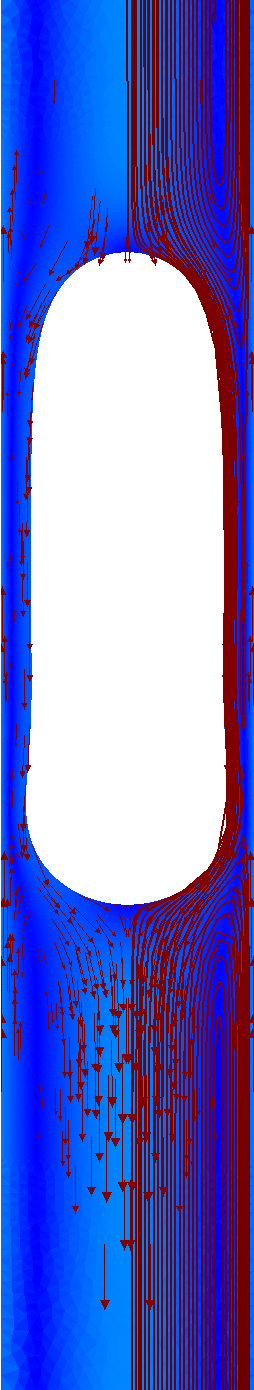} &
  	\includegraphics[width= 0.03\linewidth, angle=0]{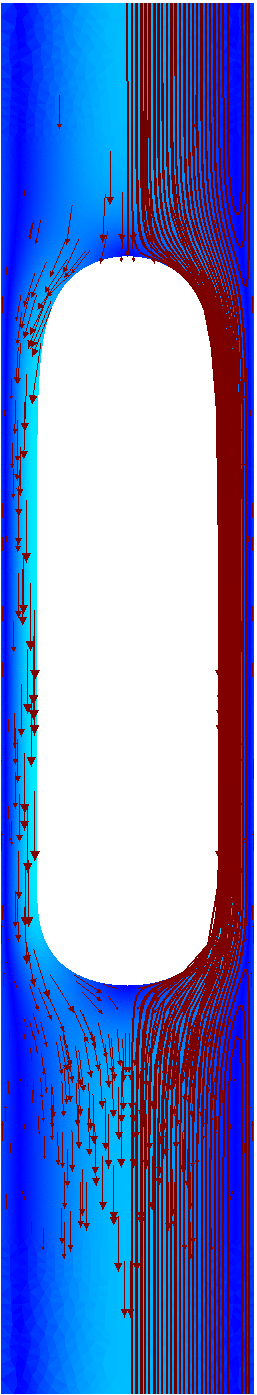} &
  	\includegraphics[width= 0.03\linewidth, angle=0]{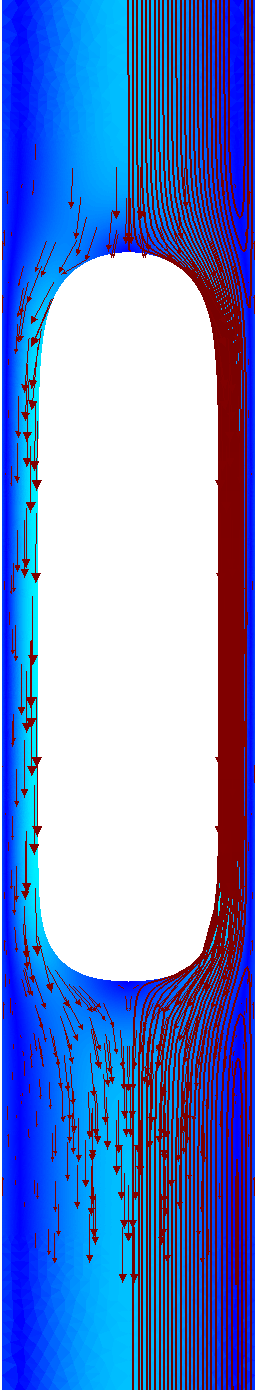} &
  	\includegraphics[width= 0.03\linewidth, angle=0]{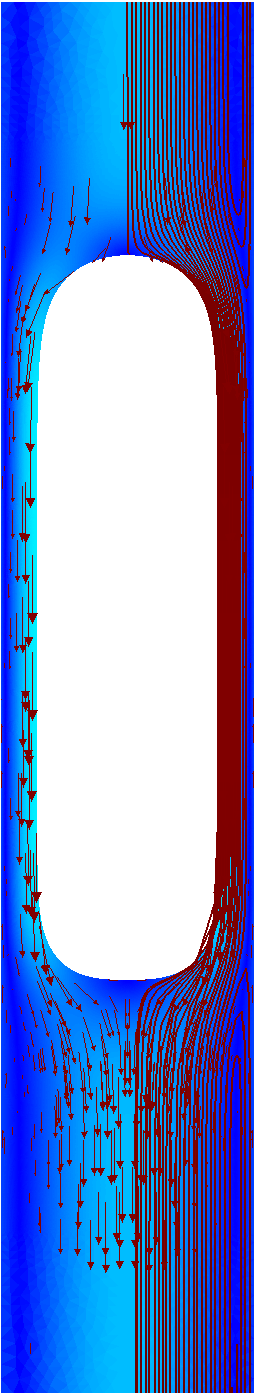} &
  	\includegraphics[width= 0.03\linewidth, angle=0]{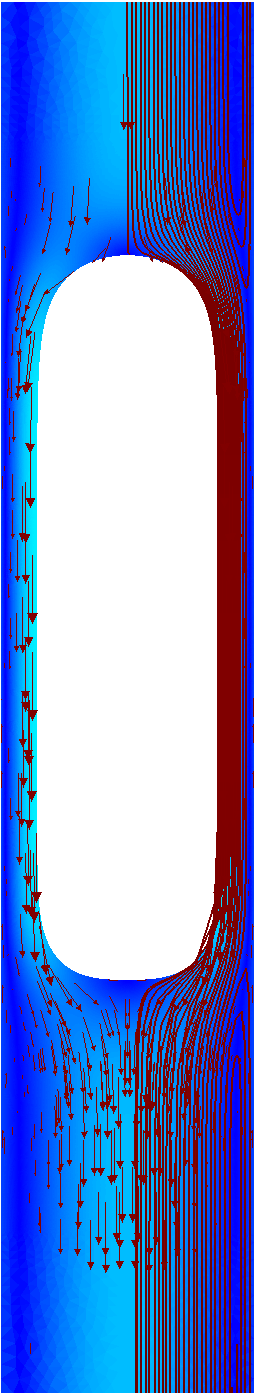} &
  	\includegraphics[width= 0.03\linewidth, angle=0]{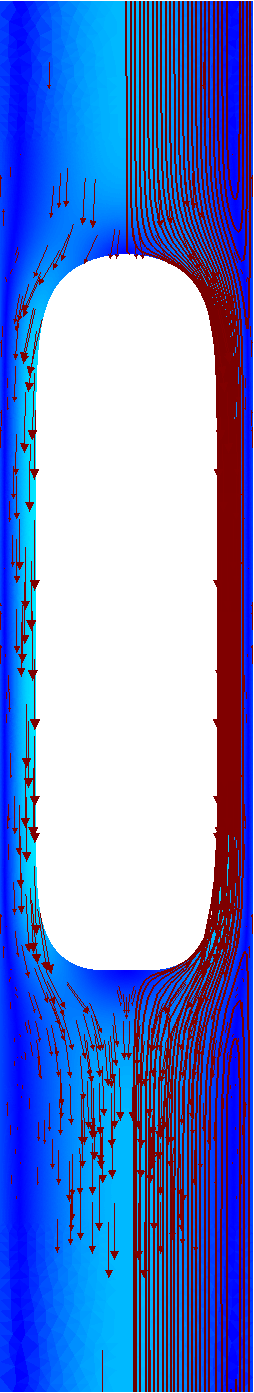} &
  	\includegraphics[width= 0.03\linewidth, angle=0]{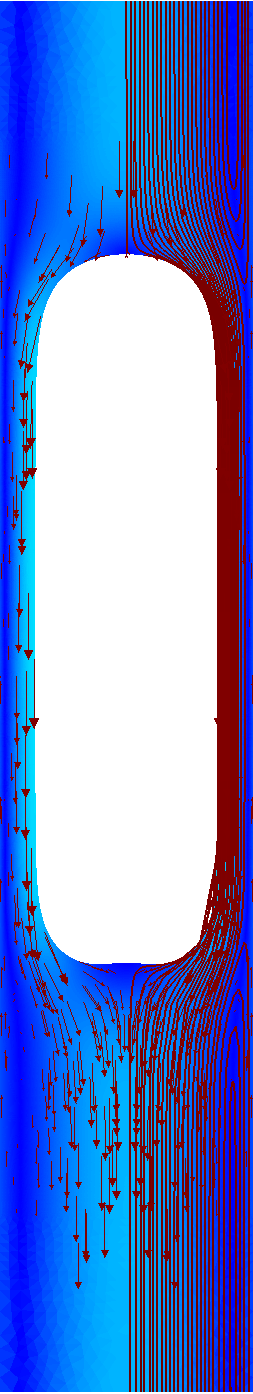} &
  	\includegraphics[width= 0.03\linewidth, angle=0]{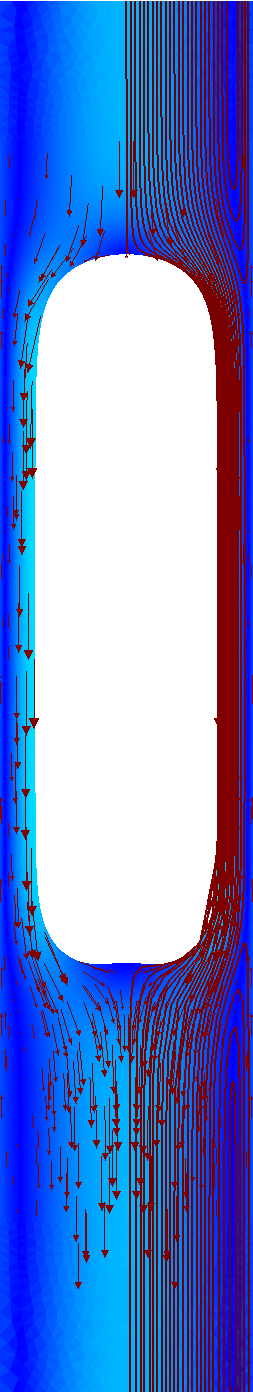} &
  	\includegraphics[width=0.08 \linewidth, angle=0]{Figures/Eo000}
 \\
\includegraphics[width=0.07 \linewidth, angle=0]{Figures/Eo000} & 
 \includegraphics[width= 0.07\linewidth, angle=0]{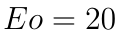} &  \includegraphics[width= 0.07 \linewidth, angle=0]{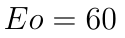} & \includegraphics[width= 0.07 \linewidth, angle=0]{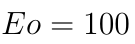} & \includegraphics[width= 0.07 \linewidth, angle=0]{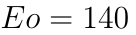} &
\includegraphics[width= 0.07 \linewidth, angle=0]{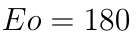} & \includegraphics[width= 0.07 \linewidth, angle=0]{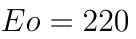} & \includegraphics[width= 0.07 \linewidth, angle=0]{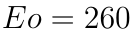} & \includegraphics[width= 0.07 \linewidth, angle=0]{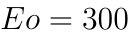} & \includegraphics[width=0.07 \linewidth, angle=0]{Figures/Eo000} 
\end{tabular}
\caption{The effect of $U_m$ and $Eo$ on the steady bubble shapes and flow fields with $Nf = 40$. In each panel, the streamlines and vector fields are superimposed on velocity magnitude pseudocolour plot on the right and left sides of the symmetry axis, respectively.}
\label{fig:ss_phasefield_steady_state_flow_Nf_40}
\end{figure} 

\begin{figure}
\centering
    \begin{subfigure}{\linewidth}
	\centering\includegraphics[width = 0.65\linewidth, angle=0]{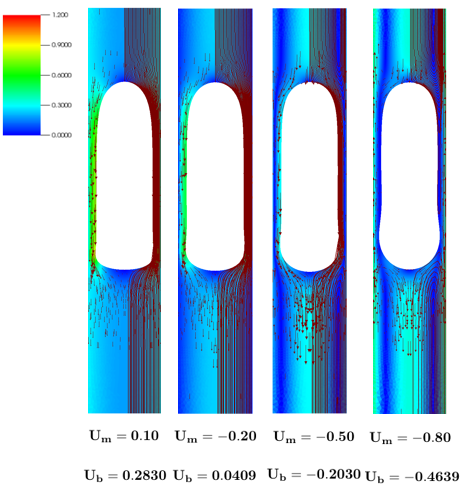}
	\caption{}
	\label{fig:ss_flowfield_Nf_80Eo20}
	\end{subfigure}
	\quad
	\begin{subfigure}{\linewidth}
	    \centering\includegraphics[width = 0.65\linewidth, angle=0]{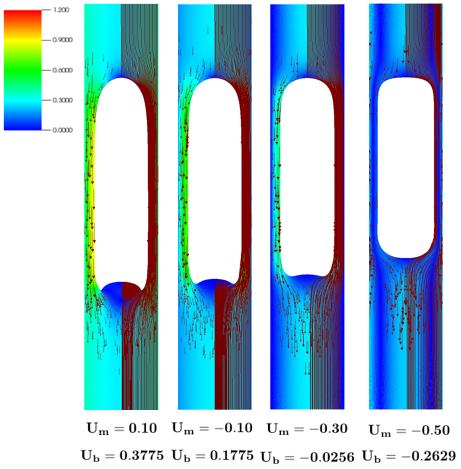}
	\caption{}
	\label{fig:ss_flowfield_Nf_60Eo220}
	\end{subfigure}
	\caption{Steady-state bubble shapes in flowing liquids: (a) effect of $U_m$ for  $Nf = 80$ and $ Eo = 20$; (b) effect of $U_m$ for $Nf=60$ and $Eo = 220$. In each panel, we show the streamlines and vector fields superimposed on the velocity magnitude pseudocolour plot on the right and left sides of the symmetry axis, respectively. 
For each case, we provide numerical predictions of the bubble rise speed, $U_b$.
	}
\end{figure}

\begin{figure}
\centering
\begin{tabular}{cccccccccc} 
\includegraphics[width=0.15 \linewidth, angle=0]{Figures/UM10U} &
    \includegraphics[width= 0.03\linewidth, angle=0]{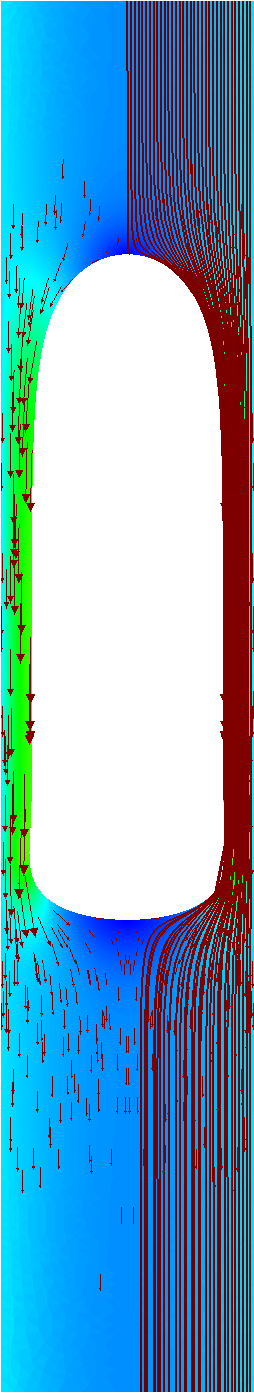} &
  	\includegraphics[width= 0.03\linewidth, angle=0]{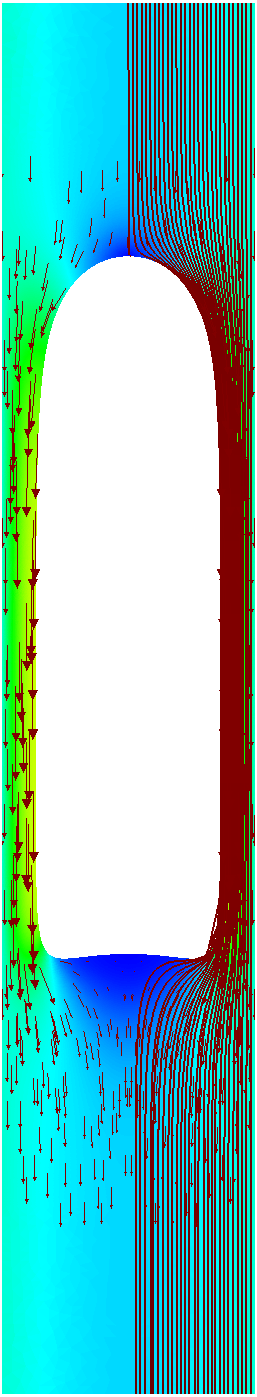} &
  	\includegraphics[width= 0.03\linewidth, angle=0]{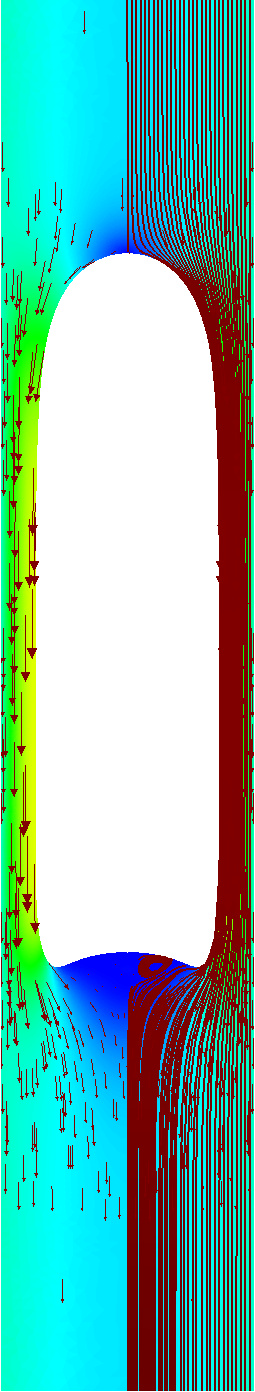} &
  	\includegraphics[width= 0.03\linewidth, angle=0]{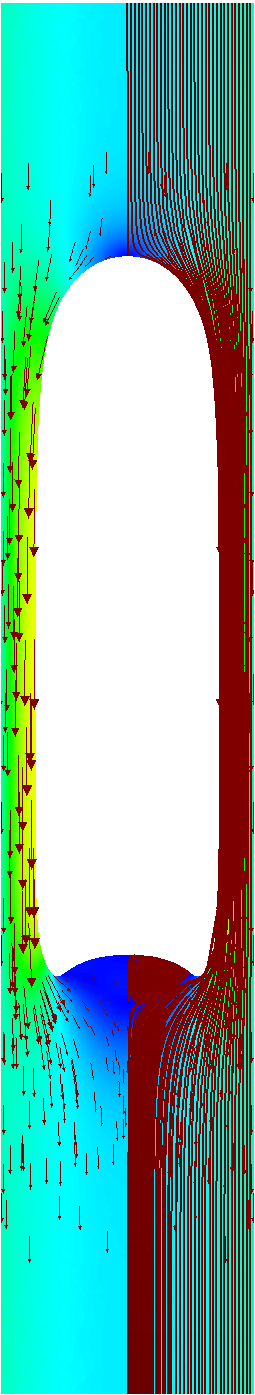} &
  	\includegraphics[width= 0.03\linewidth, angle=0]{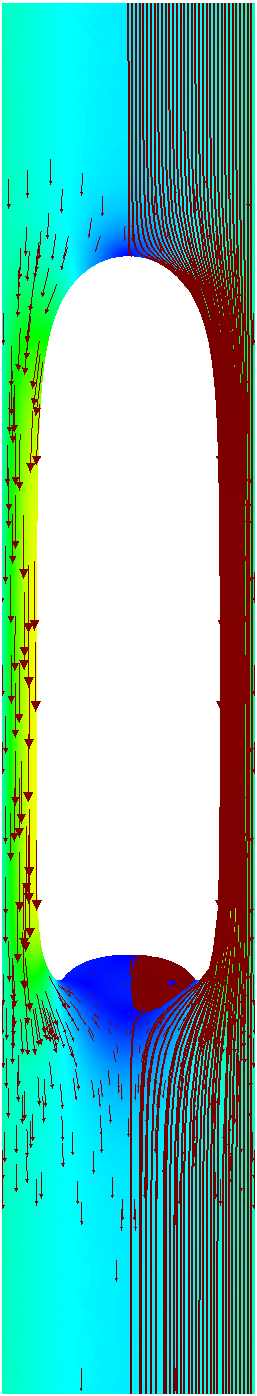} &
  	\includegraphics[width= 0.03\linewidth, angle=0]{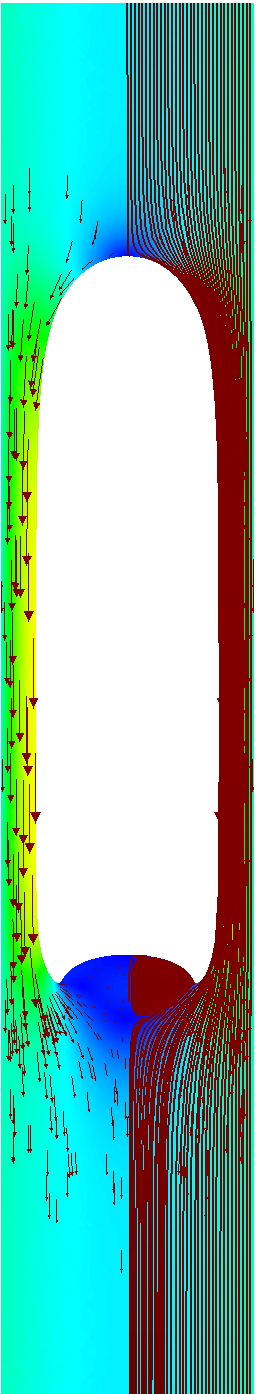} &
  	\includegraphics[width= 0.03\linewidth, angle=0]{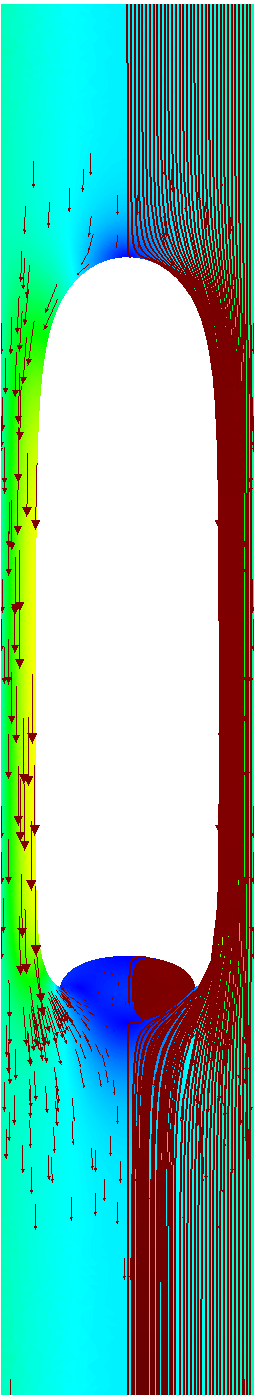} &
  	\includegraphics[width= 0.03\linewidth, angle=0]{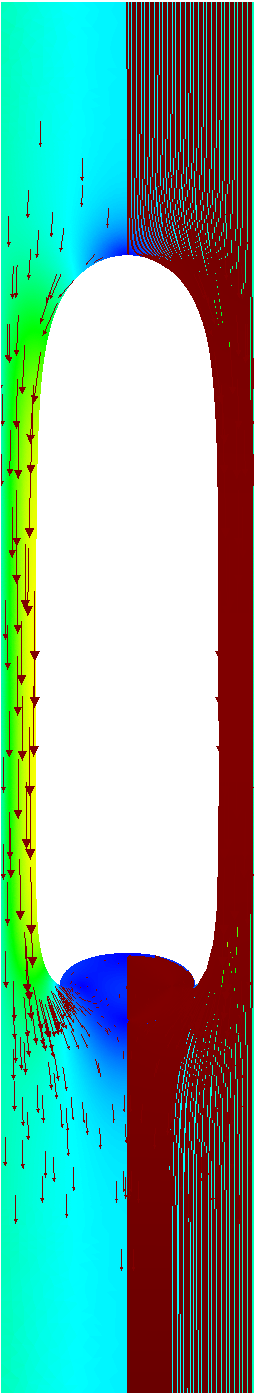} &
  	\includegraphics[width=0.08 \linewidth, angle=0]{Figures/colourmap}
\\ 
\includegraphics[width=0.10 \linewidth, angle=0]{Figures/UM00S} &
\includegraphics[width= 0.03\linewidth, angle=0]{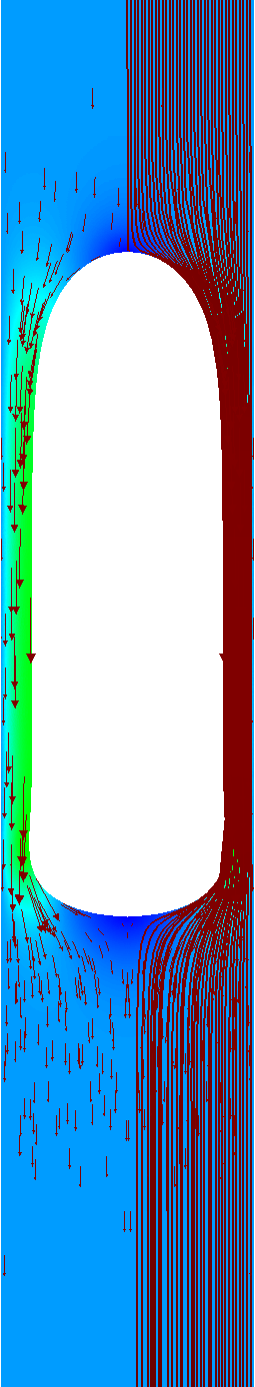} & \includegraphics[width= 0.03\linewidth, angle=0]{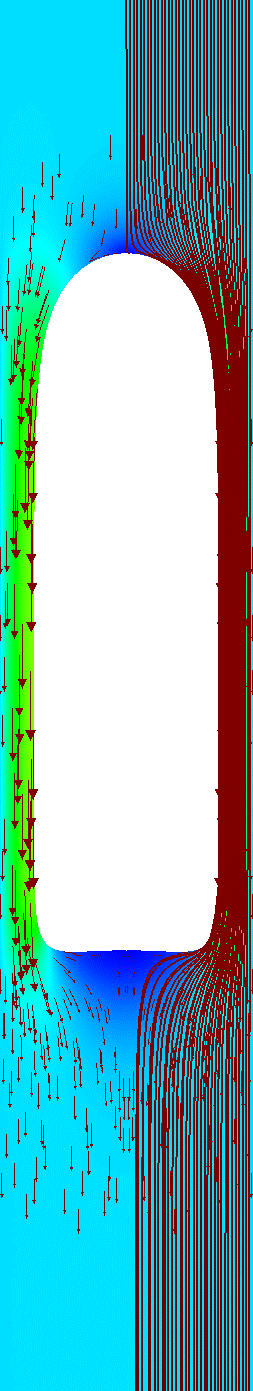} & \includegraphics[width= 0.03\linewidth, angle=0]{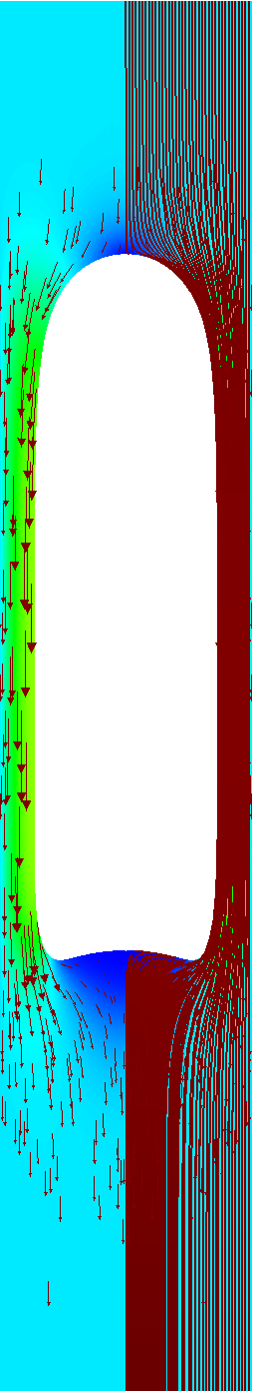} & \includegraphics[width= 0.03\linewidth, angle=0]{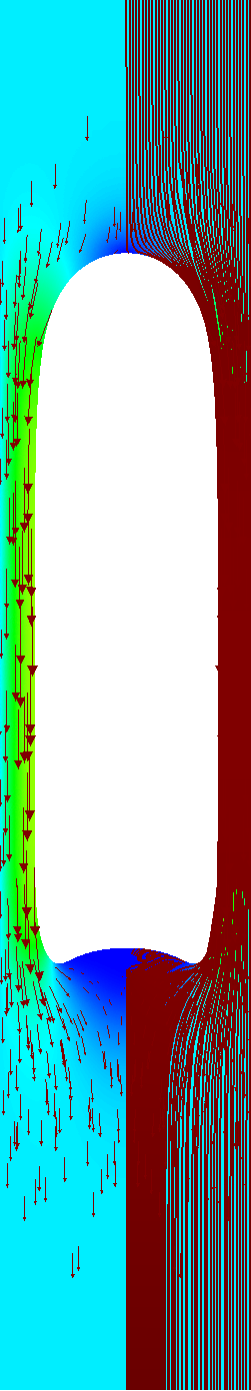} &
\includegraphics[width= 0.03\linewidth, angle=0]{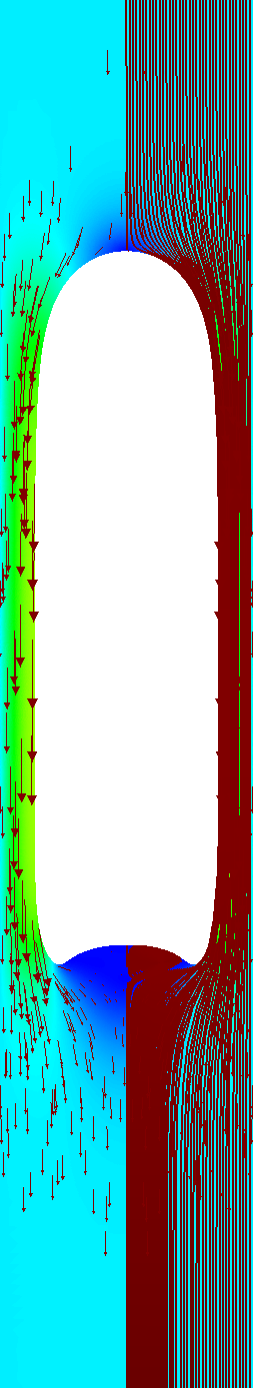} & \includegraphics[width= 0.03\linewidth, angle=0]{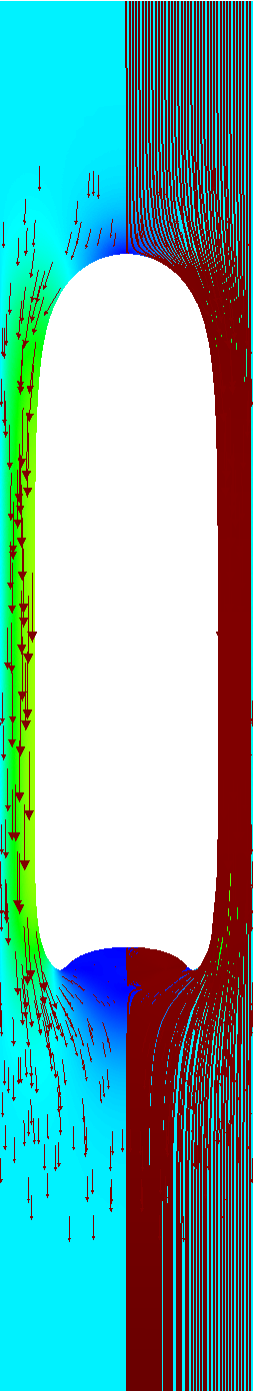} & \includegraphics[width= 0.03\linewidth, angle=0]{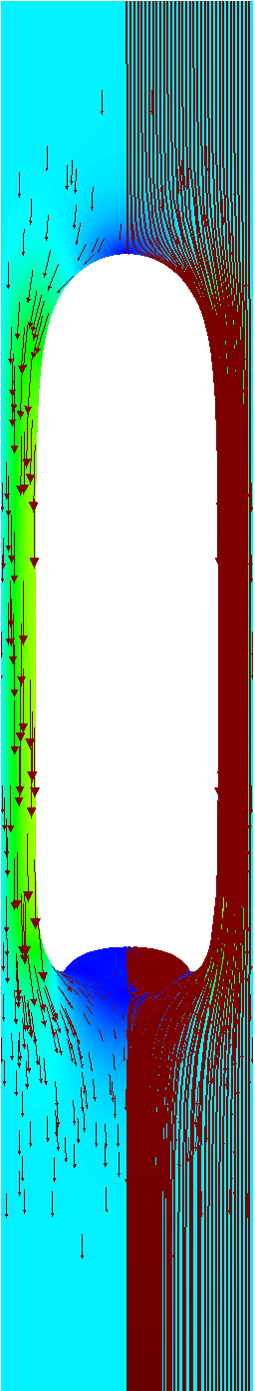} & \includegraphics[width= 0.03\linewidth, angle=0]{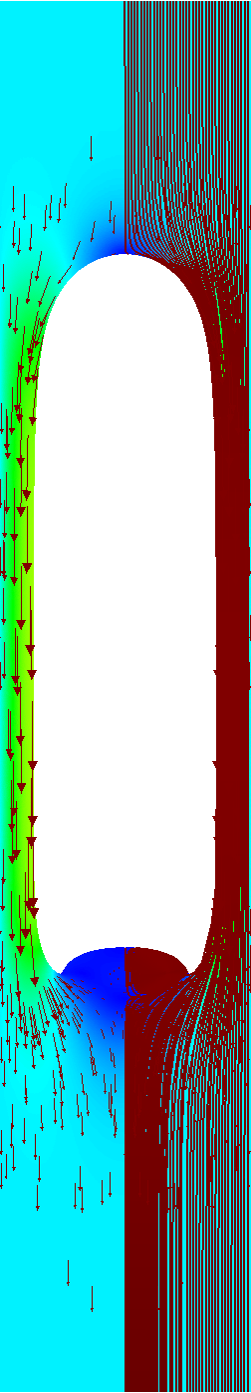} &
\includegraphics[width=0.08 \linewidth, angle=0]{Figures/Eo000}
\\ 
\includegraphics[width=0.15 \linewidth, angle=0]{Figures/UM10D} &
    \includegraphics[width= 0.03\linewidth, angle=0]{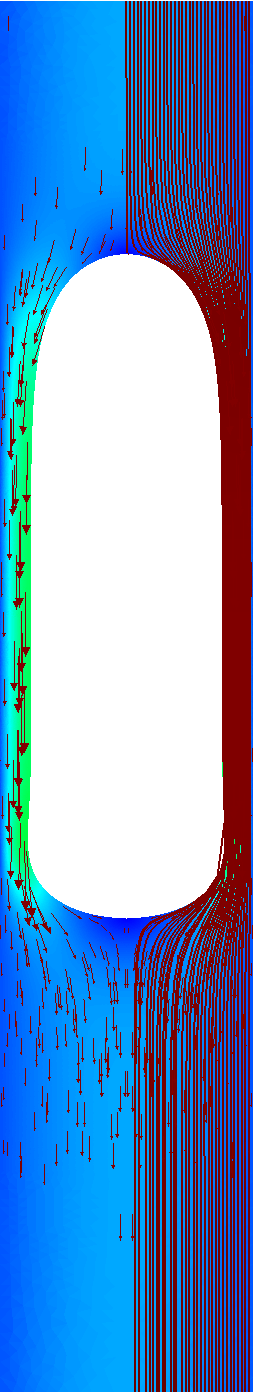} &
  	\includegraphics[width= 0.03\linewidth, angle=0]{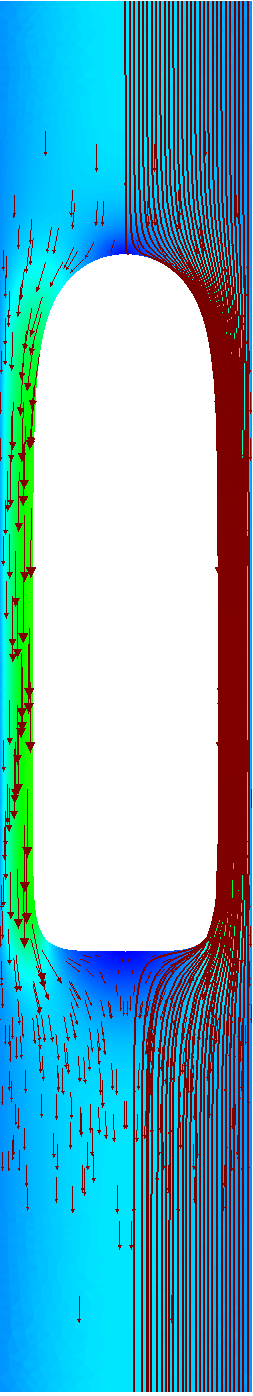} &
  	\includegraphics[width= 0.03\linewidth, angle=0]{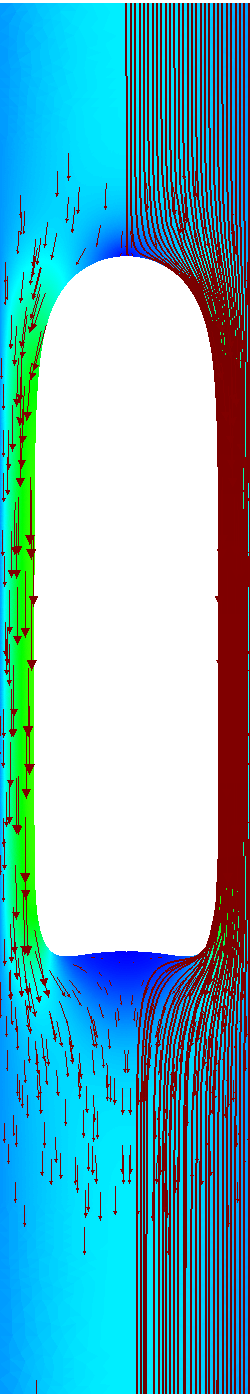} &
  	\includegraphics[width= 0.03\linewidth, angle=0]{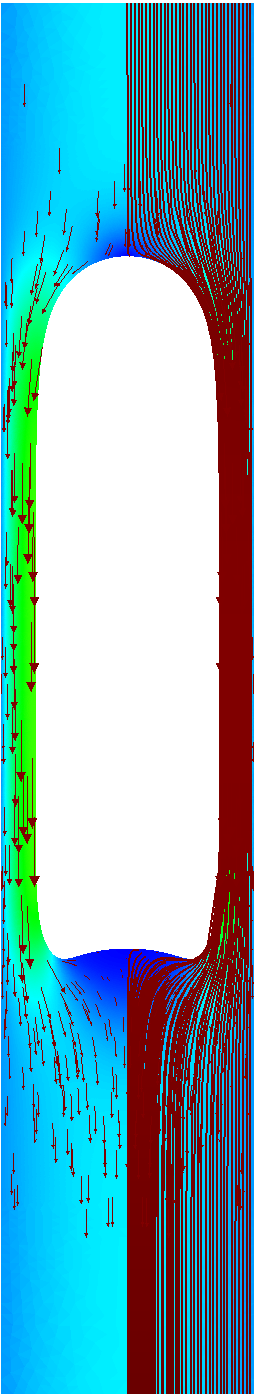} &
  	\includegraphics[width= 0.03\linewidth, angle=0]{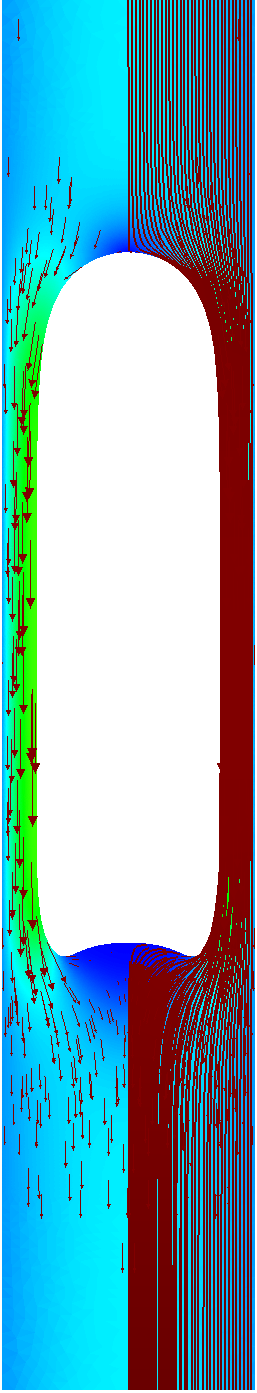} &
  	\includegraphics[width= 0.03\linewidth, angle=0]{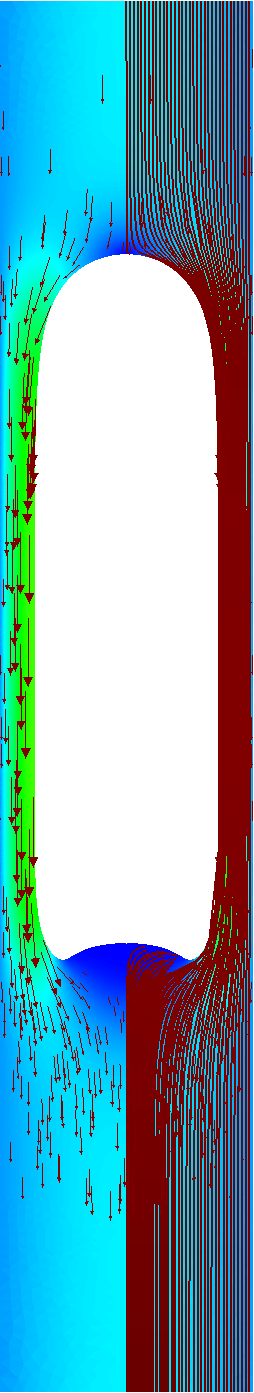} &
  	\includegraphics[width= 0.03\linewidth, angle=0]{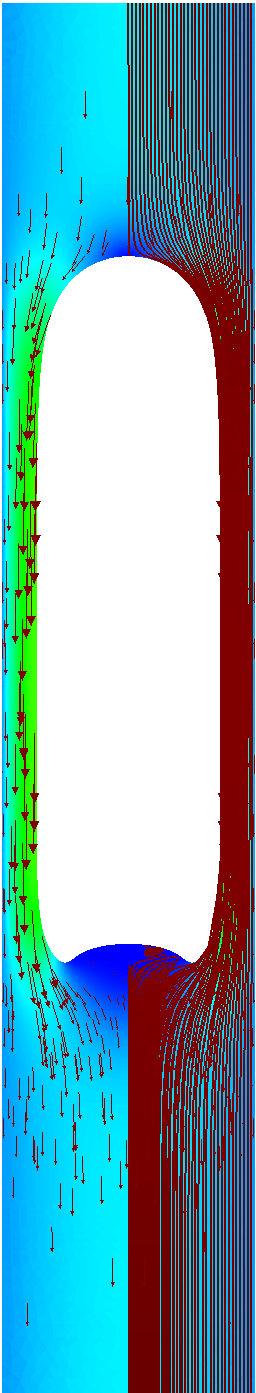} &
  	\includegraphics[width= 0.03\linewidth, angle=0]{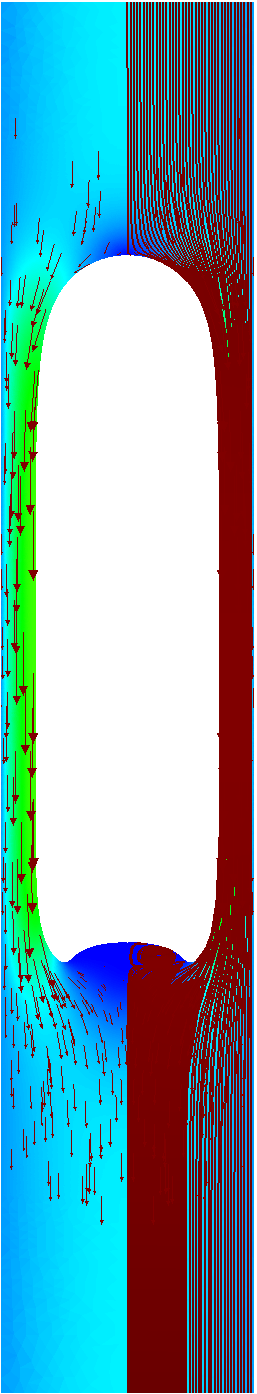} &
  	\includegraphics[width=0.08 \linewidth, angle=0]{Figures/Eo000}
\\ 
\includegraphics[width=0.15 \linewidth, angle=0]{Figures/UM20D} &
    \includegraphics[width= 0.03\linewidth, angle=0]{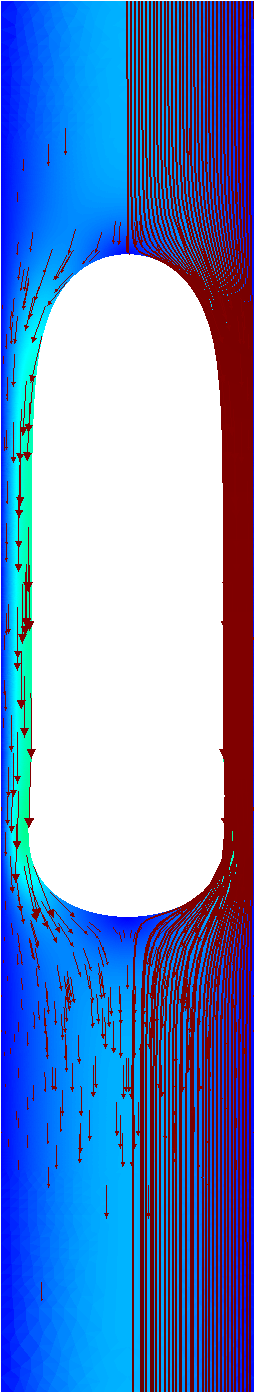} &
  	\includegraphics[width= 0.03\linewidth, angle=0]{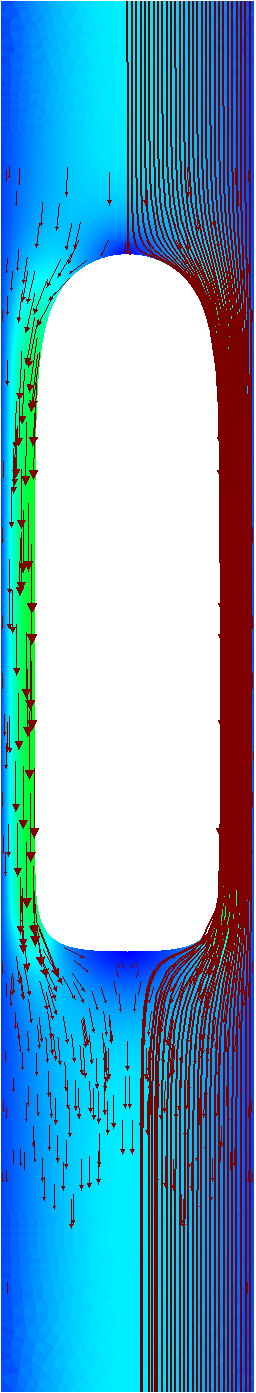} &
  	\includegraphics[width= 0.03\linewidth, angle=0]{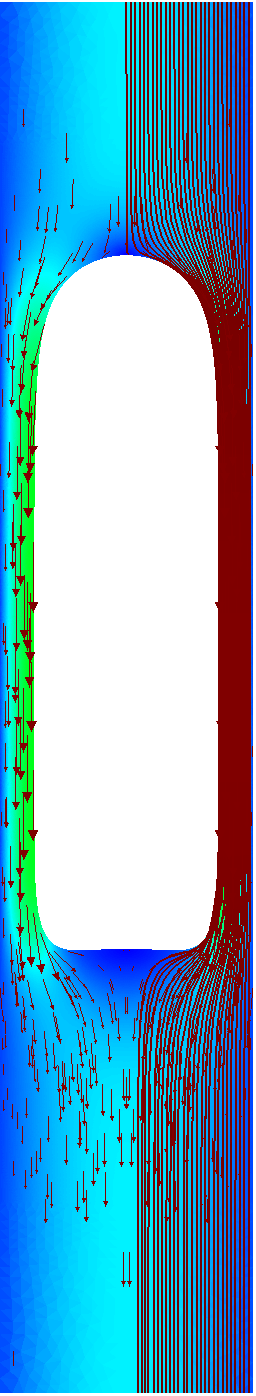} &
  	\includegraphics[width= 0.03\linewidth, angle=0]{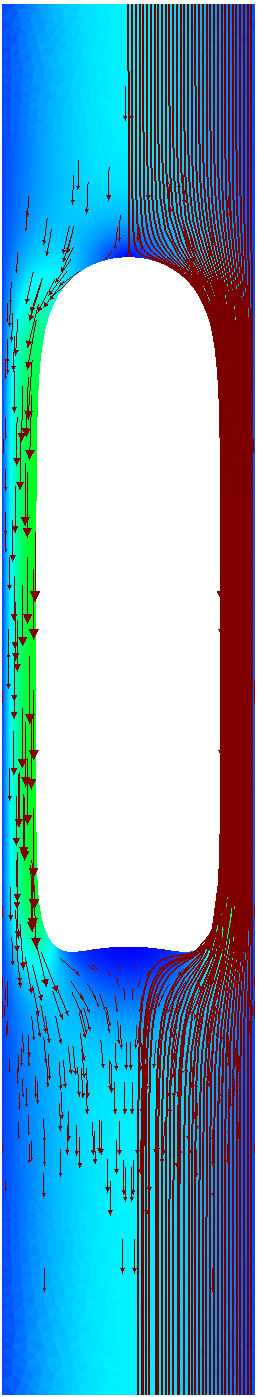} &
  	\includegraphics[width= 0.03\linewidth, angle=0]{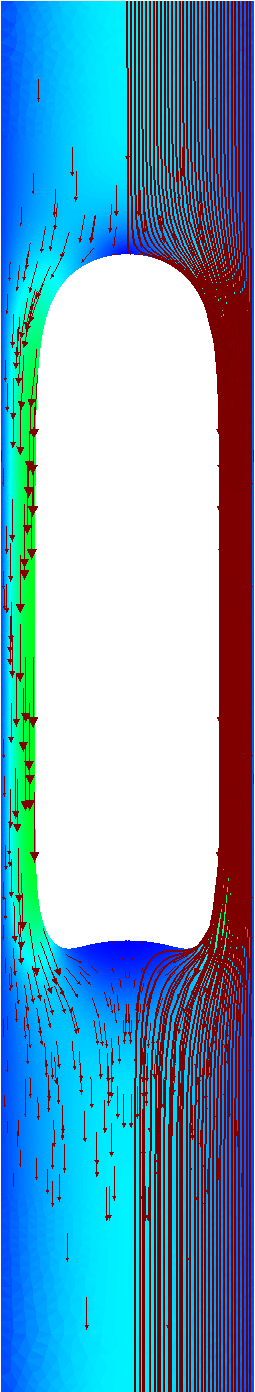} &
  	\includegraphics[width= 0.03\linewidth, angle=0]{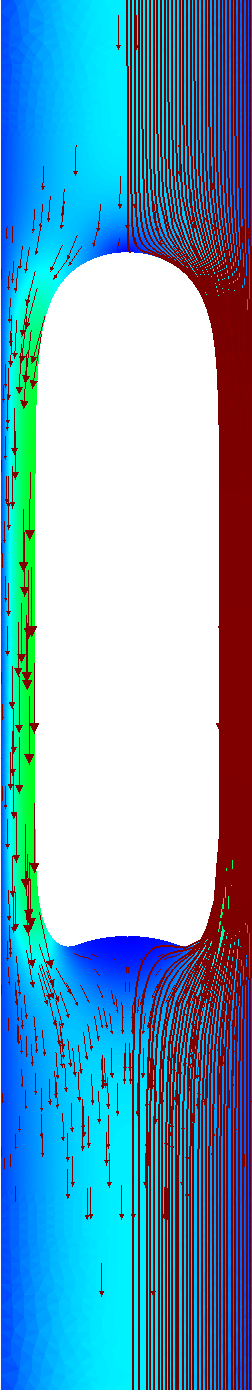} &
  	\includegraphics[width= 0.03\linewidth, angle=0]{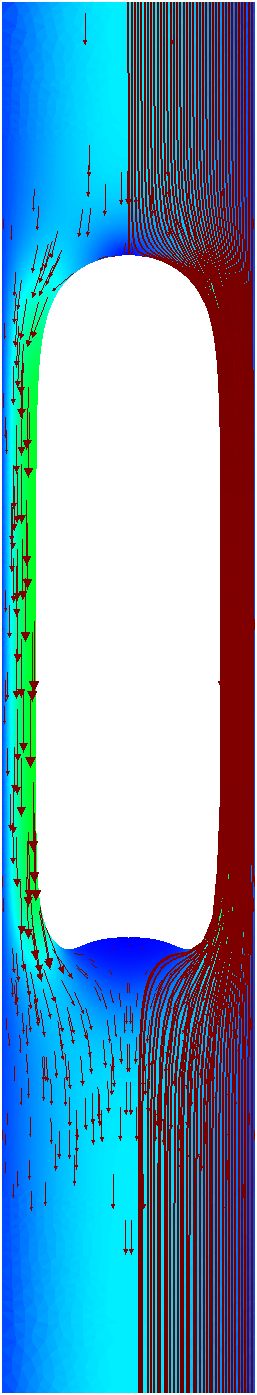} &
  	\includegraphics[width= 0.03\linewidth, angle=0]{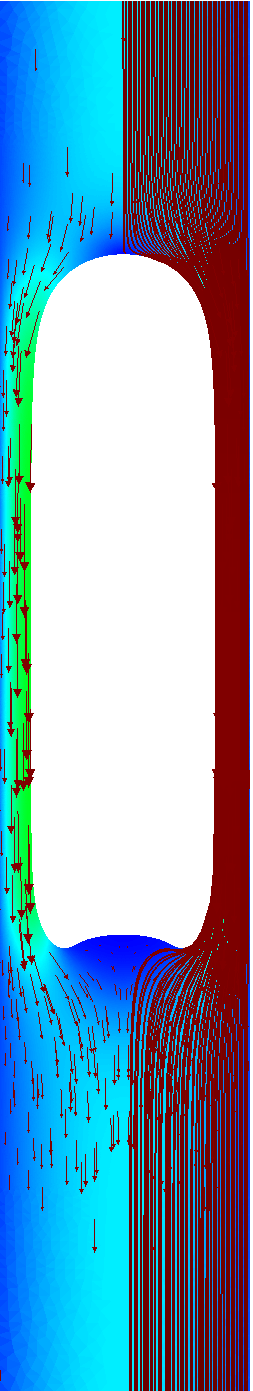} &
  	\includegraphics[width=0.08 \linewidth, angle=0]{Figures/Eo000}
\\ 
\includegraphics[width=0.15 \linewidth, angle=0]{Figures/UM30D} &
    \includegraphics[width= 0.03\linewidth, angle=0]{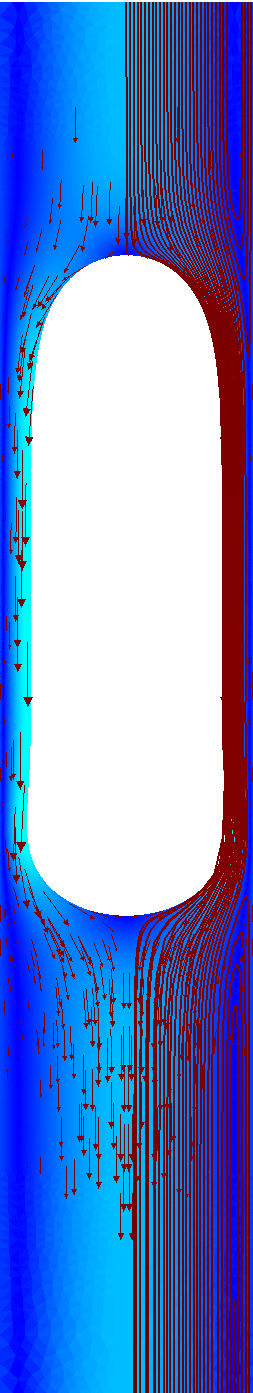} &
  	\includegraphics[width= 0.03\linewidth, angle=0]{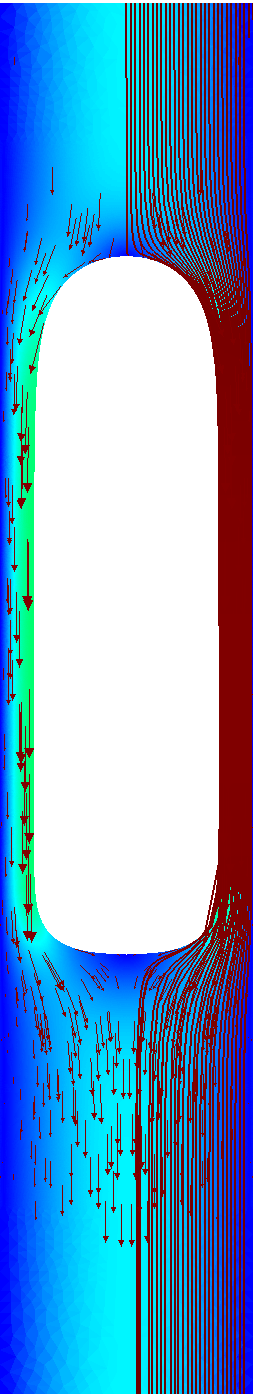} &
  	\includegraphics[width= 0.03\linewidth, angle=0]{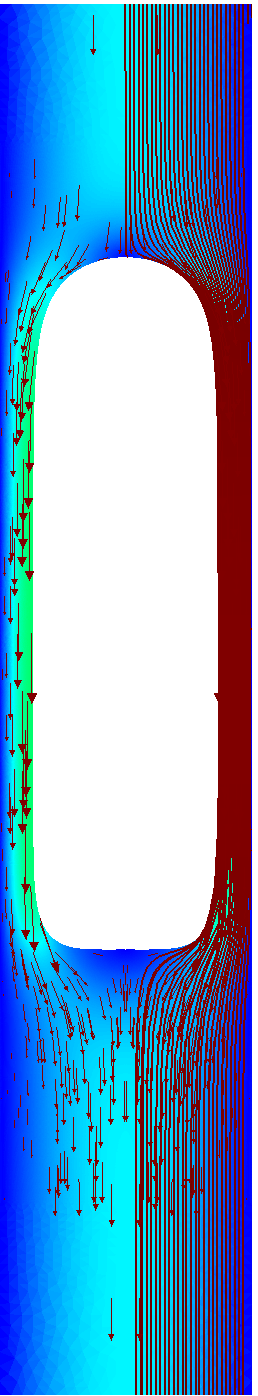} &
  	\includegraphics[width= 0.03\linewidth, angle=0]{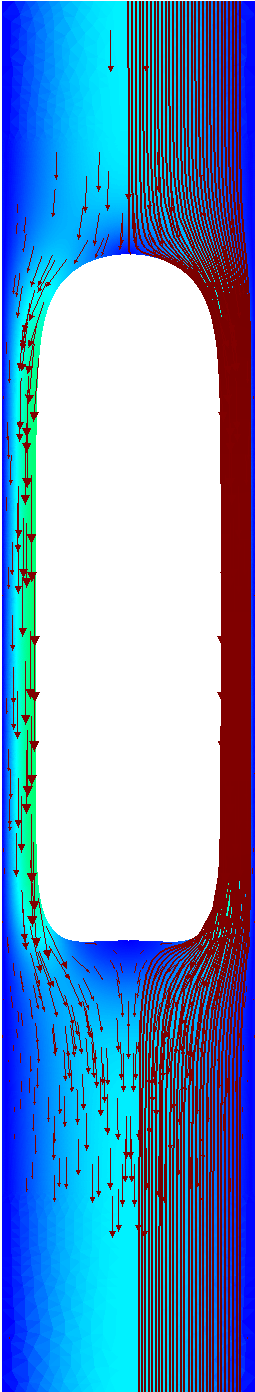} &
  	\includegraphics[width= 0.03\linewidth, angle=0]{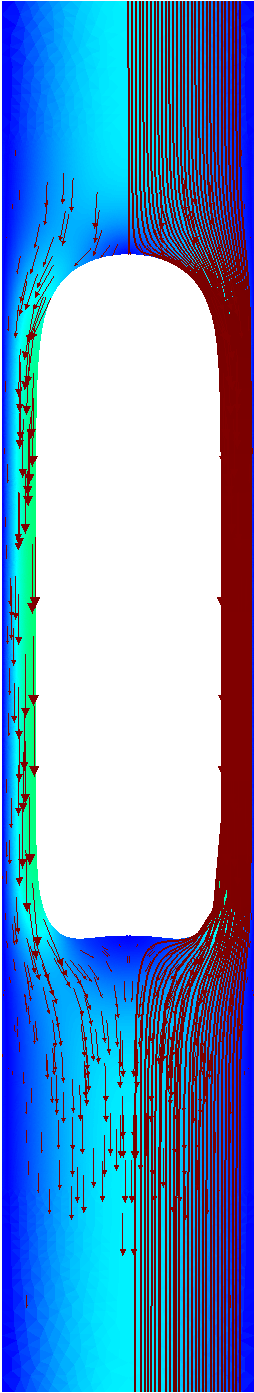} &
  	\includegraphics[width= 0.03\linewidth, angle=0]{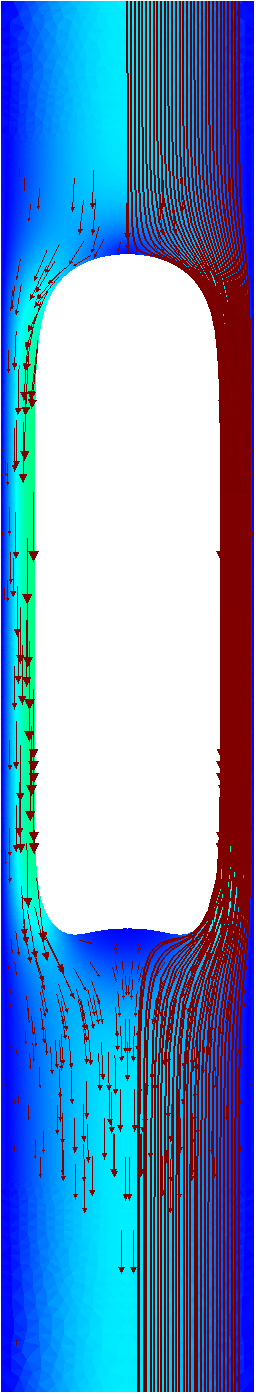} &
  	\includegraphics[width= 0.03\linewidth, angle=0]{FlowThesisFig/FlowNf60Eo260ud30} &
  	\includegraphics[width= 0.03\linewidth, angle=0]{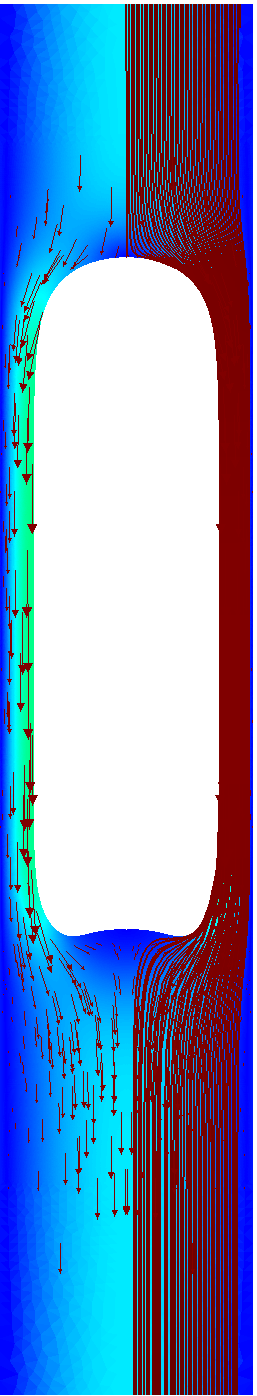} &
  	\includegraphics[width=0.08 \linewidth, angle=0]{Figures/Eo000}
\\ 
\includegraphics[width=0.15 \linewidth, angle=0]{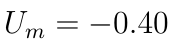} &
    \includegraphics[width= 0.03\linewidth, angle=0]{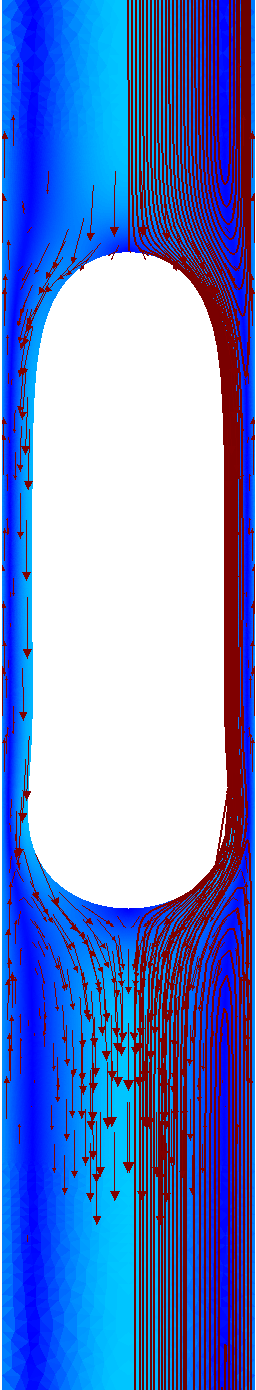} &
  	\includegraphics[width= 0.03\linewidth, angle=0]{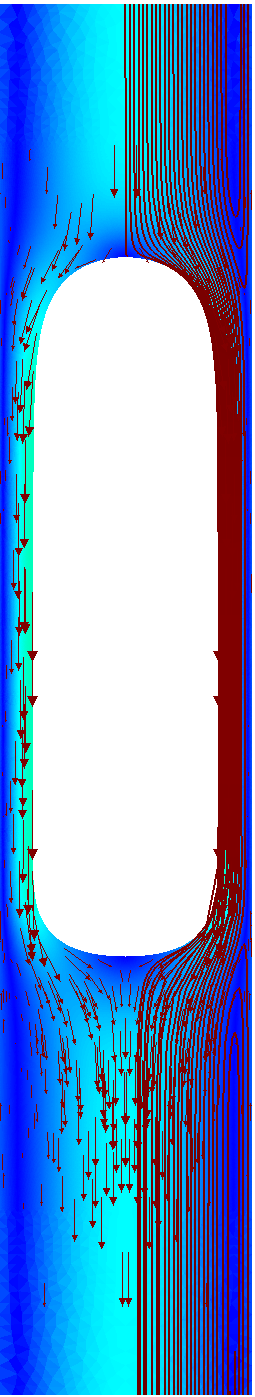} &
  	\includegraphics[width= 0.03\linewidth, angle=0]{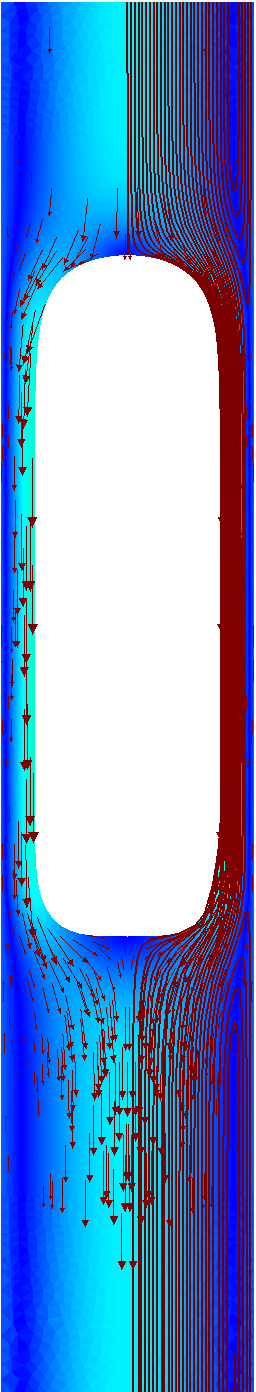} &
  	\includegraphics[width= 0.03\linewidth, angle=0]{FlowThesisFig/FlowNf60Eo140ud40} &
  	\includegraphics[width= 0.03\linewidth, angle=0]{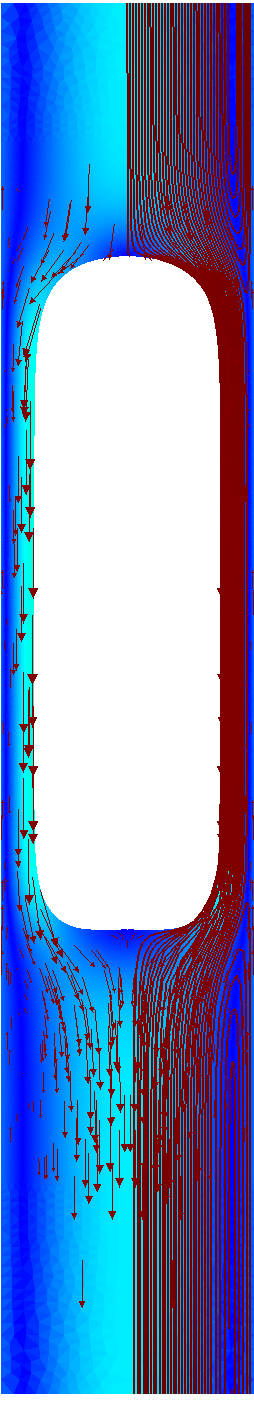} &
  	\includegraphics[width= 0.03\linewidth, angle=0]{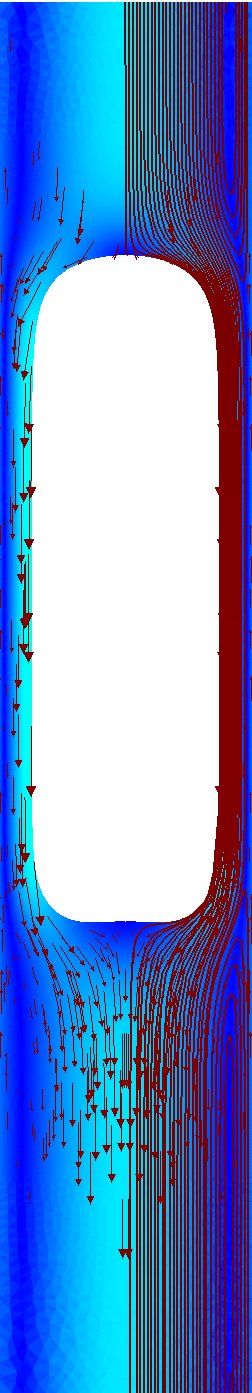} &
  	\includegraphics[width= 0.03\linewidth, angle=0]{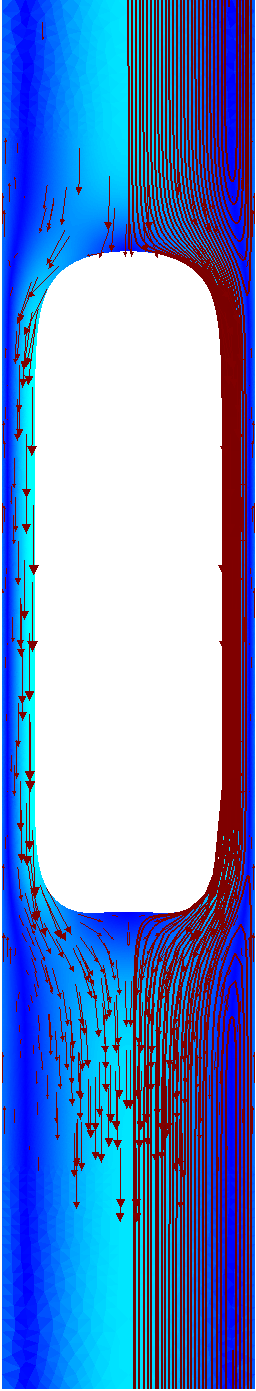} &
  	\includegraphics[width= 0.03\linewidth, angle=0]{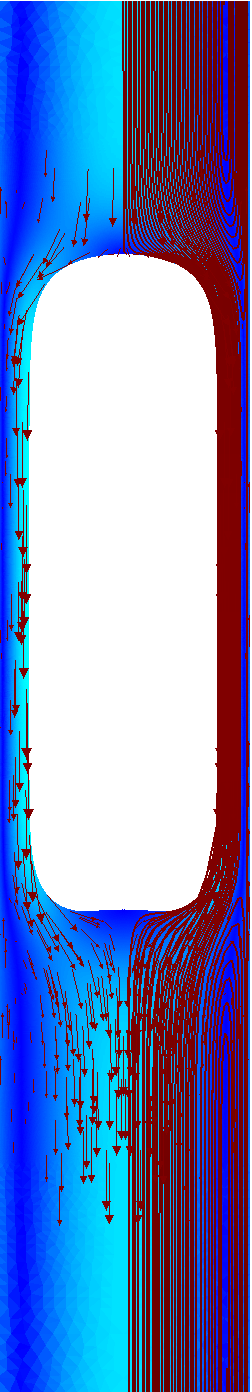} &
  	\includegraphics[width=0.08 \linewidth, angle=0]{Figures/Eo000}
\\ 
\includegraphics[width=0.15 \linewidth, angle=0]{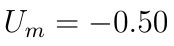} &
    \includegraphics[width= 0.03\linewidth, angle=0]{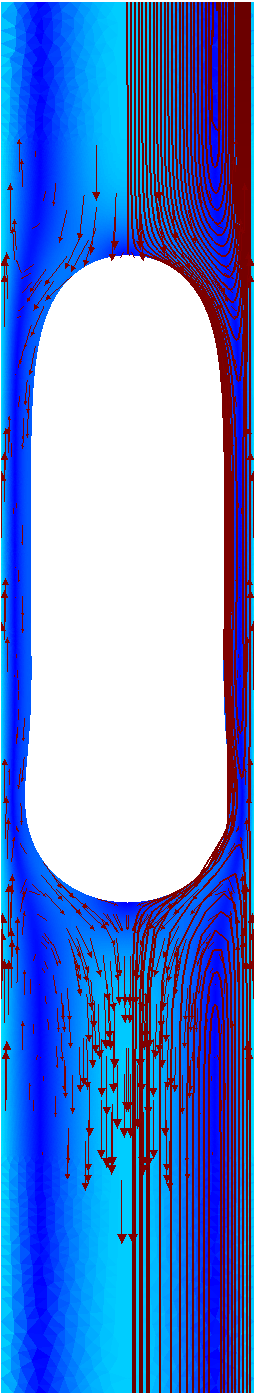} &
  	\includegraphics[width= 0.03\linewidth, angle=0]{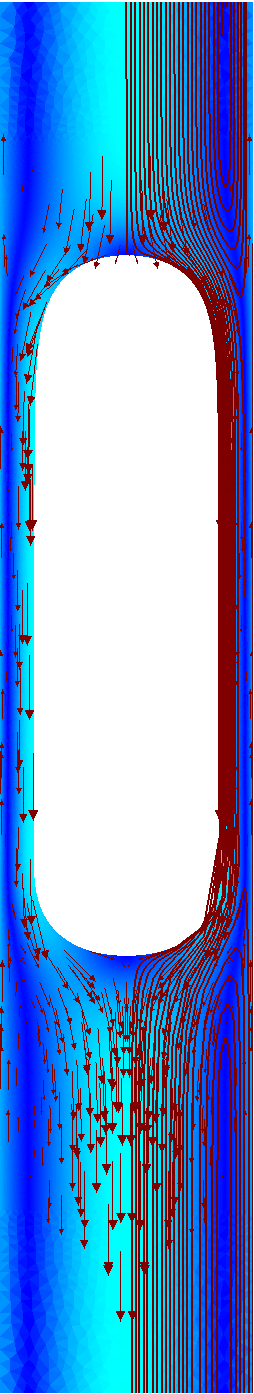} &
  	\includegraphics[width= 0.03\linewidth, angle=0]{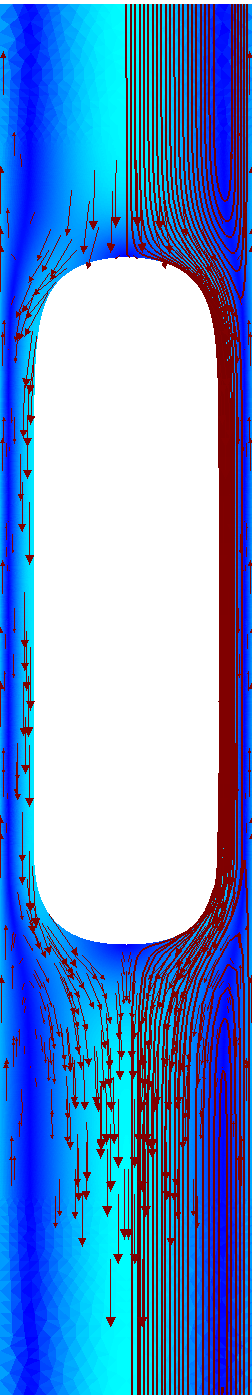} &
  	\includegraphics[width= 0.03\linewidth, angle=0]{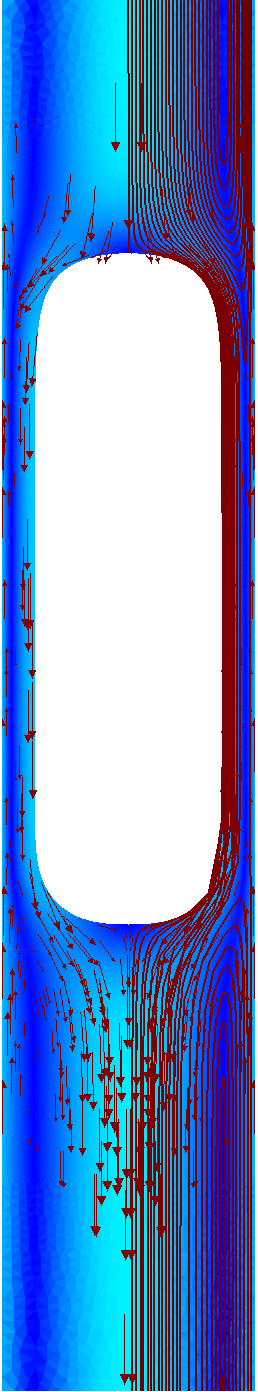} &
  	\includegraphics[width= 0.03\linewidth, angle=0]{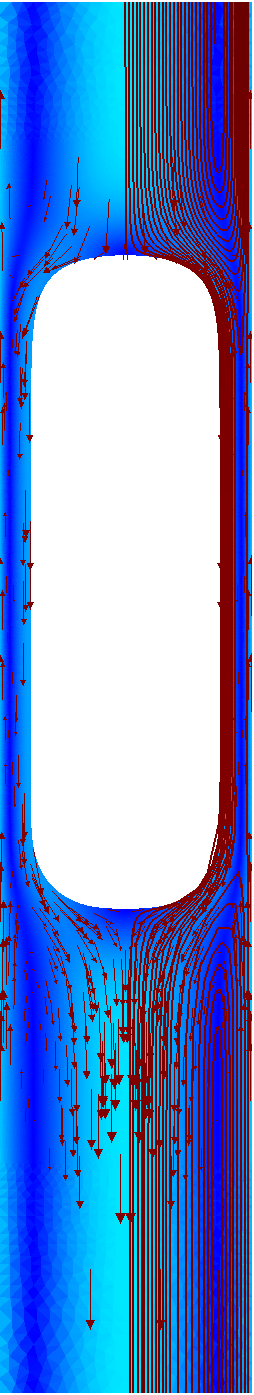} &
  	\includegraphics[width= 0.03\linewidth, angle=0]{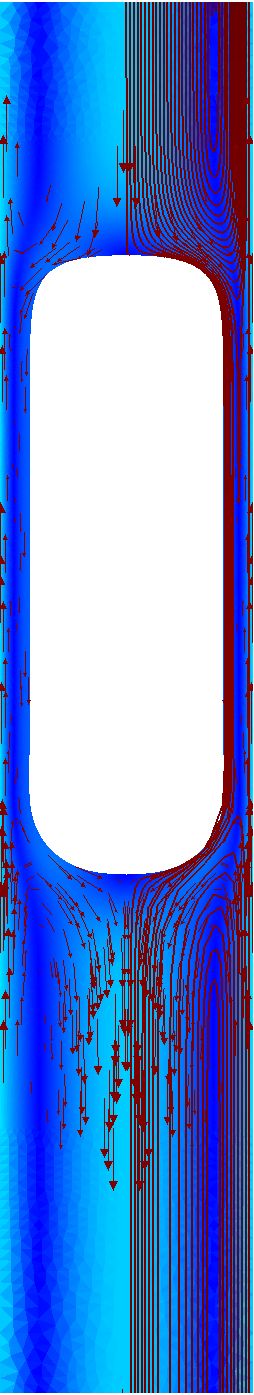} &
  	\includegraphics[width=0.03 \linewidth, angle=0]{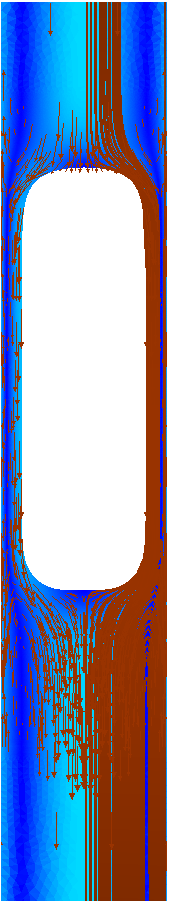} &
  	\includegraphics[width=0.03 \linewidth, angle=0]{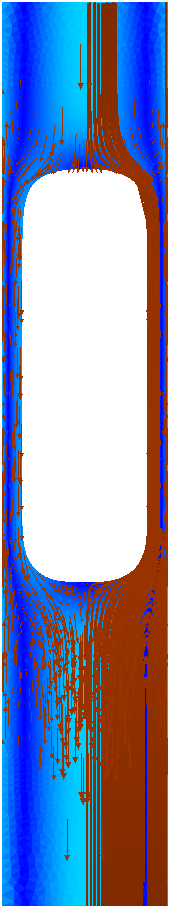} &
  	\includegraphics[width=0.08 \linewidth, angle=0]{Figures/Eo000} 
\\
\includegraphics[width=0.07 \linewidth, angle=0]{Figures/Eo000} & 
 \includegraphics[width= 0.07\linewidth, angle=0]{Figures/Eo20} &  \includegraphics[width= 0.07 \linewidth, angle=0]{Figures/Eo60} & \includegraphics[width= 0.07 \linewidth, angle=0]{Figures/Eo100} & \includegraphics[width= 0.07 \linewidth, angle=0]{Figures/Eo140} &
\includegraphics[width= 0.07 \linewidth, angle=0]{Figures/Eo180} & \includegraphics[width= 0.07 \linewidth, angle=0]{Figures/Eo220} & \includegraphics[width= 0.07 \linewidth, angle=0]{Figures/Eo260} & \includegraphics[width= 0.07 \linewidth, angle=0]{Figures/Eo300} & \includegraphics[width=0.07 \linewidth, angle=0]{Figures/Eo000} 
\end{tabular}
\caption{The effect of $U_m$ and $Eo$ on the steady bubble shapes and flow fields with $Nf = 60$. In each panel, the streamlines and vector fields are superimposed on velocity magnitude pseudocolour plot on the right and left sides of the symmetry axis, respectively.}
\label{fig:ss_phasefield_steady_state_flow_Nf_60}
\end{figure}

\begin{figure}
\centering
\begin{tabular}{cccccccccc} 
\includegraphics[width=0.15 \linewidth, angle=0]{Figures/UM10U} &
    \includegraphics[width= 0.03\linewidth, angle=0]{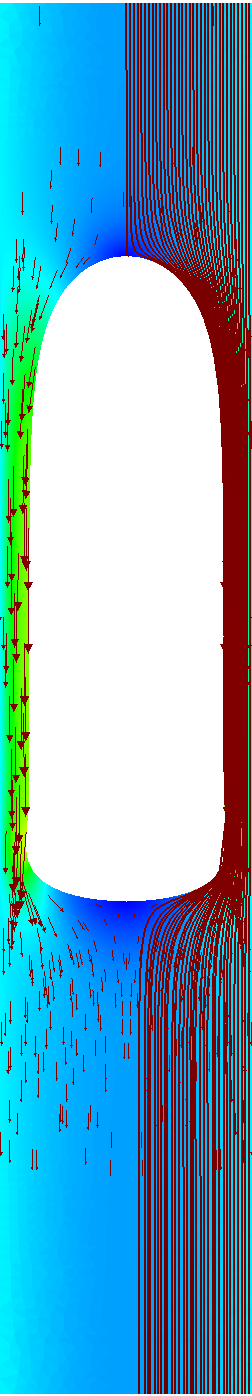} &
  	\includegraphics[width= 0.03\linewidth, angle=0]{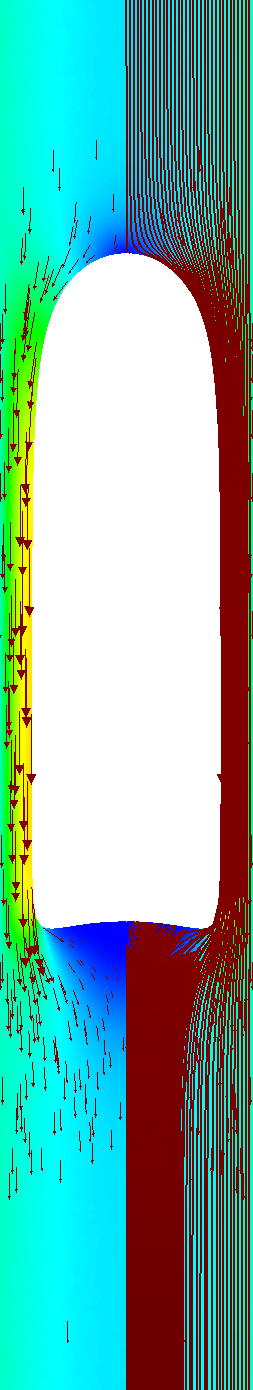} &
  	\includegraphics[width= 0.03\linewidth, angle=0]{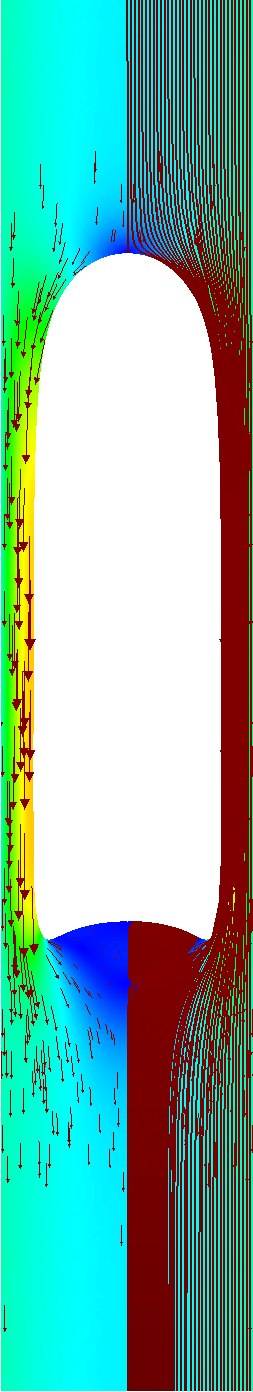} &
  	\includegraphics[width= 0.03\linewidth, angle=0]{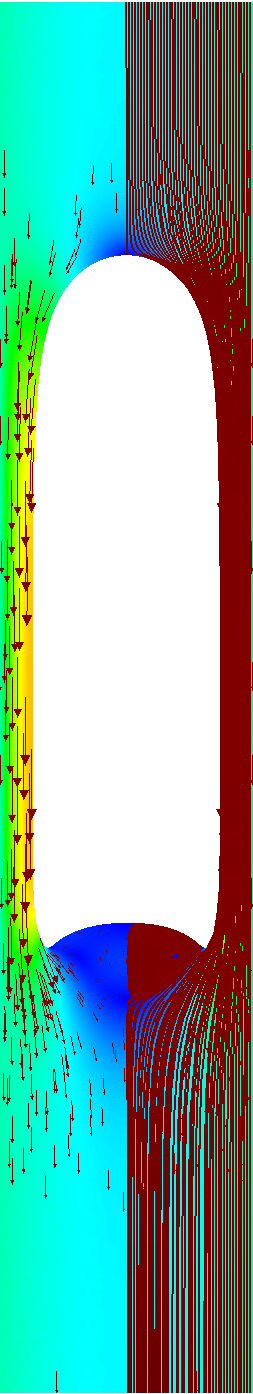} &
  	\includegraphics[width= 0.03\linewidth, angle=0]{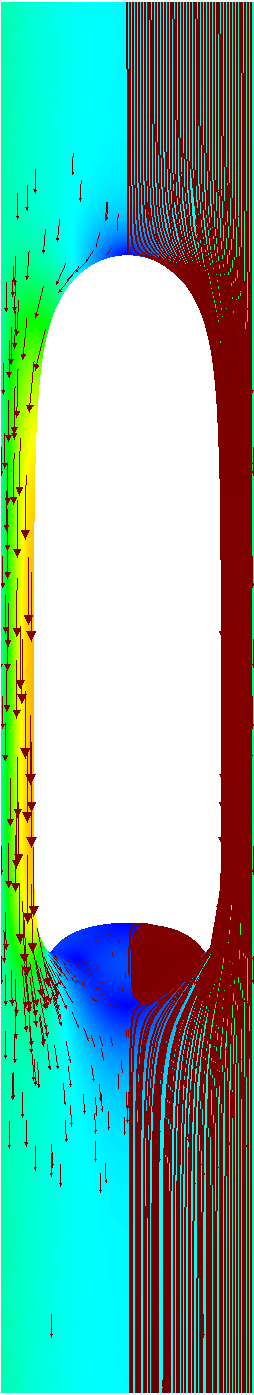} &
  	\includegraphics[width= 0.03\linewidth, angle=0]{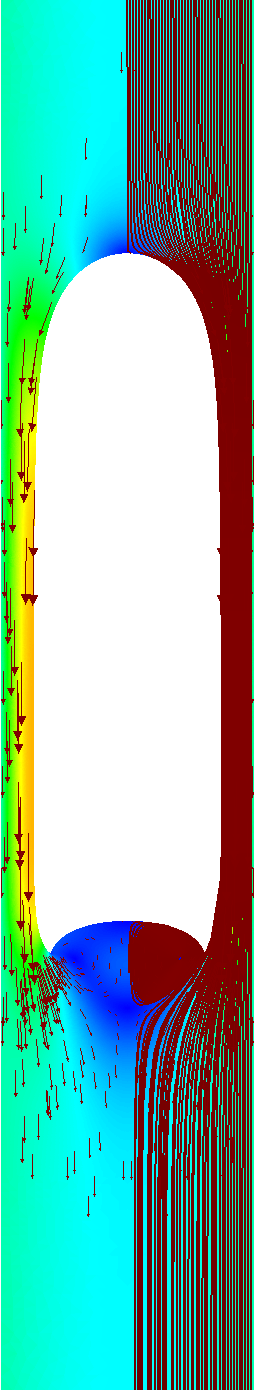} &
  	\includegraphics[width= 0.03\linewidth, angle=0]{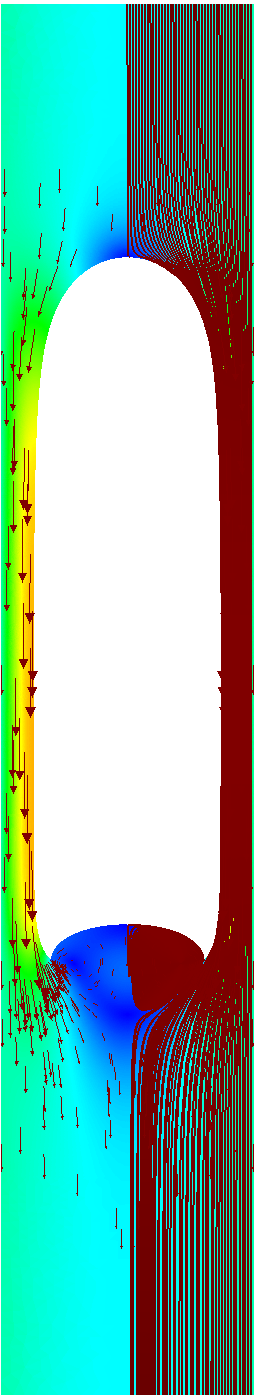} &
  	\includegraphics[width= 0.03\linewidth, angle=0]{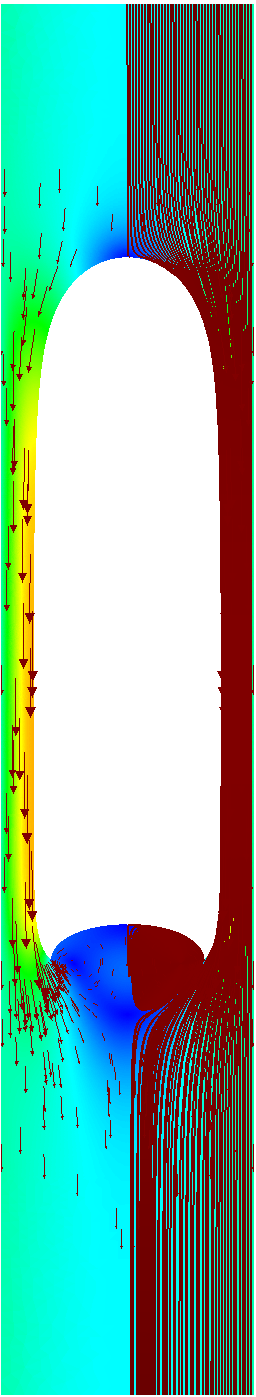} &
  	\includegraphics[width=0.08 \linewidth, angle=0]{Figures/colourmap}
\\ 
\includegraphics[width=0.10 \linewidth, angle=0]{Figures/UM00S} &
\includegraphics[width= 0.03\linewidth, angle=0]{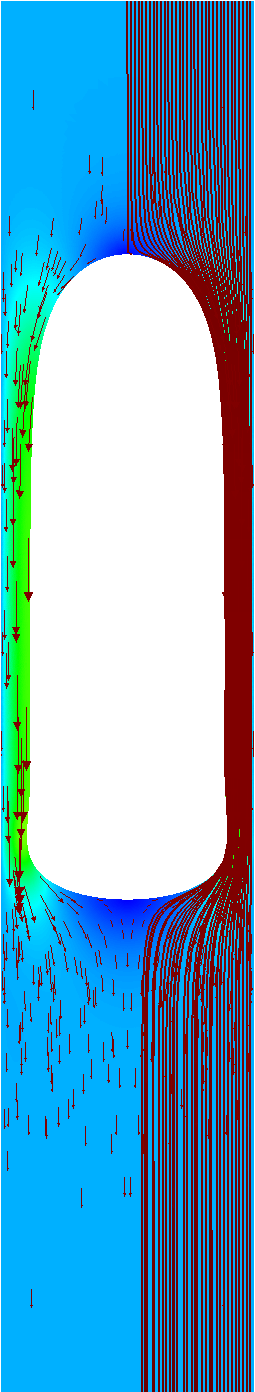} & \includegraphics[width= 0.03\linewidth, angle=0]{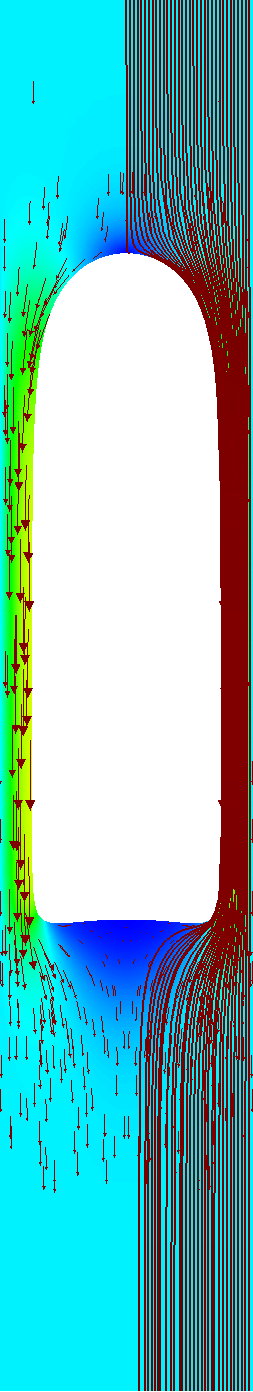} & \includegraphics[width= 0.03\linewidth, angle=0]{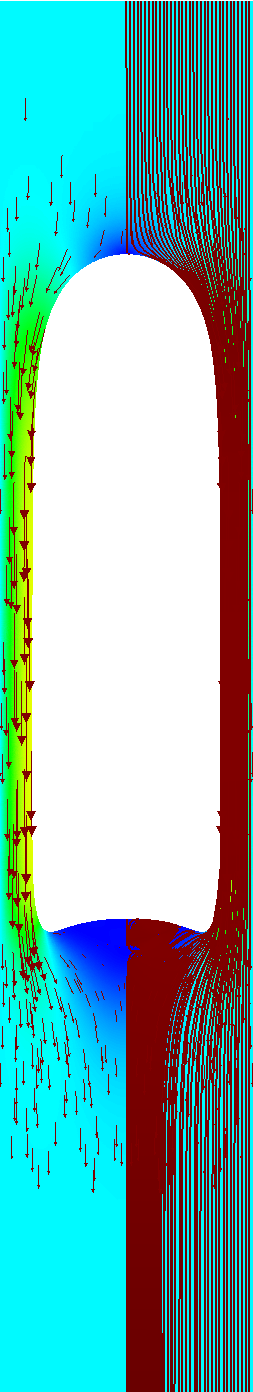} & \includegraphics[width= 0.03\linewidth, angle=0]{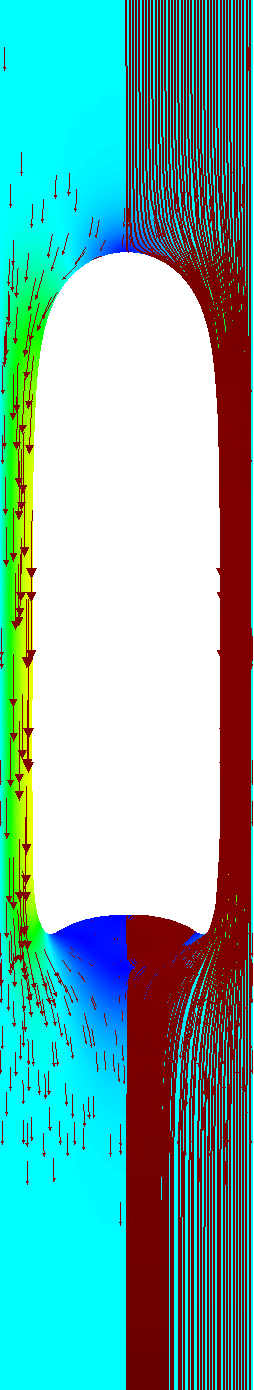} &
\includegraphics[width= 0.03\linewidth, angle=0]{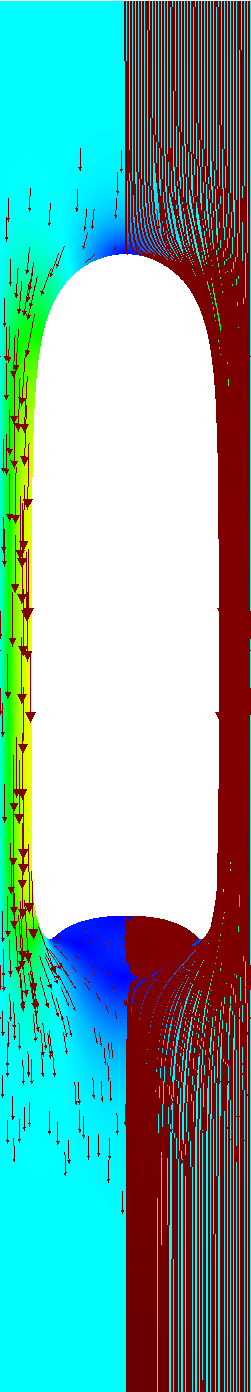} & \includegraphics[width= 0.03\linewidth, angle=0]{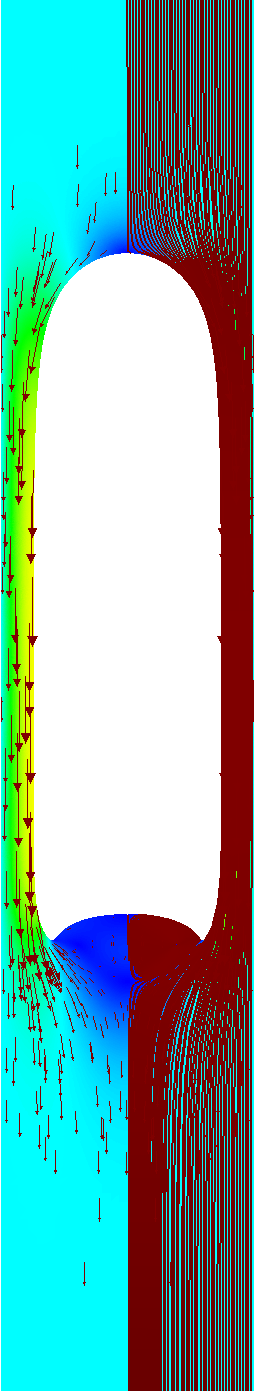} & \includegraphics[width= 0.03\linewidth, angle=0]{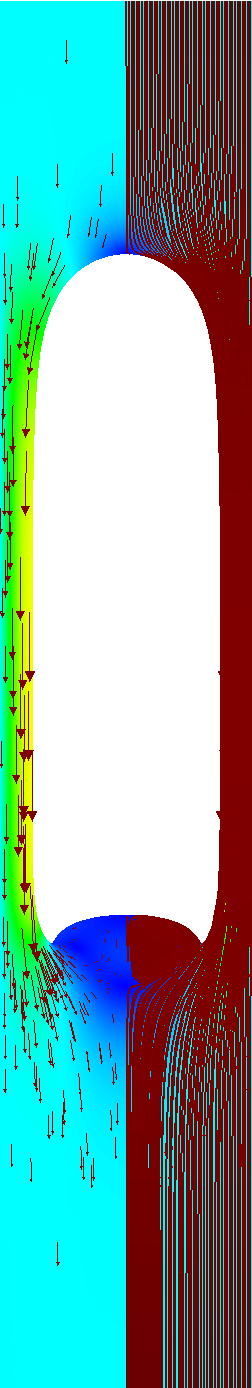} & \includegraphics[width= 0.03\linewidth, angle=0]{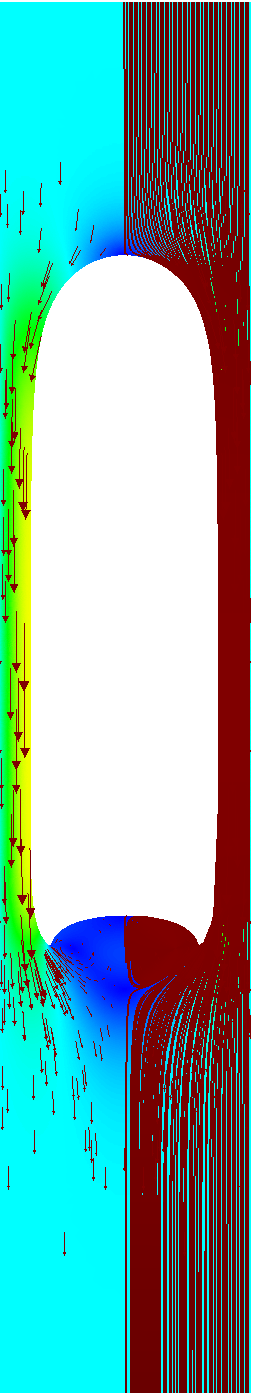} &
\includegraphics[width=0.08 \linewidth, angle=0]{Figures/Eo000}
\\ 
\includegraphics[width=0.15 \linewidth, angle=0]{Figures/UM10D} &
    \includegraphics[width= 0.03\linewidth, angle=0]{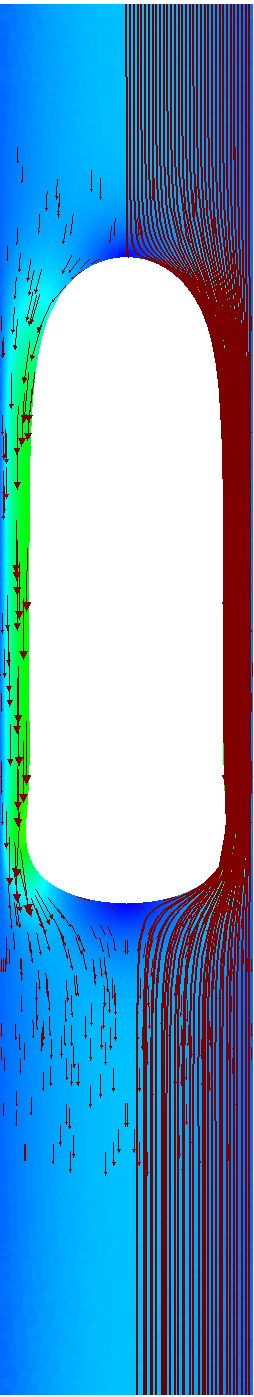} &
  	\includegraphics[width= 0.03\linewidth, angle=0]{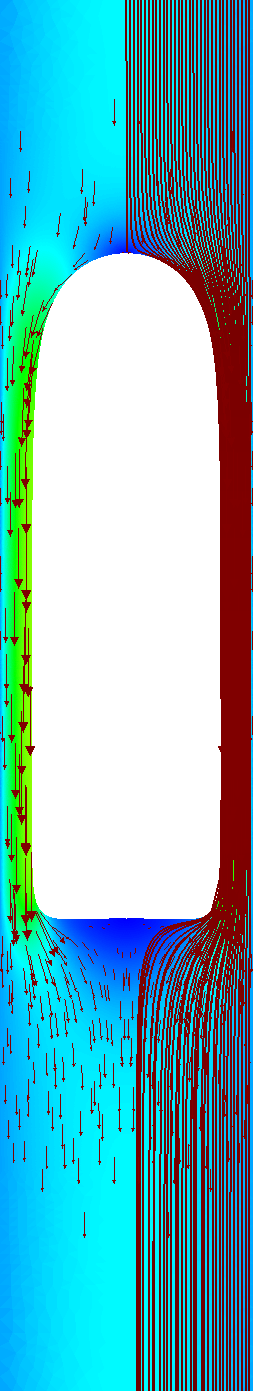} &
  	\includegraphics[width= 0.03\linewidth, angle=0]{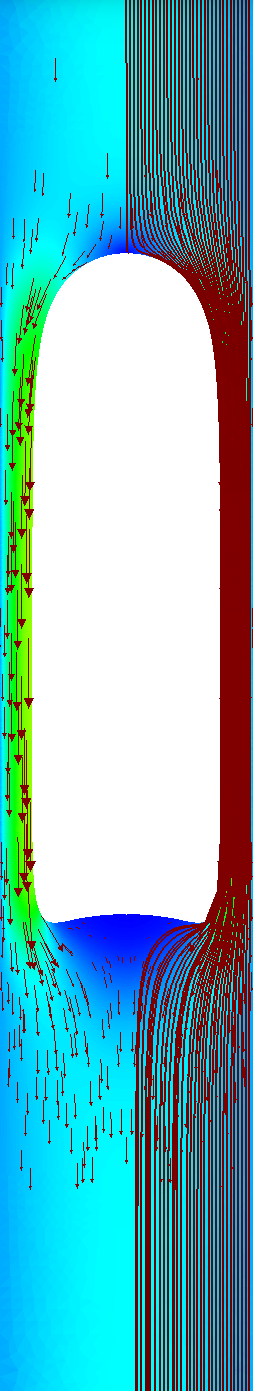} &
  	\includegraphics[width= 0.03\linewidth, angle=0]{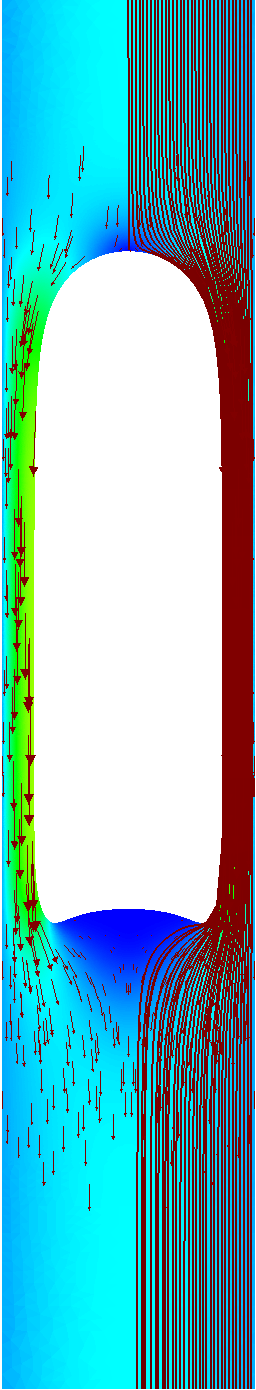} &
  	\includegraphics[width= 0.03\linewidth, angle=0]{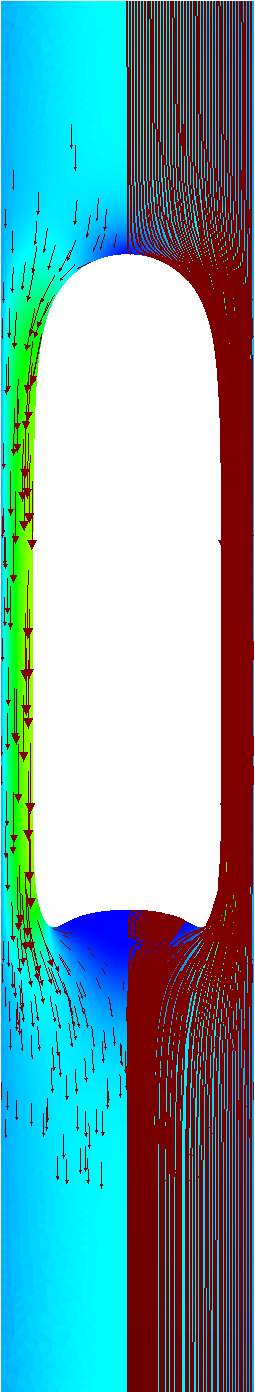} &
  	\includegraphics[width= 0.03\linewidth, angle=0]{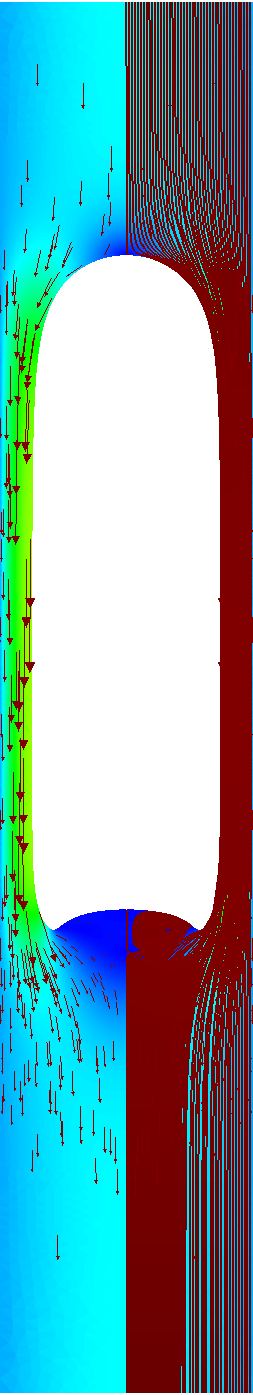} &
  	\includegraphics[width= 0.03\linewidth, angle=0]{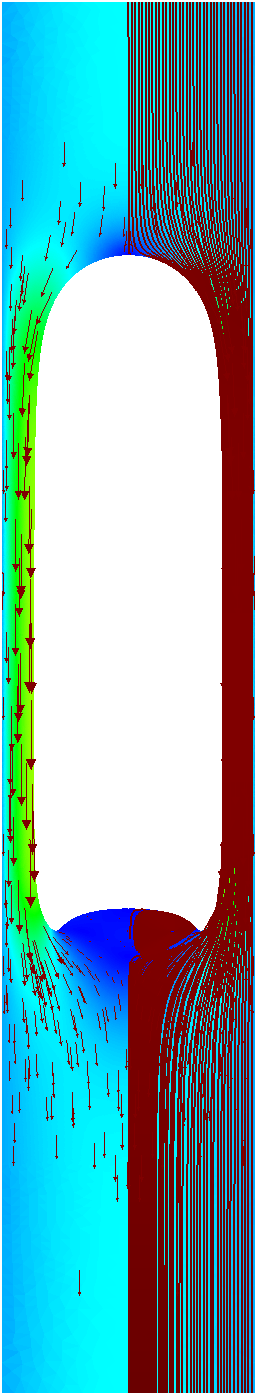} &
  	\includegraphics[width= 0.03\linewidth, angle=0]{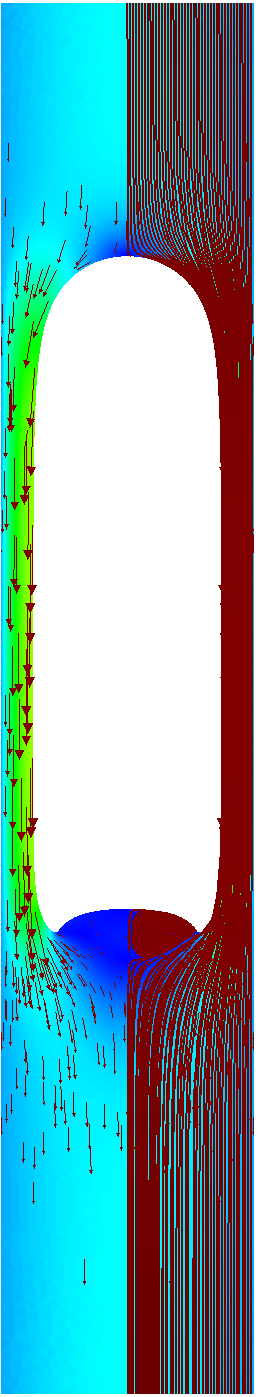} &
  	\includegraphics[width=0.08 \linewidth, angle=0]{Figures/Eo000}
\\ 
\includegraphics[width=0.15 \linewidth, angle=0]{Figures/UM20D} &
    \includegraphics[width= 0.03\linewidth, angle=0]{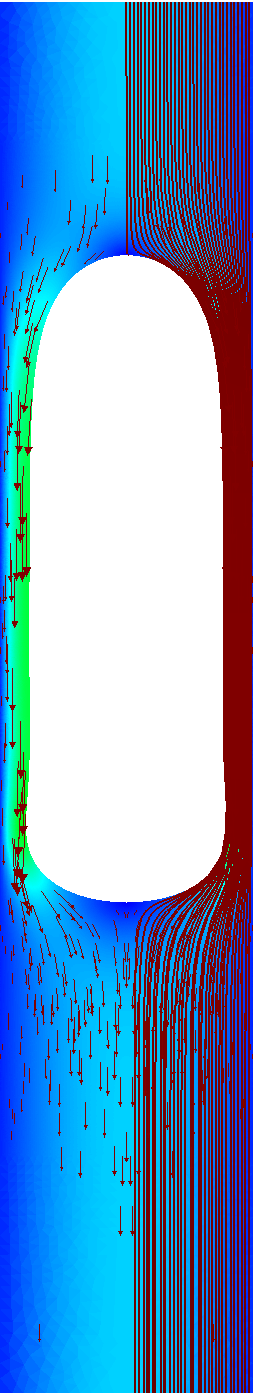} &
  	\includegraphics[width= 0.03\linewidth, angle=0]{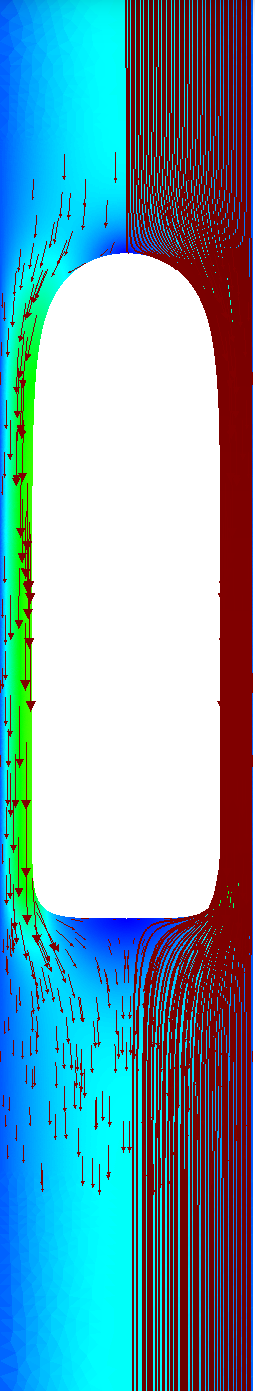} &
  	\includegraphics[width= 0.03\linewidth, angle=0]{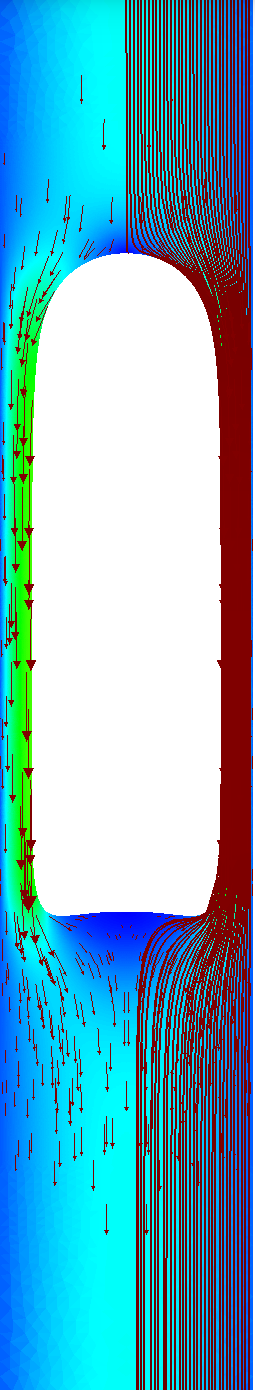} &
  	\includegraphics[width= 0.03\linewidth, angle=0]{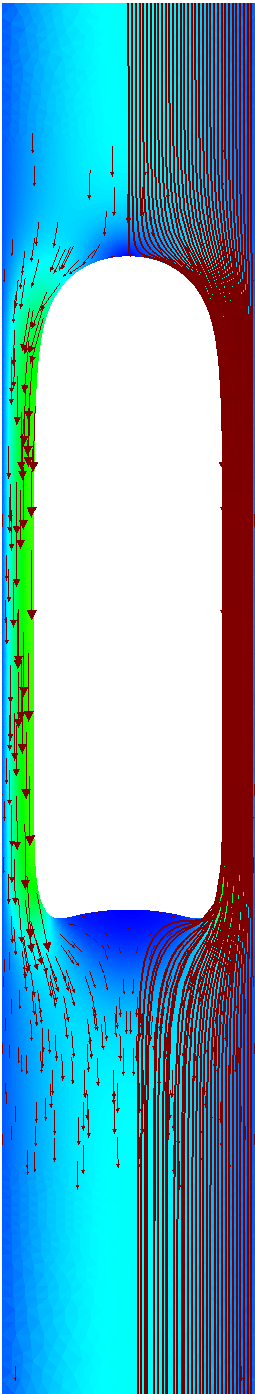} &
  	\includegraphics[width= 0.03\linewidth, angle=0]{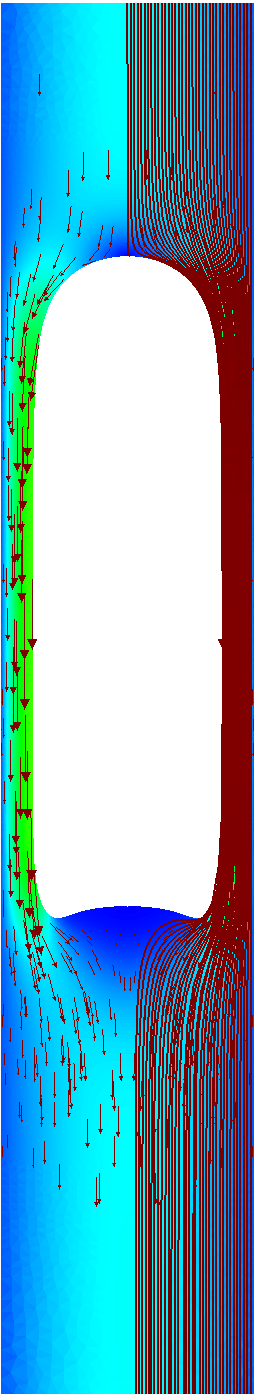} &
  	\includegraphics[width= 0.03\linewidth, angle=0]{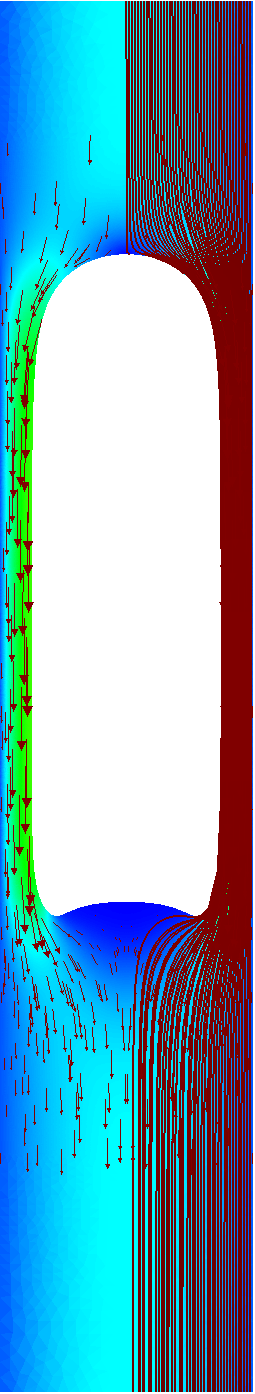} &
  	\includegraphics[width= 0.03\linewidth, angle=0]{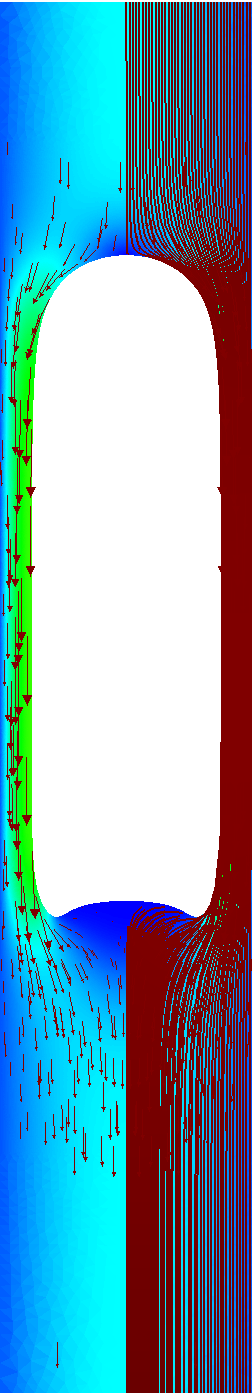} &
  	\includegraphics[width= 0.03\linewidth, angle=0]{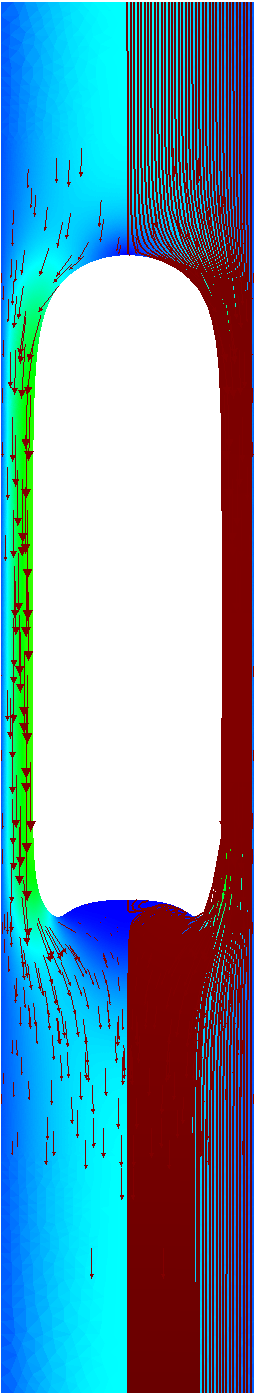} &
  	\includegraphics[width=0.08 \linewidth, angle=0]{Figures/Eo000}
\\ 
\includegraphics[width=0.15 \linewidth, angle=0]{Figures/UM30D} &
    \includegraphics[width= 0.03\linewidth, angle=0]{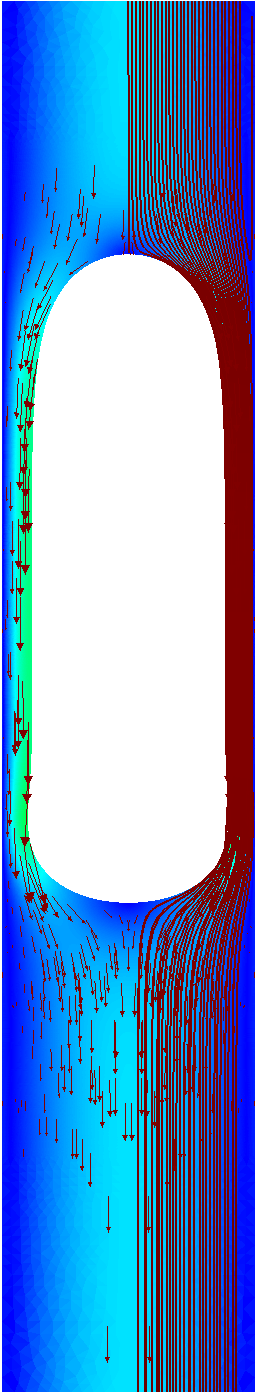} &
  	\includegraphics[width= 0.03\linewidth, angle=0]{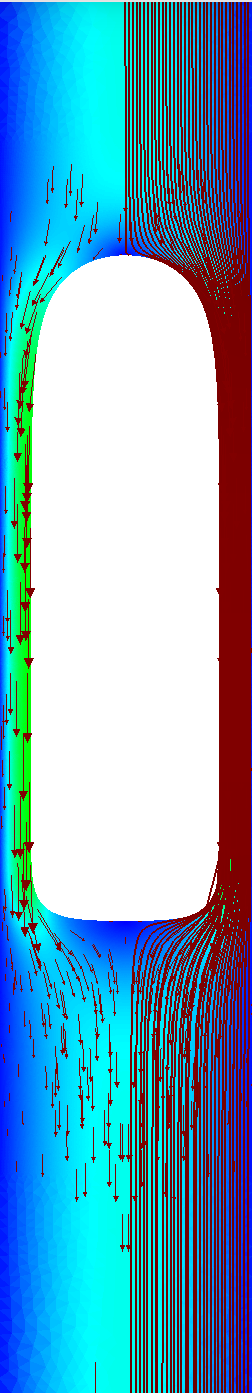} &
  	\includegraphics[width= 0.03\linewidth, angle=0]{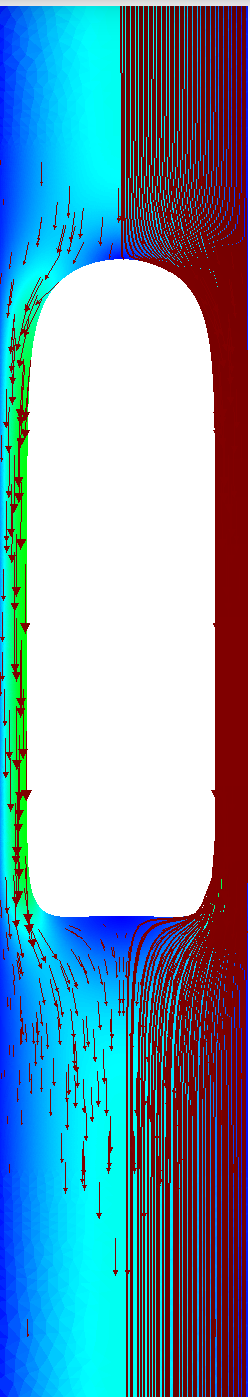} &
  	\includegraphics[width= 0.03\linewidth, angle=0]{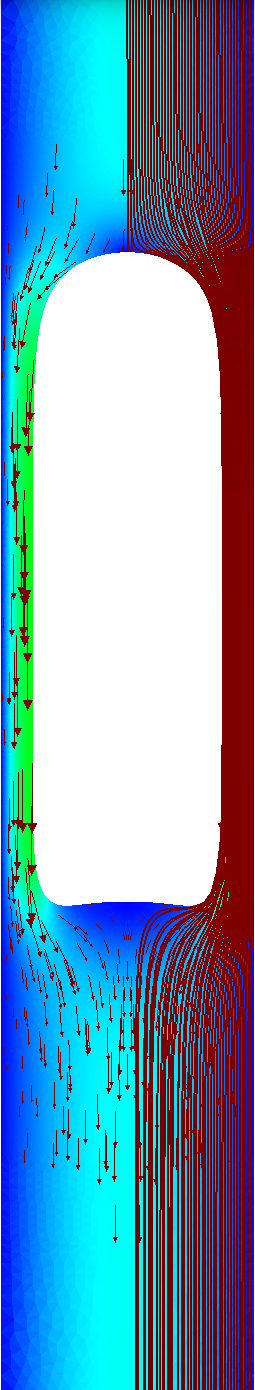} &
  	\includegraphics[width= 0.03\linewidth, angle=0]{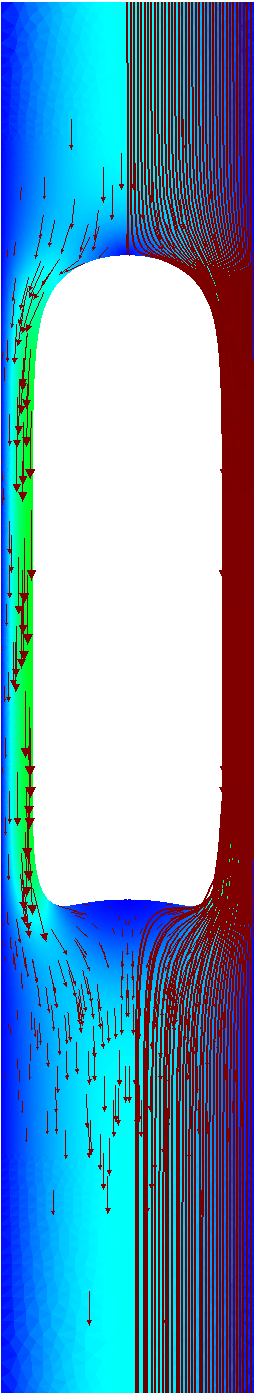} &
  	\includegraphics[width= 0.03\linewidth, angle=0]{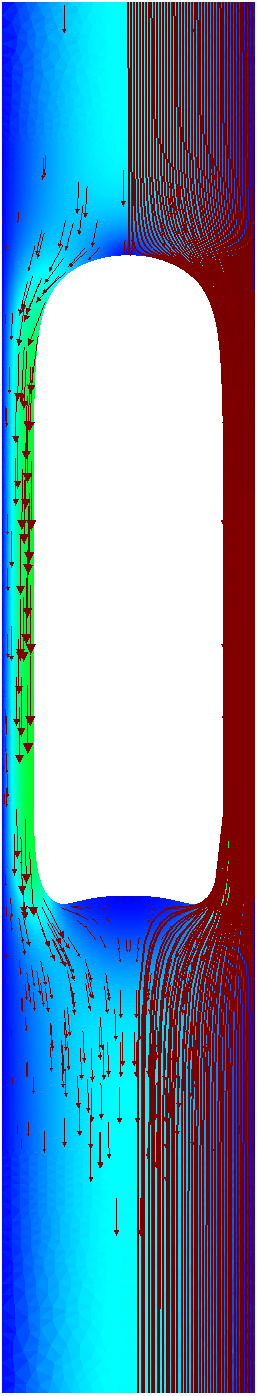} &
  	\includegraphics[width= 0.03\linewidth, angle=0]{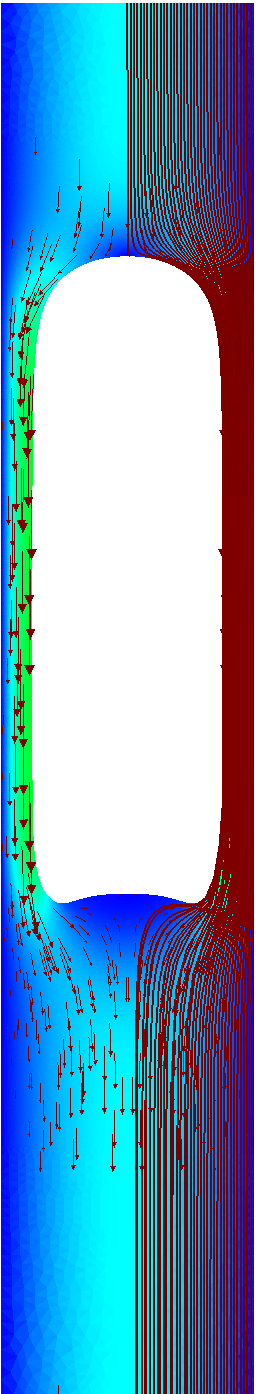} &
  	\includegraphics[width= 0.03\linewidth, angle=0]{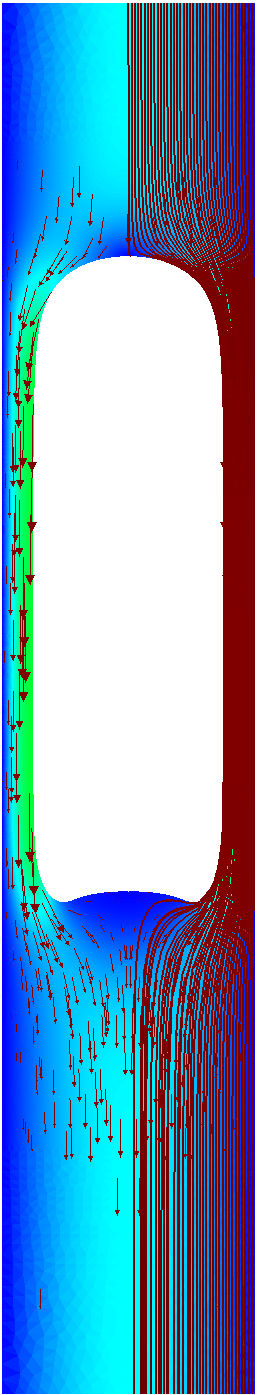} &
  	\includegraphics[width=0.08 \linewidth, angle=0]{Figures/Eo000}
\\ 
\includegraphics[width=0.15 \linewidth, angle=0]{Figures/UM40D} &
    \includegraphics[width= 0.03\linewidth, angle=0]{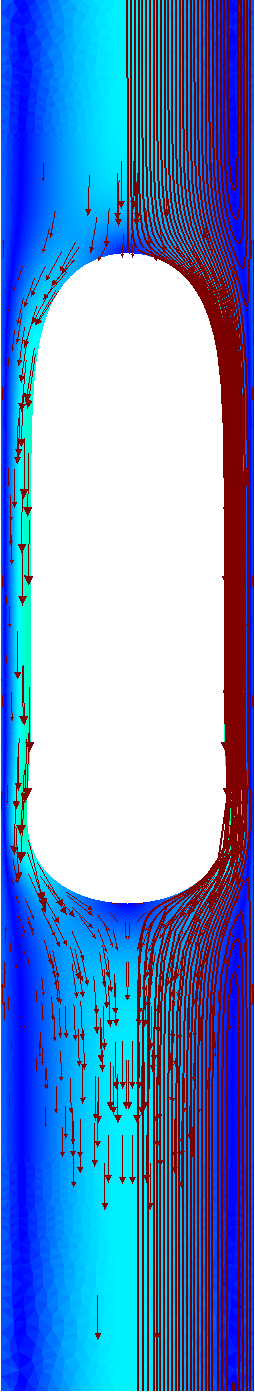} &
  	\includegraphics[width= 0.03\linewidth, angle=0]{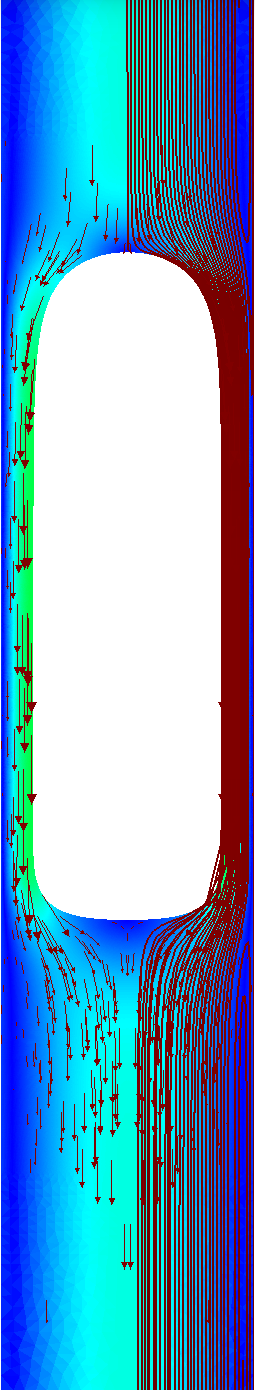} &
  	\includegraphics[width= 0.03\linewidth, angle=0]{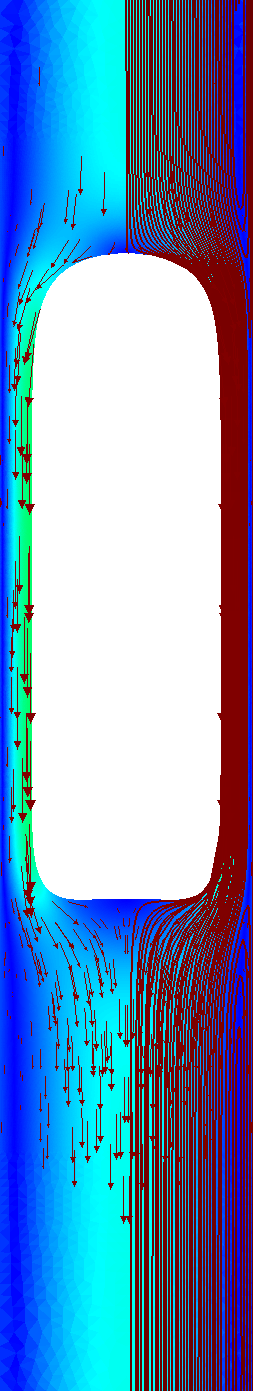} &
  	\includegraphics[width= 0.03\linewidth, angle=0]{FlowThesisFig/FlowNf80Eo140ud40} &
  	\includegraphics[width= 0.03\linewidth, angle=0]{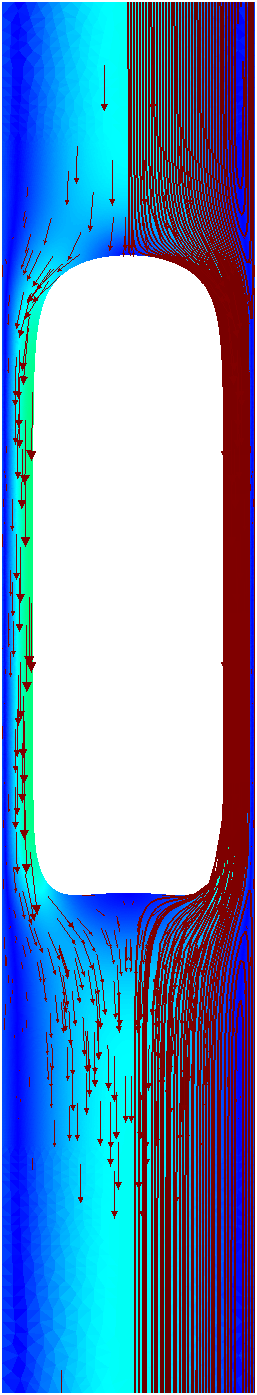} &
  	\includegraphics[width= 0.03\linewidth, angle=0]{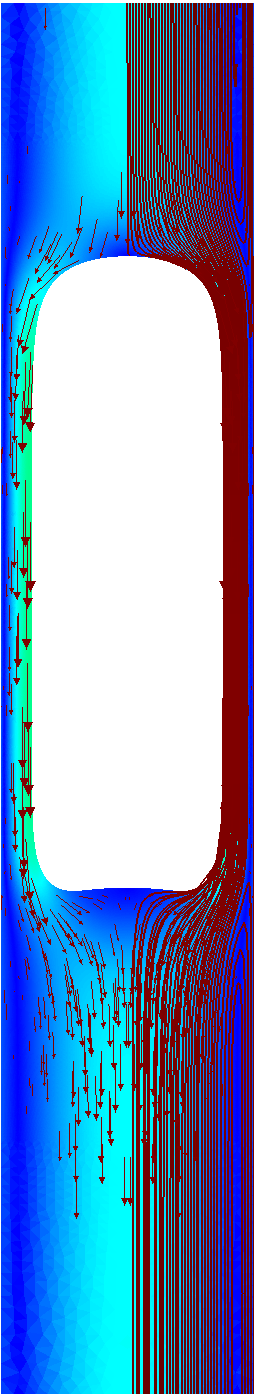} &
  	\includegraphics[width= 0.03\linewidth, angle=0]{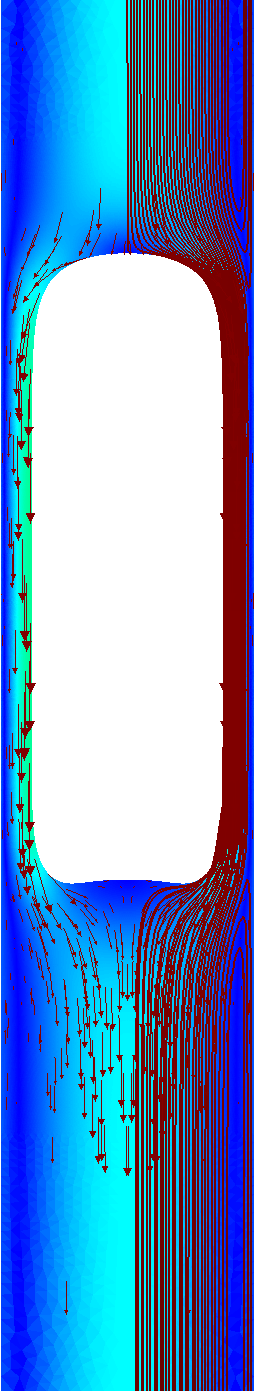} &
  	\includegraphics[width= 0.03\linewidth, angle=0]{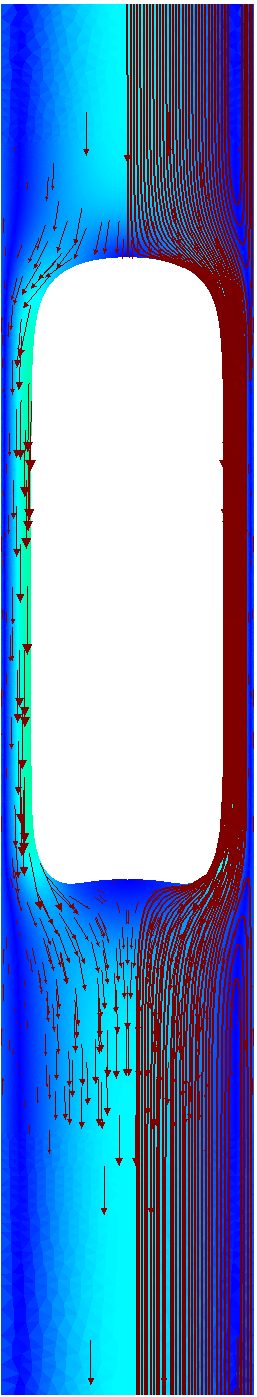} &
  	\includegraphics[width=0.08 \linewidth, angle=0]{Figures/Eo000}
\\ 
\includegraphics[width=0.15 \linewidth, angle=0]{Figures/UM50D} &
    \includegraphics[width= 0.03\linewidth, angle=0]{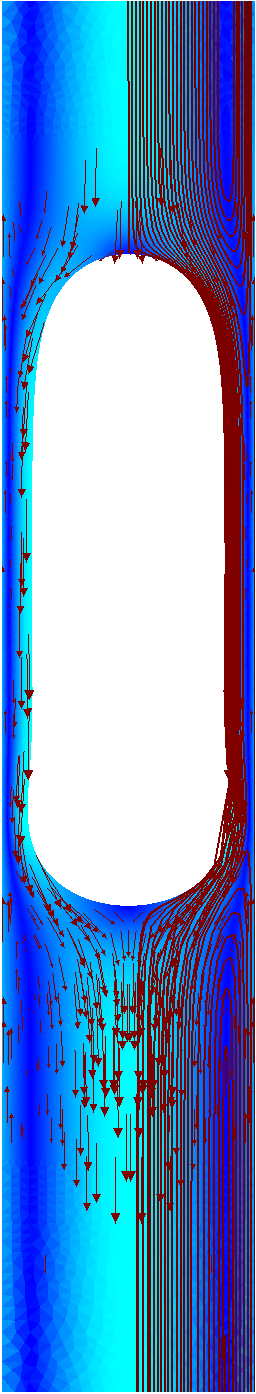} &
  	\includegraphics[width= 0.03\linewidth, angle=0]{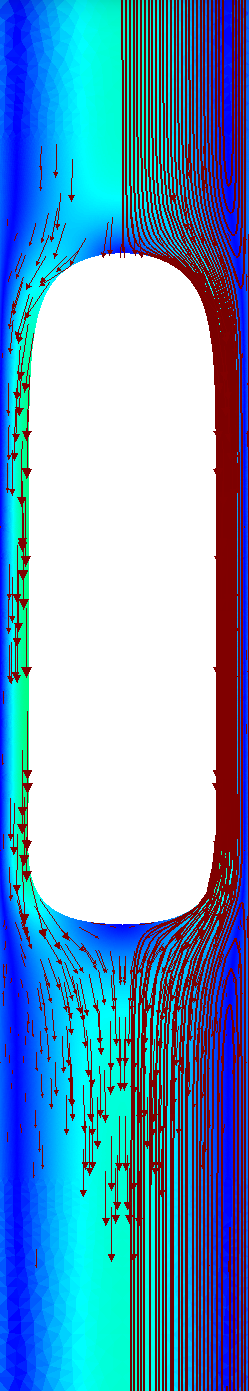} &
  	\includegraphics[width= 0.03\linewidth, angle=0]{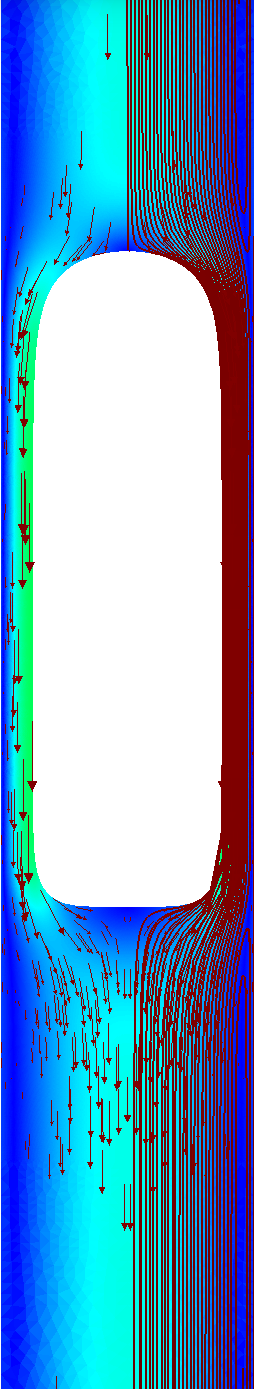} &
  	\includegraphics[width= 0.03\linewidth, angle=0]{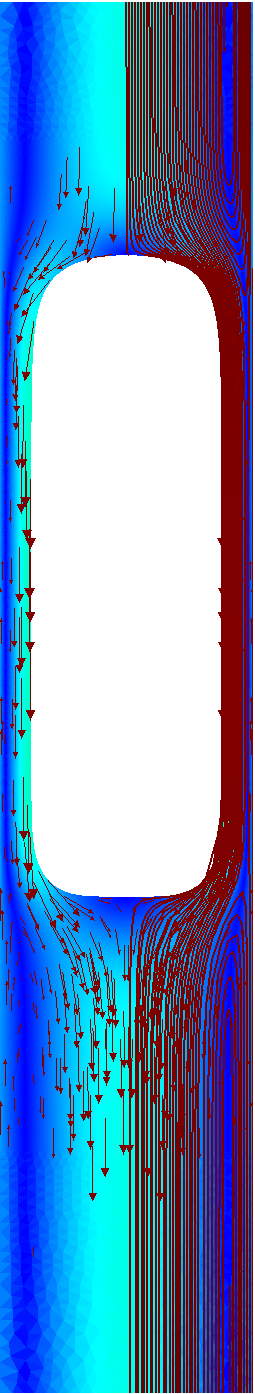} &
  	\includegraphics[width= 0.03\linewidth, angle=0]{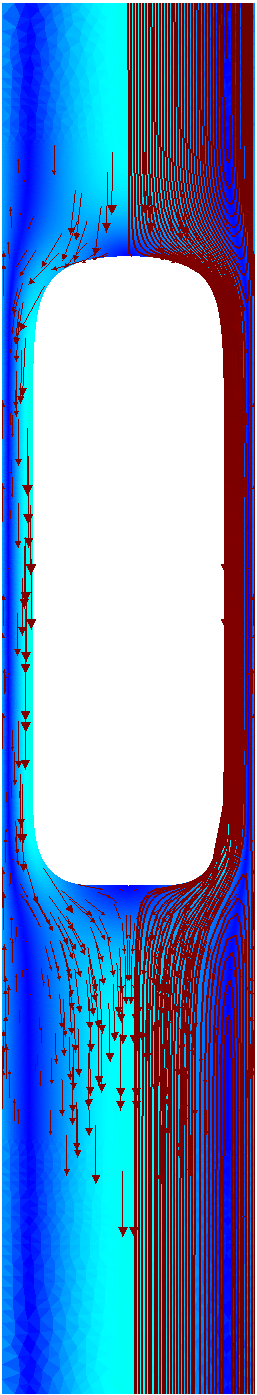} &
  	\includegraphics[width= 0.03\linewidth, angle=0]{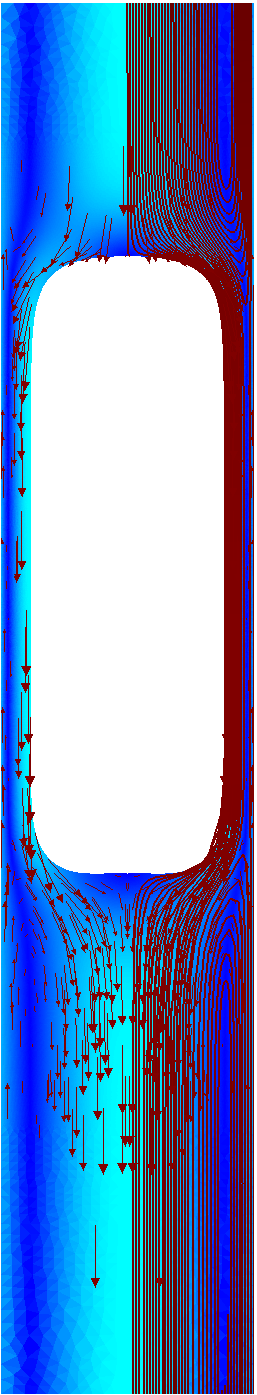} &
  	\includegraphics[width= 0.03\linewidth, angle=0]{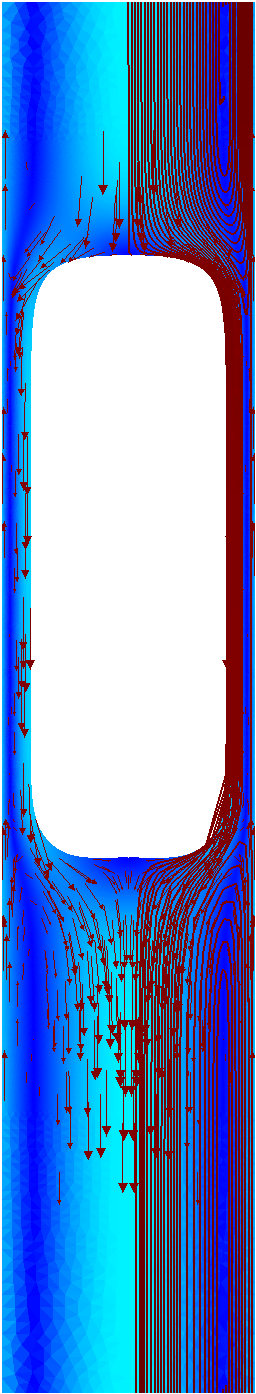} &
  	\includegraphics[width= 0.03\linewidth, angle=0]{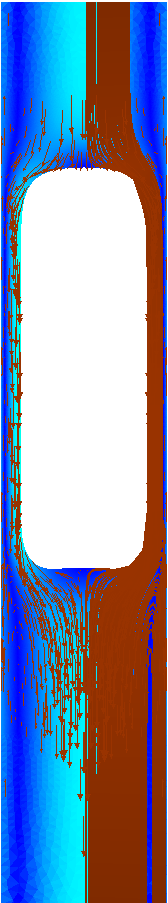} &
  	\includegraphics[width=0.08 \linewidth, angle=0]{Figures/Eo000} 
\\
\includegraphics[width=0.07 \linewidth, angle=0]{Figures/Eo000} & 
 \includegraphics[width= 0.07\linewidth, angle=0]{Figures/Eo20} &  \includegraphics[width= 0.07 \linewidth, angle=0]{Figures/Eo60} & \includegraphics[width= 0.07 \linewidth, angle=0]{Figures/Eo100} & \includegraphics[width= 0.07 \linewidth, angle=0]{Figures/Eo140} &
\includegraphics[width= 0.07 \linewidth, angle=0]{Figures/Eo180} & \includegraphics[width= 0.07 \linewidth, angle=0]{Figures/Eo220} & \includegraphics[width= 0.07 \linewidth, angle=0]{Figures/Eo260} & \includegraphics[width= 0.07 \linewidth, angle=0]{Figures/Eo300} & \includegraphics[width=0.07 \linewidth, angle=0]{Figures/Eo000} 
\end{tabular}
\caption{The effect of $U_m$ and $Eo$ on the steady bubble shapes and flow fields with $Nf = 80$. In each panel, the streamlines and vector fields are superimposed on velocity magnitude pseudocolour plot on the right and left sides of the symmetry axis, respectively.}
\label{fig:ss_phasefield_steady_state_flow_Nf_80}
\end{figure}

\begin{figure}
\centering
\begin{tabular}{cccccccccc} 
\includegraphics[width=0.10 \linewidth, angle=0]{Figures/UM00S} &
\includegraphics[width= 0.03\linewidth, angle=0]{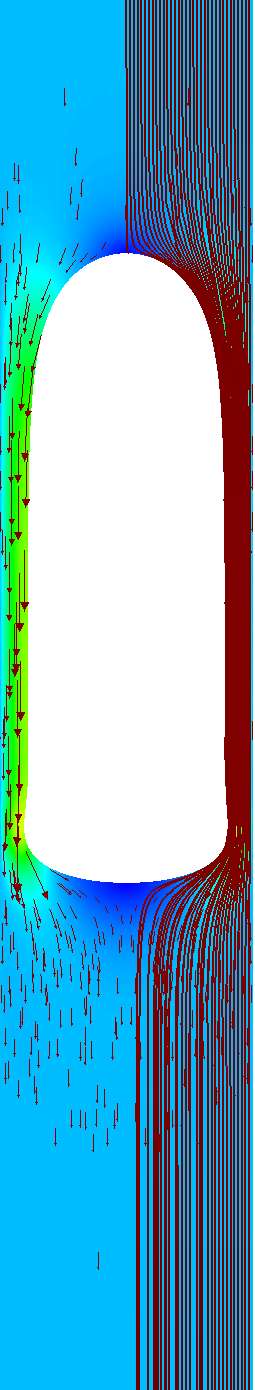} & \includegraphics[width= 0.03\linewidth, angle=0]{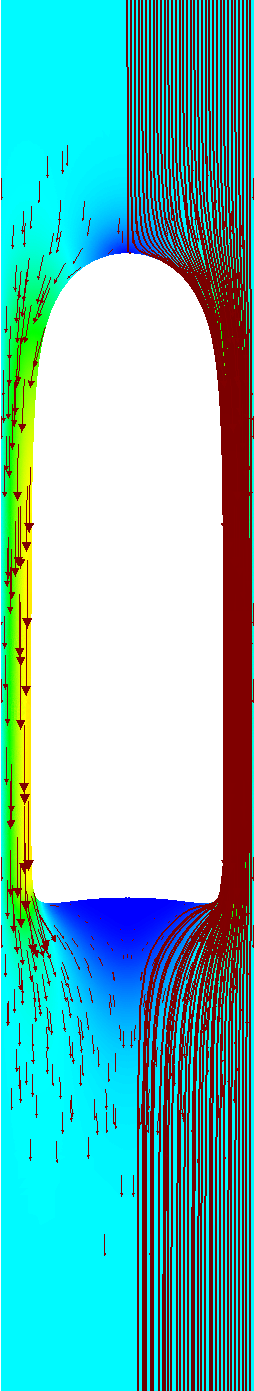} & \includegraphics[width= 0.03\linewidth, angle=0]{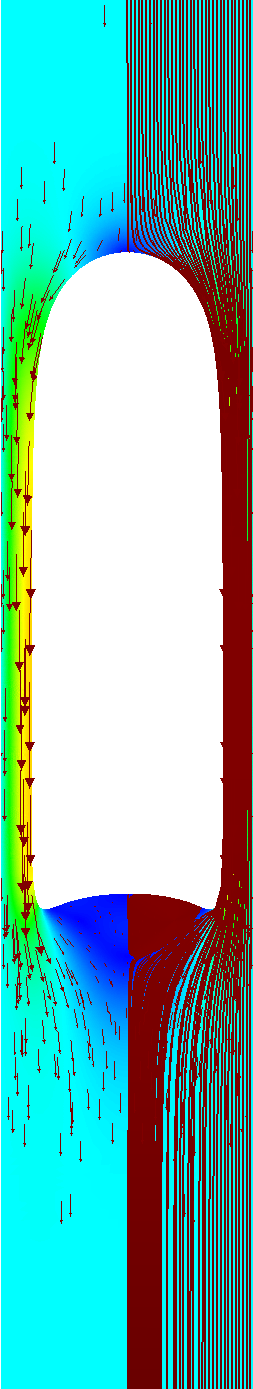} & \includegraphics[width= 0.03\linewidth, angle=0]{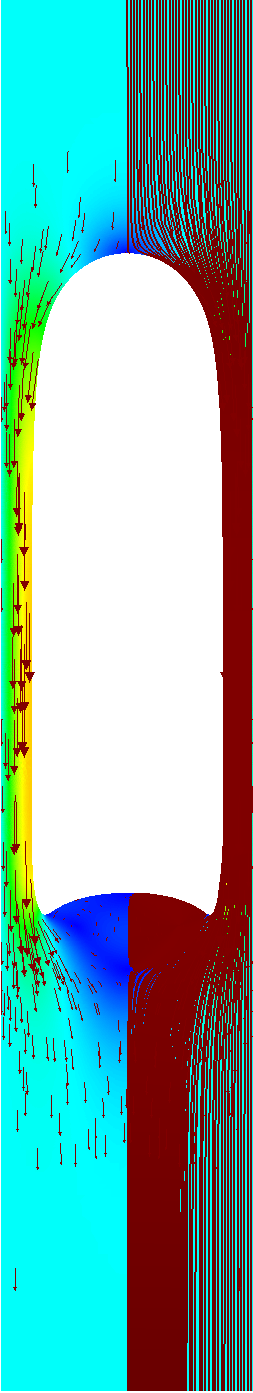} &
\includegraphics[width= 0.03\linewidth, angle=0]{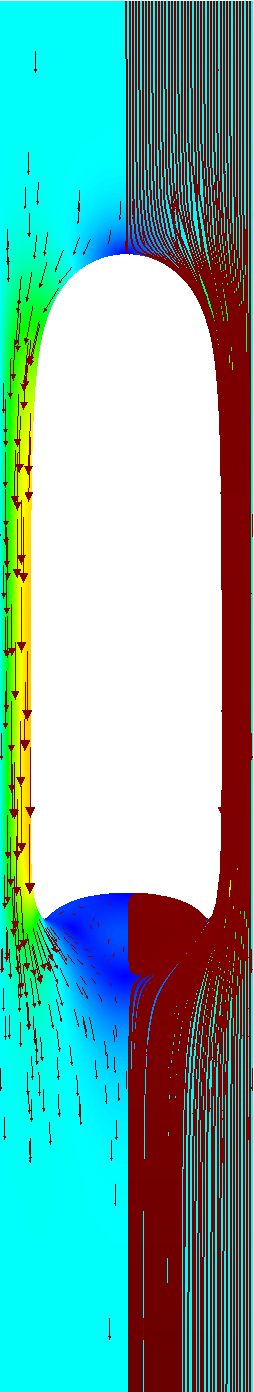} & \includegraphics[width= 0.03\linewidth, angle=0]{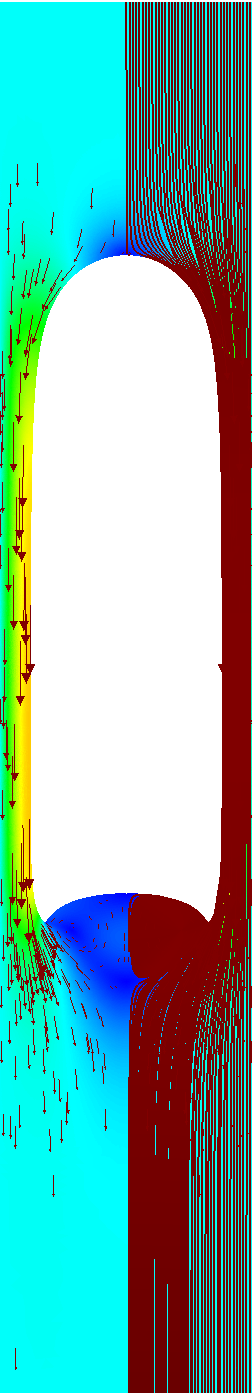} & \includegraphics[width= 0.03\linewidth, angle=0]{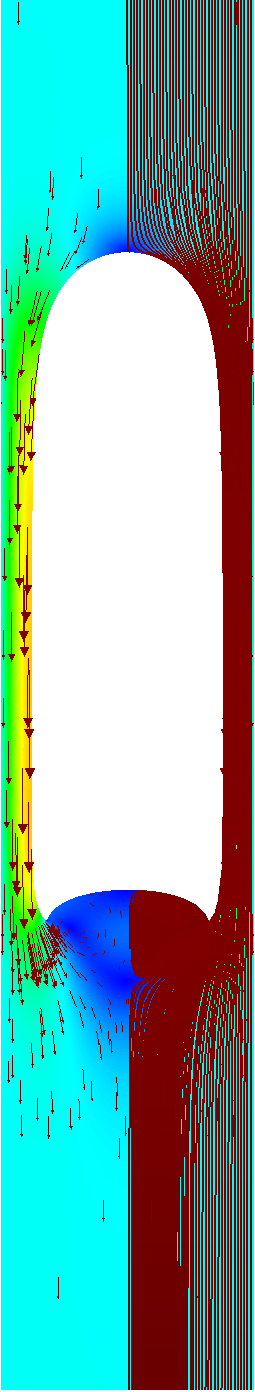} & \includegraphics[width= 0.03\linewidth, angle=0]{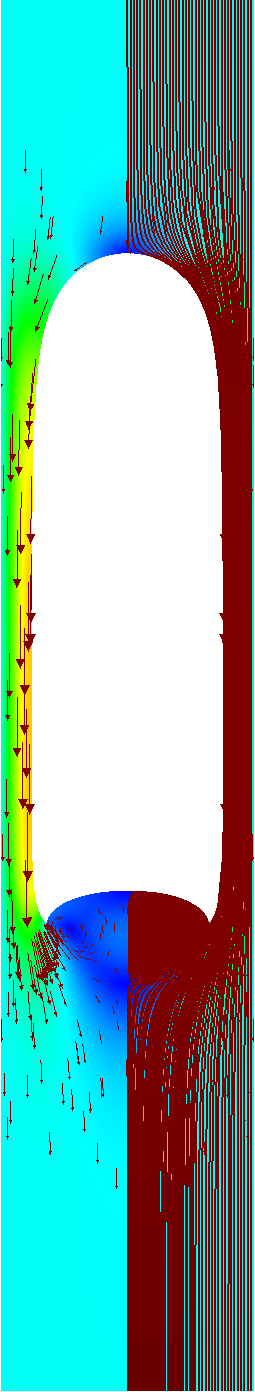} &
\includegraphics[width=0.08 \linewidth, angle=0]{Figures/colourmap}
\\ 
\includegraphics[width=0.15 \linewidth, angle=0]{Figures/UM10D} &
    \includegraphics[width= 0.03\linewidth, angle=0]{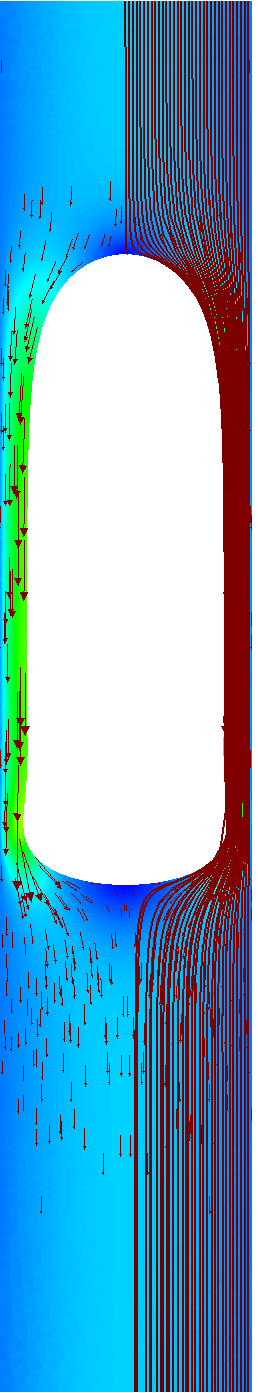} &
  	\includegraphics[width= 0.03\linewidth, angle=0]{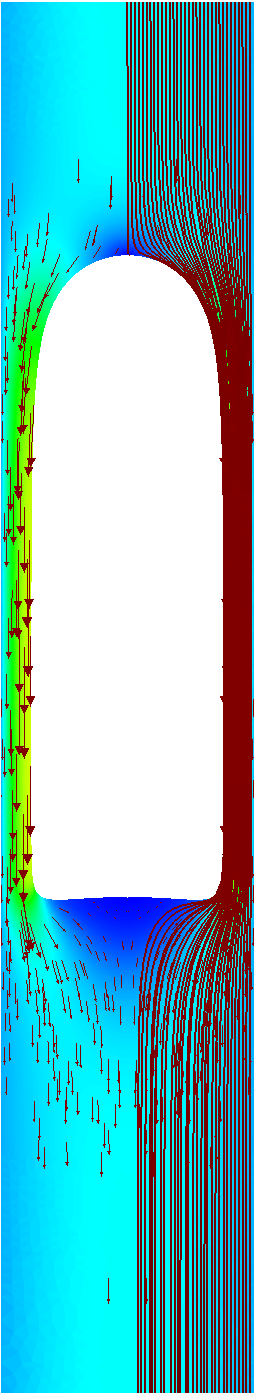} &
  	\includegraphics[width= 0.03\linewidth, angle=0]{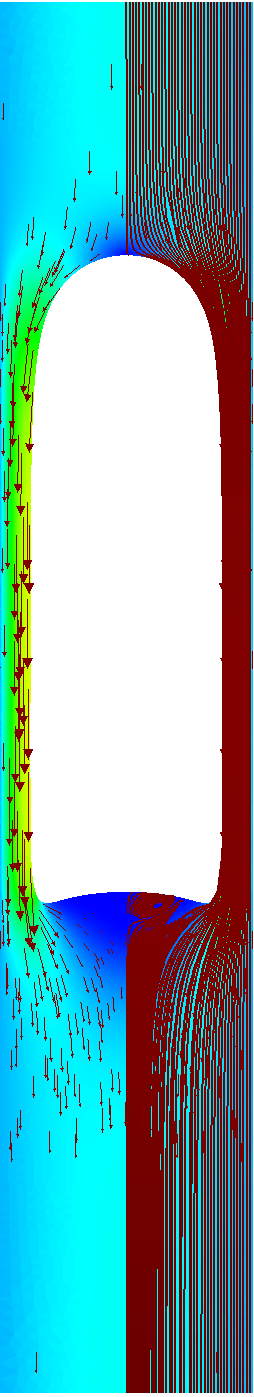} &
  	\includegraphics[width= 0.03\linewidth, angle=0]{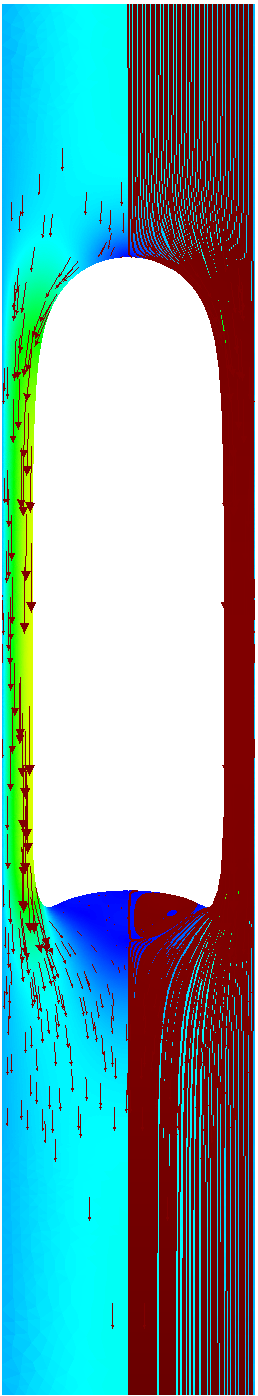} &
  	\includegraphics[width= 0.03\linewidth, angle=0]{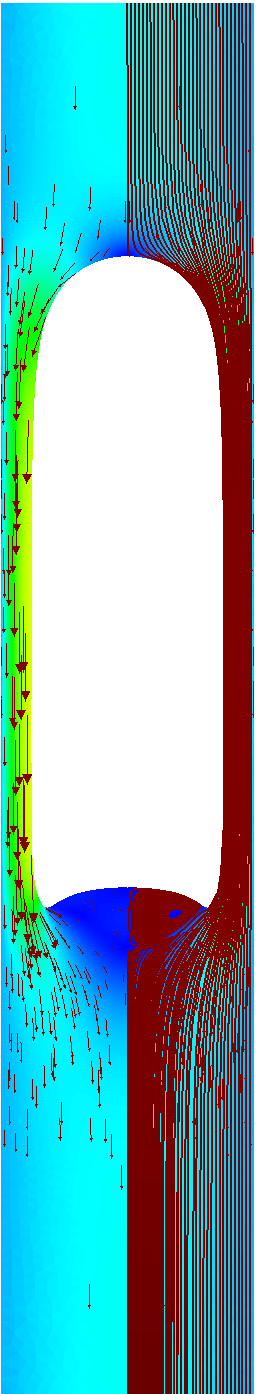} &
  	\includegraphics[width= 0.03\linewidth, angle=0]{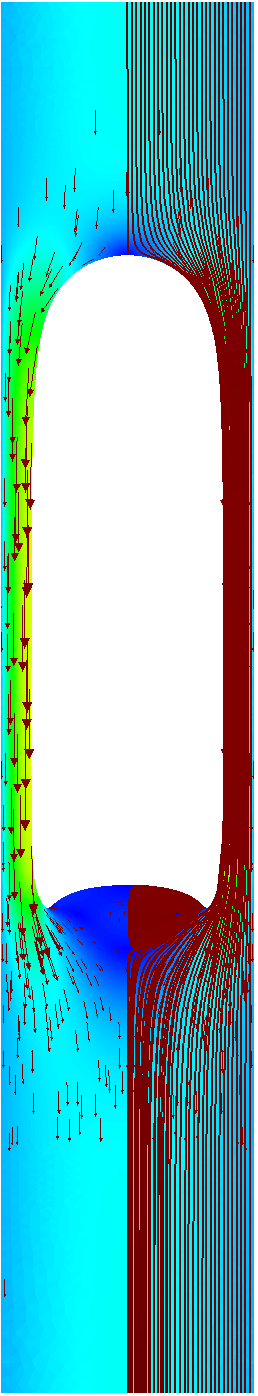} &
  	\includegraphics[width= 0.03\linewidth, angle=0]{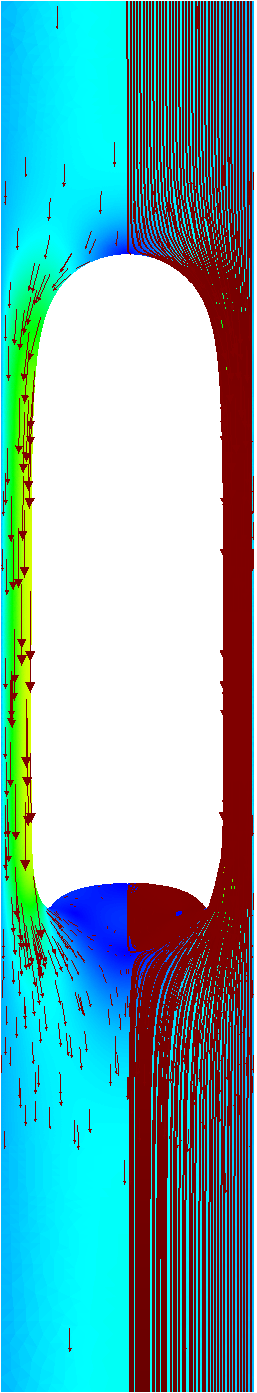} &
  	\includegraphics[width= 0.03\linewidth, angle=0]{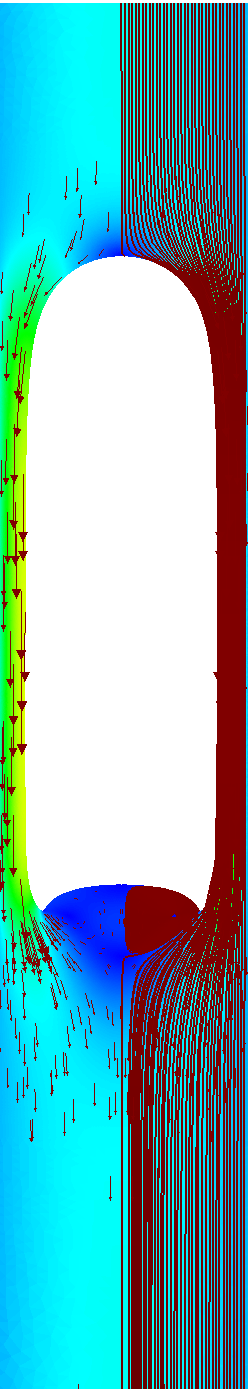} &
  	\includegraphics[width=0.08 \linewidth, angle=0]{Figures/Eo000}
\\ 
\includegraphics[width=0.15 \linewidth, angle=0]{Figures/UM20D} &
    \includegraphics[width= 0.03\linewidth, angle=0]{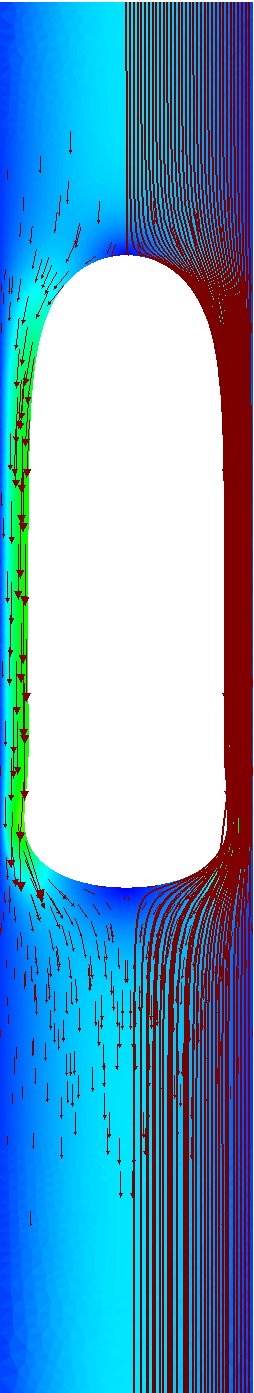} &
  	\includegraphics[width= 0.03\linewidth, angle=0]{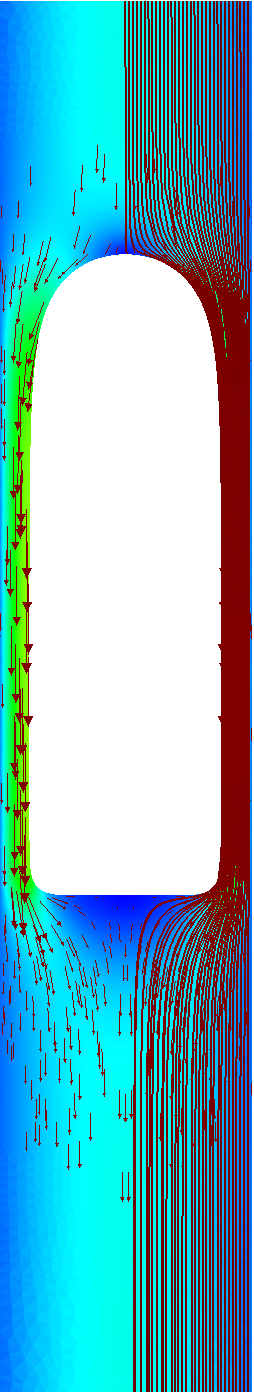} &
  	\includegraphics[width= 0.03\linewidth, angle=0]{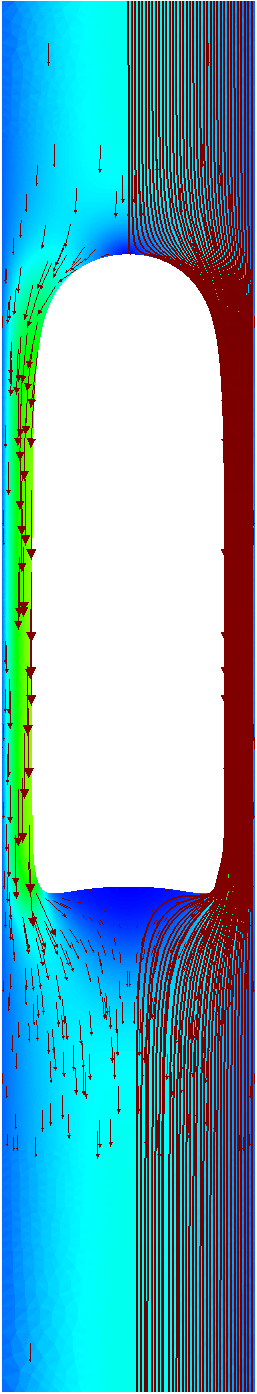} &
  	\includegraphics[width= 0.03\linewidth, angle=0]{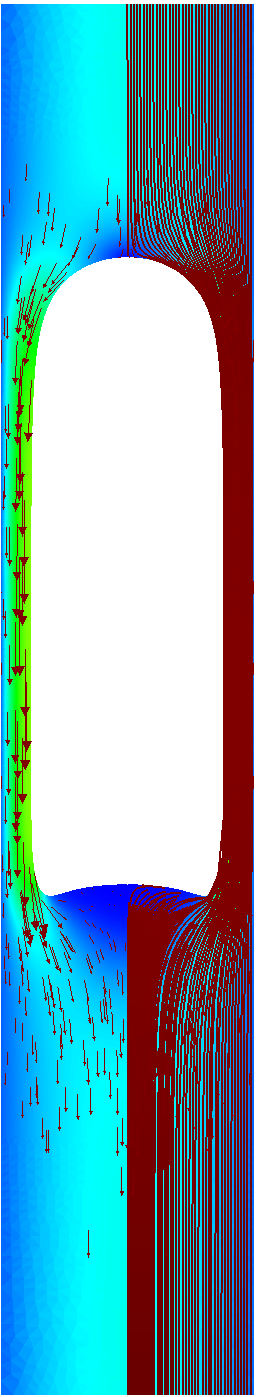} &
  	\includegraphics[width= 0.03\linewidth, angle=0]{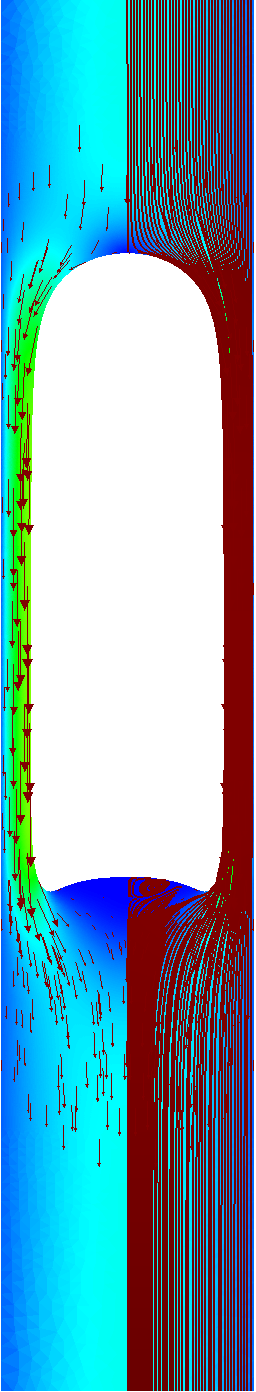} &
  	\includegraphics[width= 0.03\linewidth, angle=0]{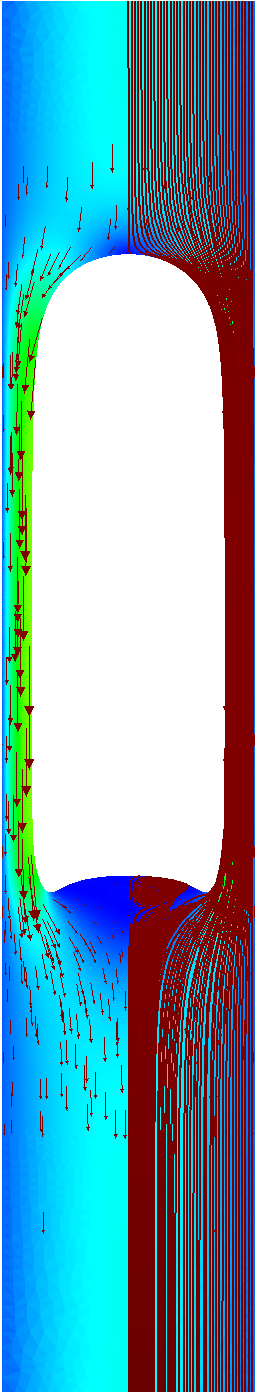} &
  	\includegraphics[width= 0.03\linewidth, angle=0]{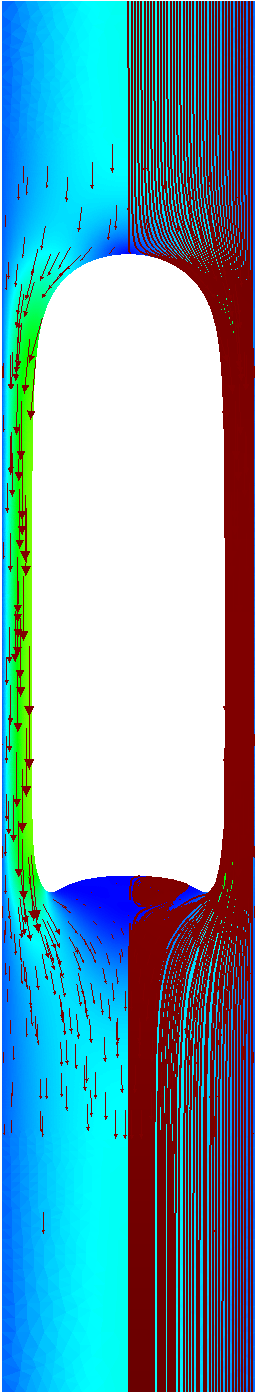} &
  	\includegraphics[width= 0.03\linewidth, angle=0]{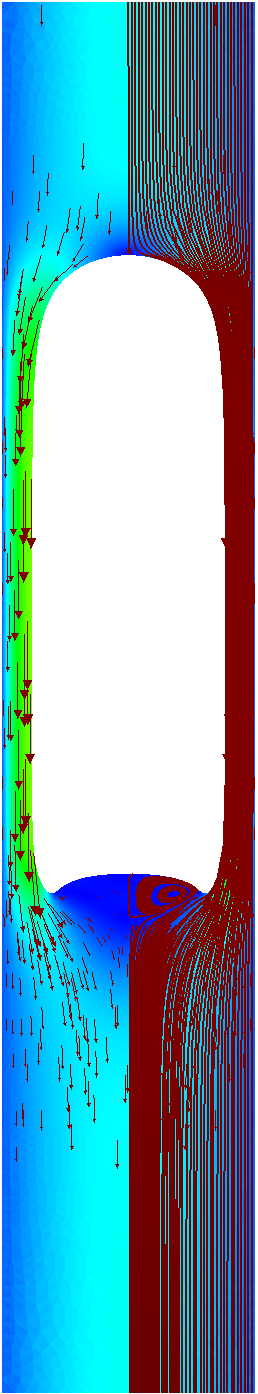} &
  	\includegraphics[width=0.08 \linewidth, angle=0]{Figures/Eo000}
\\ 
\includegraphics[width=0.15 \linewidth, angle=0]{Figures/UM30D} &
    \includegraphics[width= 0.03\linewidth, angle=0]{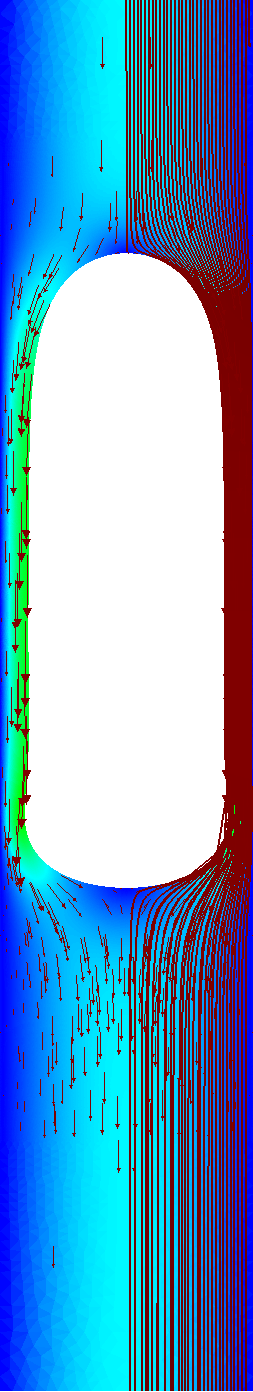} &
  	\includegraphics[width= 0.03\linewidth, angle=0]{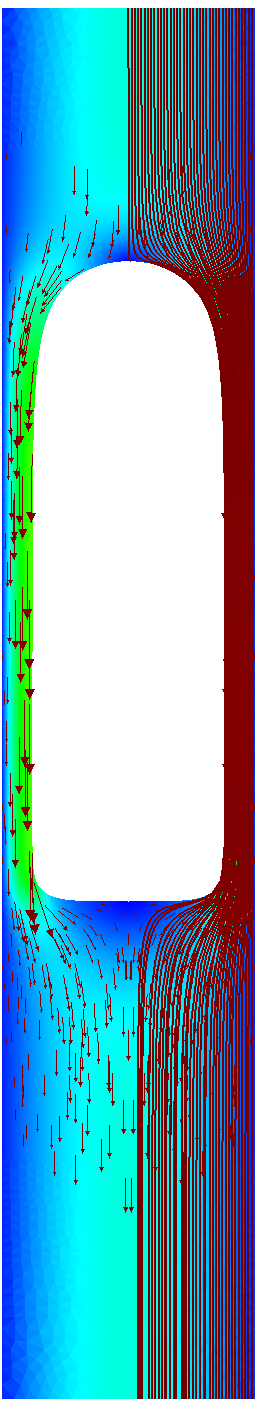} &
  	\includegraphics[width= 0.03\linewidth, angle=0]{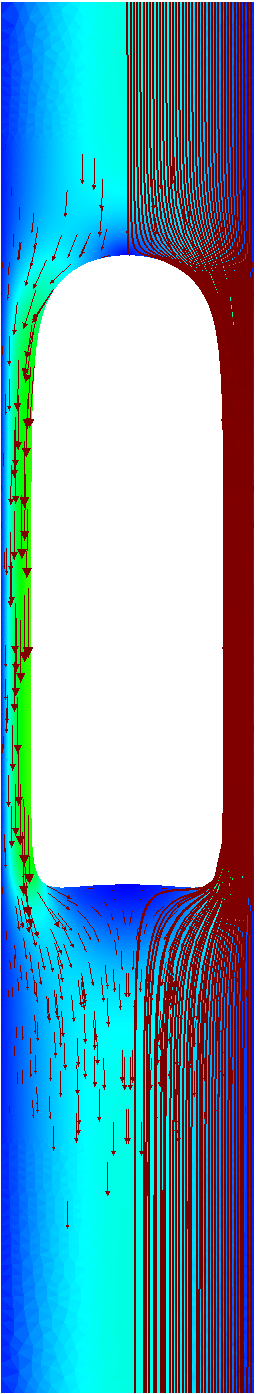} &
  	\includegraphics[width= 0.03\linewidth, angle=0]{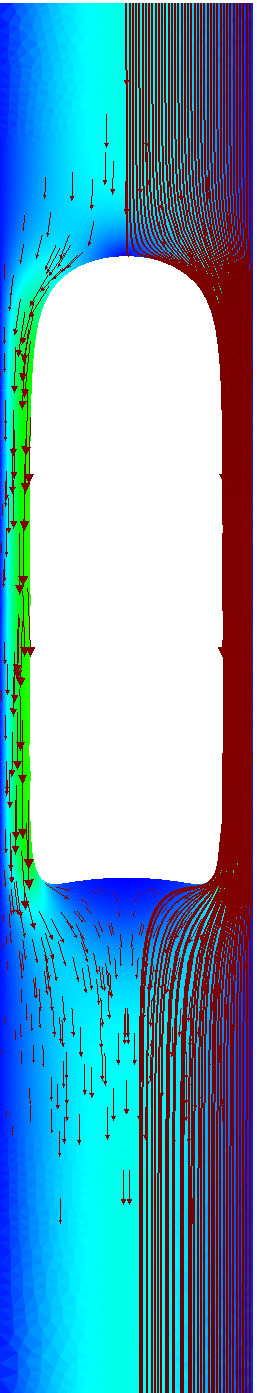} &
  	\includegraphics[width= 0.03\linewidth, angle=0]{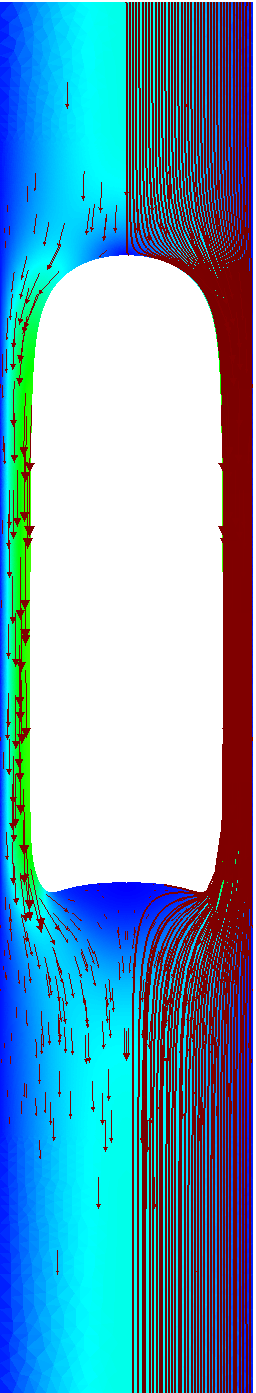} &
  	\includegraphics[width= 0.03\linewidth, angle=0]{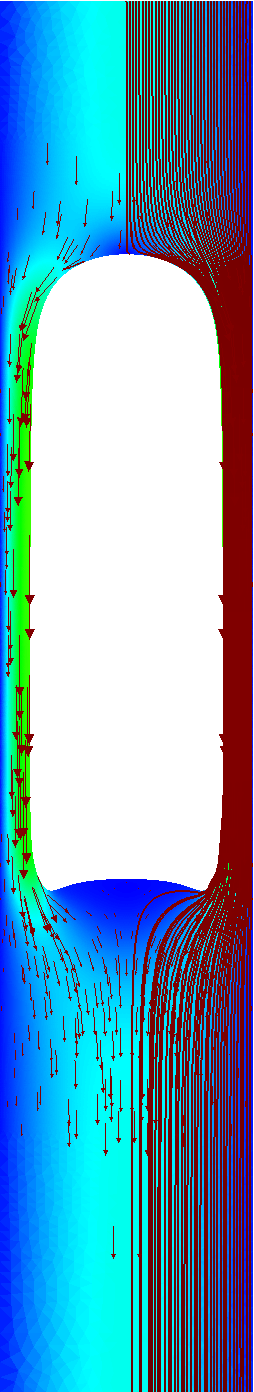} &
  	\includegraphics[width= 0.03\linewidth, angle=0]{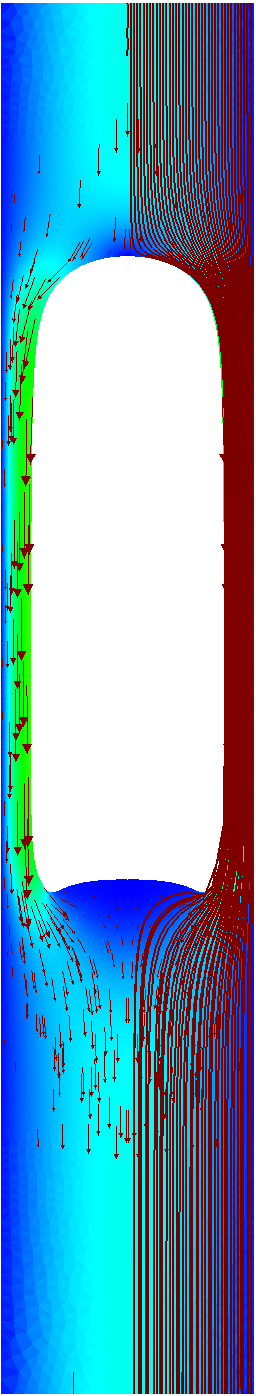} &
  	\includegraphics[width= 0.03\linewidth, angle=0]{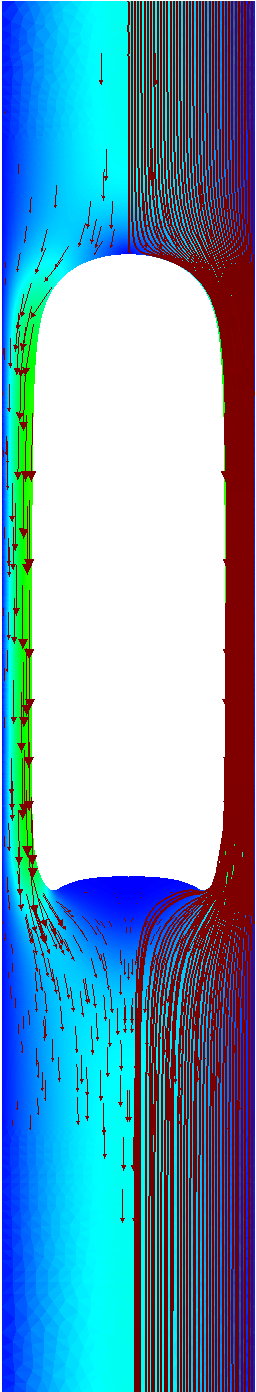} &
  	\includegraphics[width=0.08 \linewidth, angle=0]{Figures/Eo000}
\\ 
\includegraphics[width=0.15 \linewidth, angle=0]{Figures/UM40D} &
    \includegraphics[width= 0.03\linewidth, angle=0]{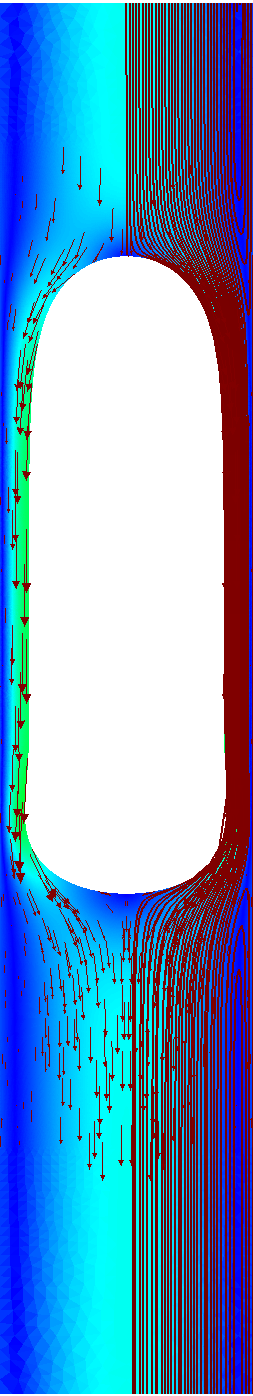} &
  	\includegraphics[width= 0.03\linewidth, angle=0]{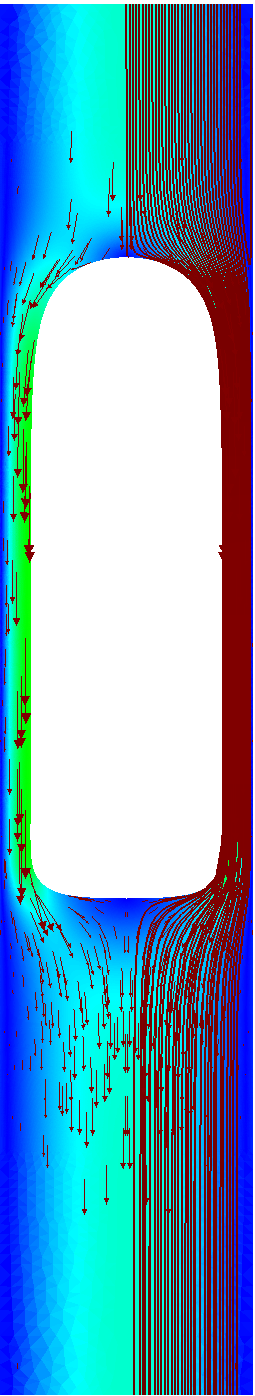} &
  	\includegraphics[width= 0.03\linewidth, angle=0]{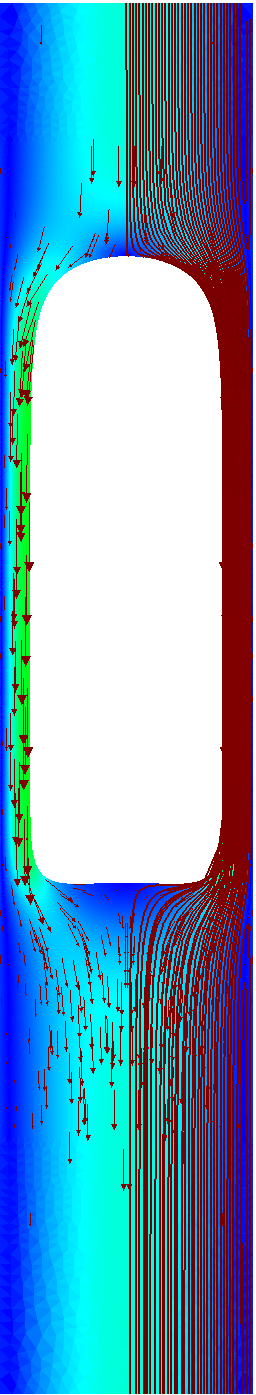} &
  	\includegraphics[width= 0.03\linewidth, angle=0]{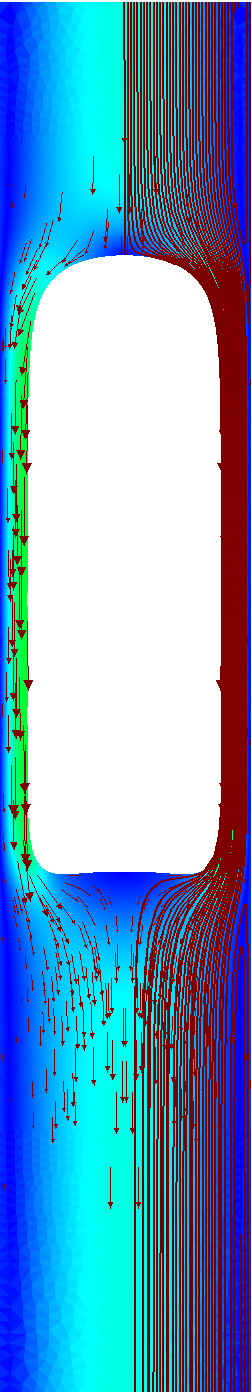} &
  	\includegraphics[width= 0.03\linewidth, angle=0]{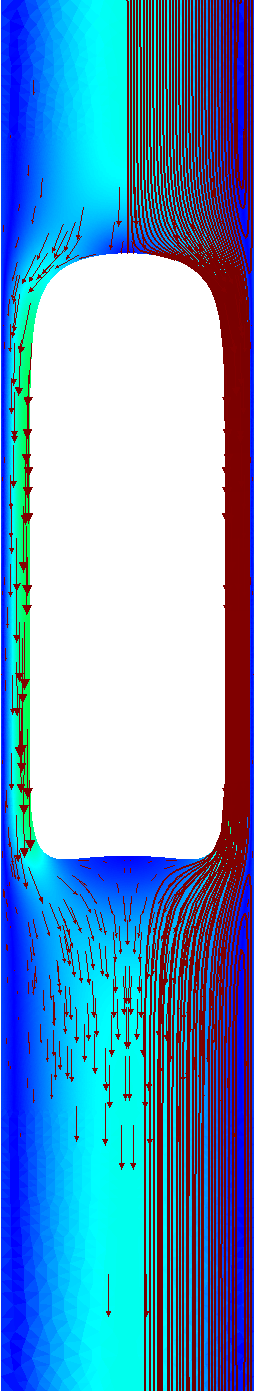} &
  	\includegraphics[width= 0.03\linewidth, angle=0]{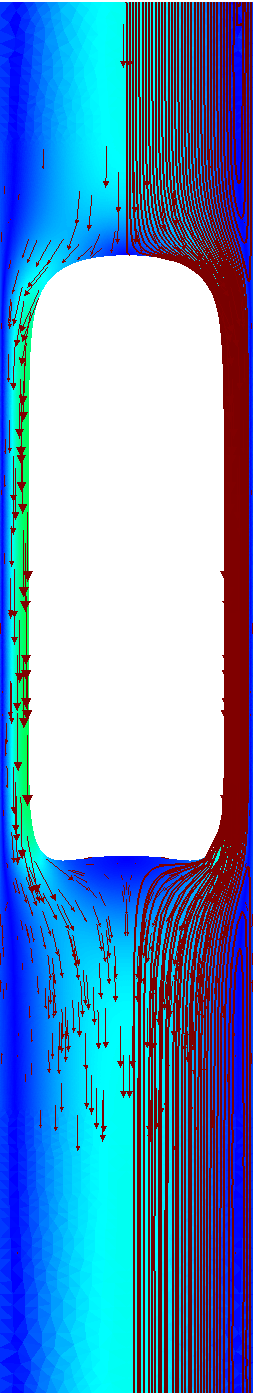} &
  	\includegraphics[width= 0.03\linewidth, angle=0]{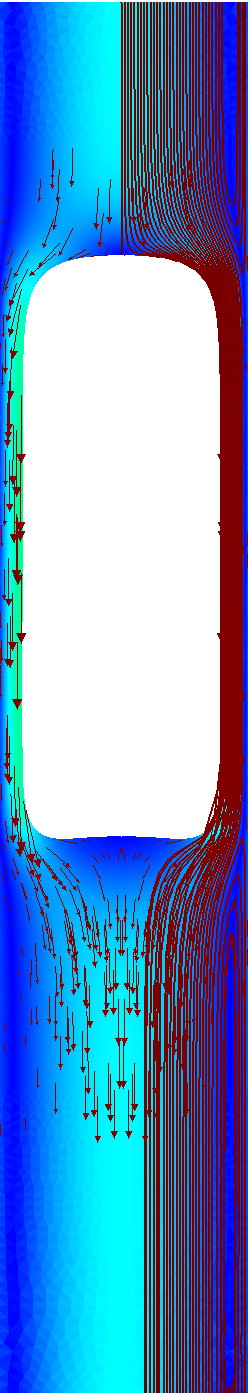} &
  	\includegraphics[width= 0.03\linewidth, angle=0]{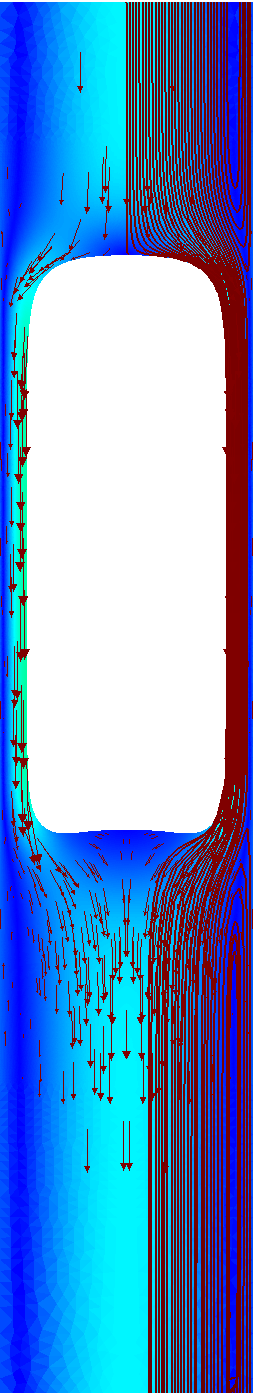} &
  	\includegraphics[width=0.08 \linewidth, angle=0]{Figures/Eo000}
\\ 
\includegraphics[width=0.15 \linewidth, angle=0]{Figures/UM50D} &
    \includegraphics[width= 0.03\linewidth, angle=0]{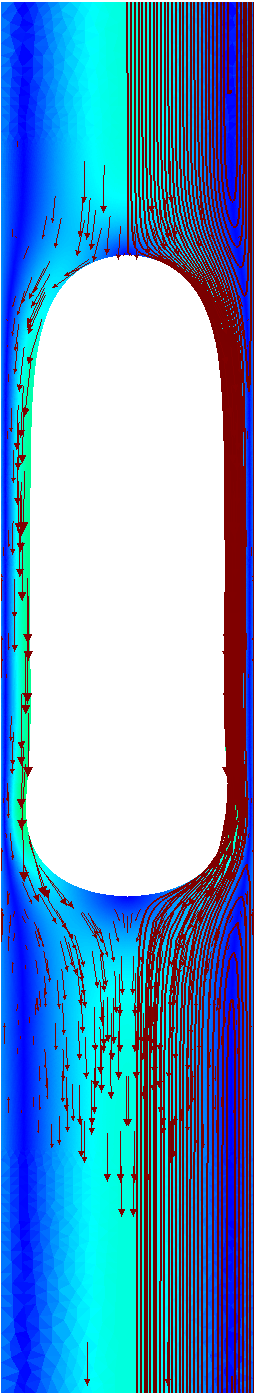} &
  	\includegraphics[width= 0.03\linewidth, angle=0]{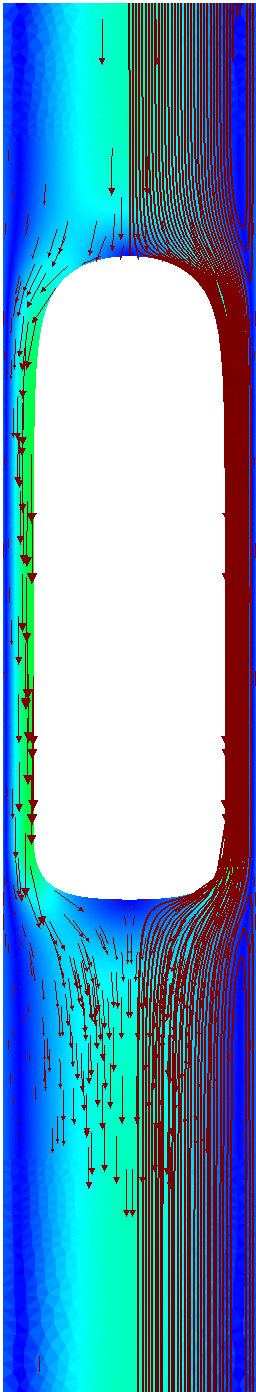} &
  	\includegraphics[width= 0.03\linewidth, angle=0]{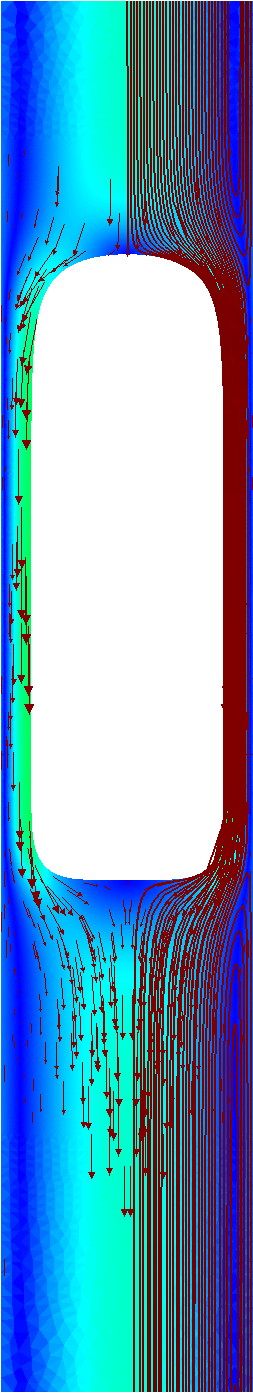} &
  	\includegraphics[width= 0.03\linewidth, angle=0]{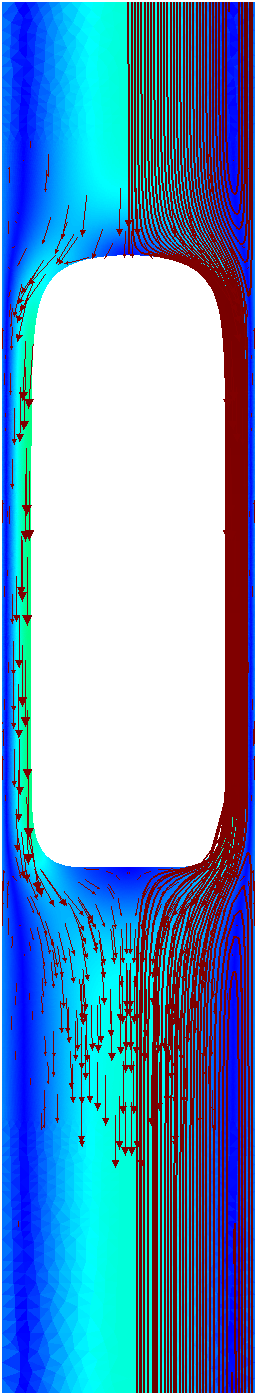} &
  	\includegraphics[width= 0.03\linewidth, angle=0]{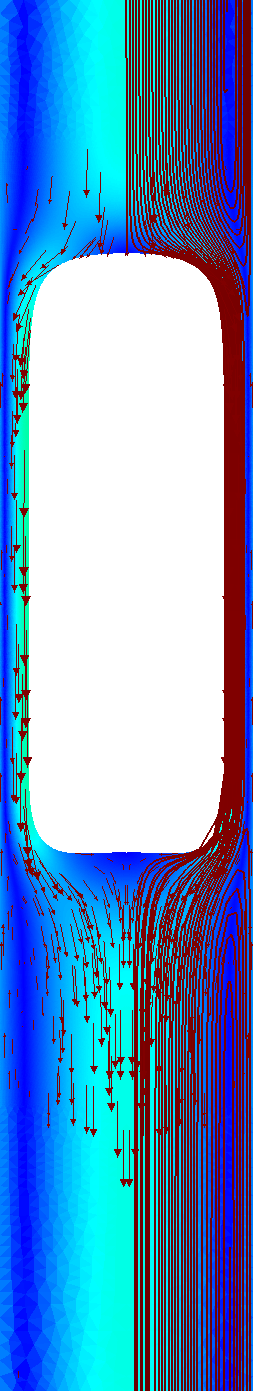} &
  	\includegraphics[width= 0.03\linewidth, angle=0]{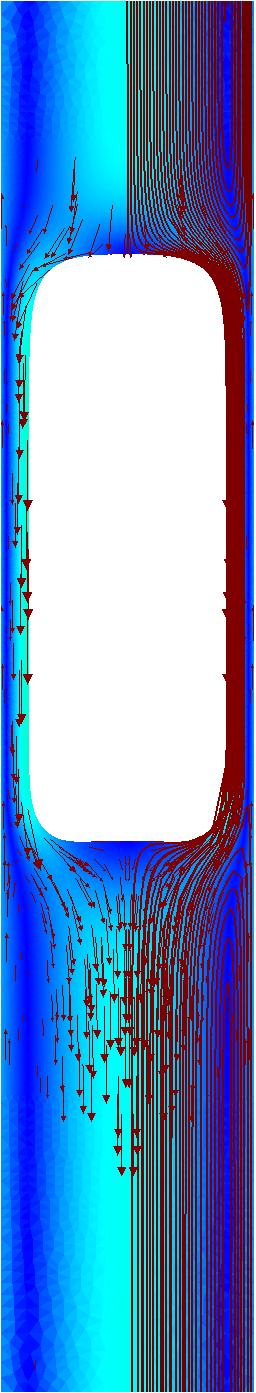} &
  	\includegraphics[width= 0.03\linewidth, angle=0]{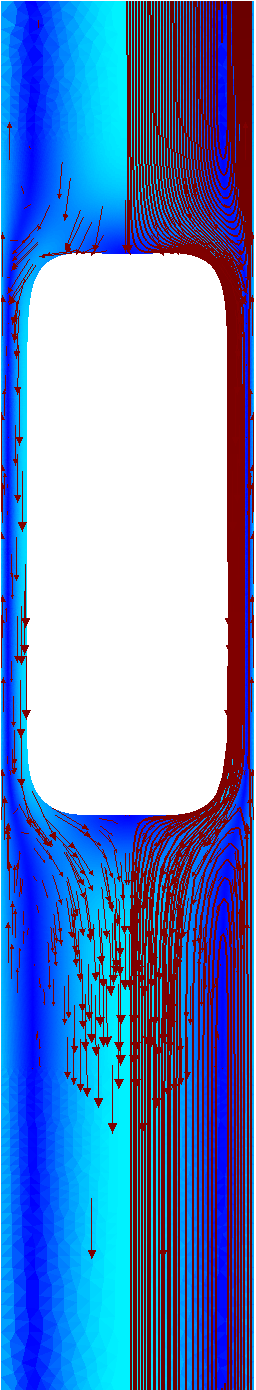}  &
  	\includegraphics[width= 0.03\linewidth, angle=0]{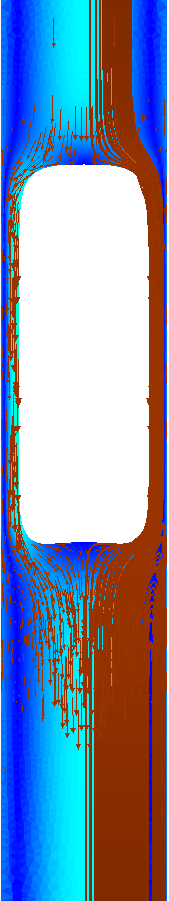}  &
  	\includegraphics[width=0.08 \linewidth, angle=0]{Figures/Eo000} 
\\
\includegraphics[width=0.07 \linewidth, angle=0]{Figures/Eo000} & 
 \includegraphics[width= 0.07\linewidth, angle=0]{Figures/Eo20} &  \includegraphics[width= 0.07 \linewidth, angle=0]{Figures/Eo60} & \includegraphics[width= 0.07 \linewidth, angle=0]{Figures/Eo100} & \includegraphics[width= 0.07 \linewidth, angle=0]{Figures/Eo140} &
\includegraphics[width= 0.07 \linewidth, angle=0]{Figures/Eo180} & \includegraphics[width= 0.07 \linewidth, angle=0]{Figures/Eo220} & \includegraphics[width= 0.07 \linewidth, angle=0]{Figures/Eo260} & \includegraphics[width= 0.07 \linewidth, angle=0]{Figures/Eo300} & \includegraphics[width=0.07 \linewidth, angle=0]{Figures/Eo000} 
\end{tabular}
\caption{The effect of $U_m$ and $Eo$ on the steady bubble shapes and flow fields with $Nf = 100$. In each panel, the streamlines and vector fields are superimposed on velocity magnitude pseudocolour plot on the right and left sides of the symmetry axis, respectively.}
\label{fig:ss_phasefield_steady_state_flow_Nf_100}
\end{figure}

We have also carried out a full parametric study of the effect of $U_m$ on the steady bubble shape and associated flow field for a wide range of $Nf$ and $Eo$. As shown in Figures \eqref{fig:ss_phasefield_steady_state_flow_Nf_40}-\eqref{fig:ss_phasefield_steady_state_flow_Nf_100}, a transition from downward to upward flow, characterised by a change in the sign of $U_m$ has a similar effect to an increase in $Eo$ for constant $Nf$ or a rise in $Nf$ with $Eo$ held fixed; this transition results in longer bubbles with more pointed noses and concave tails accompanied by wake formation for sufficiently large $Eo$ and/or $Nf$. For the lowest values of $Eo$ investigated, the bubbles develop bulges in the zone connecting the thin film and the bottom regions of the bubble which become more pronounced with increasingly negative $U_m$ values \textcolor{black}{(see Figure \ref{fig:ss_flowfield_Nf_80Eo20}). For sufficiently large and negative $U_m$, we see the emergence of bubbles with dimpled tops and/or bottoms, an indication of a steadily falling bubble, which is confirmed by the negative value of their rise velocity, see Figure \ref{fig:ss_flowfield_Nf_60Eo220}}.  

\section{Summary and conclusions} \label{sec:steady_state_summary}
Numerical solutions  of an axisymmetric Taylor bubble  moving steadily in stagnant and flowing liquids are computed by solving the steady-state Navier-Stokes equations using a Galerkin finite-element method based on kinematic update of the interface. Our validation of the numerical simulation strategy using the experimental data of \cite{Bugg_Saad_2002} shows a good agreement between the numerical results and the experiment. Utilising the strategy, we computed the steady-state shapes and evaluated the hydrodynamic features characterising the nose, film, interface, and bottom regions around the bubble for different dimensionless inverse viscosity numbers, E\"otv\"os, and Froude numbers based on the liquid centreline velocity. 

The results show that above $Eo \sim 100$, surface tension has insignificant influence on the hydrodynamic features studied. For the interval $Eo = \left(10, 30\right]$, analysis of the results indicates that the influence of increased $Nf$ results in a distinct feature that is not observed at higher $Eo$; emergence of a bulge in the film region close to the bubble bottom which becomes more pronounced and appears to propagates towards the nose as $Eo$ is decreased. Thus the intervals $Eo = \left( 20,30 \right]$ is considered as the limit below which surface tension has strong influence on Taylor bubble dynamics. Similarly, from the normalised frontal radius, we show that  interval $Nf = \left( 60,80 \right]$ can be considered as the limit below which viscous effects are significant. 

Based on our analysis of the normal stress at the interface, we deduced that it is the interaction  between the stresses due to curvature and viscosity that modifies the shape of the nose and bottom regions. In the bottom region, we made use of our results for the dependence of the bubble bottom shape and existence of the wake on $Nf$ and $Eo$ to produce a flow pattern map depicting regions of dimensionless parameters space that are associated with the presence or absence of wake formation together with the prevailing bubble bottom shape. 

Qualitative analysis of the effect of imposed liquid flow on the steady-state solution shows that the influence is more pronounced in the features that characterise the nose and bottom regions. For upward liquid flow, the nose becomes increasingly pointed and the bottom more concave as the liquid speed is increased. In contrast, increased downward liquid flow leads to the flattening of the bubble nose and increased convexity of the bubble bottom relative to the liquid. For sufficiently large speeds of downward-flowing liquids, it becomes difficult to distinguish the bubble nose and bottom regions which acquire very similar shapes as the bubble falls steadily.

Although we have obtained axisymmetric solutions for the parameter space investigated, it is uncertain whether some of the solutions, particular the ones associated with the downward-flowing liquid cases, are physically observable in experiments. In fact, experimental observations have shown that for certain downward liquid flow conditions, the shape of Taylor bubbles becomes asymmetric. 
In a companion paper, Part II of this two-part study \cite{Abubakar_Matar_2021}, we examine the linear stability of the axisymmetric steady-state solutions obtained here 
and determine the influence of $Nf$, $Eo$, and $U_m$ on the transition to asymmetry. 
In addition, we carry out an energy analysis in order to pinpoint the dominant destabilising mechanisms depending on the choice of parameter values.

\subsection*{Acknowledgements}

This work is supported by a Petroleum Technology Development Fund scholarship for HAA, and the Engineering \& Physical Sciences Research Council, United Kingdom, through the EPSRC MEMPHIS (EP/K003976/1) and PREMIERE (EP/T000414/1) Programme Grants. OKM also acknowledges funding via the PETRONAS/Royal Academy of Engineering Research Chair in Multiphase Fluid Dynamics. We also acknowledge HPC facilities provided by the Research Computing Service (RCS) of Imperial College London for the computing time.  \\

{\bf Declaration of interests:} The authors report no conflict of interest.


\appendix

\bibliographystyle{jfm}

\end{document}